\documentclass[12pt]{article}
\usepackage{amssymb, amsmath}
\usepackage{graphicx, longtable}
\usepackage{graphics}

\title{Charge-stripe phases versus a weak anisotropy \\of nearest-neighbor hopping}
\author{V. Derzhko, J. J{\c{e}}drzejewski\\
Institute of Theoretical Physics, University of Wroc{\l}aw,\\ pl.
Maksa Borna 9, 50--204 Wroc{\l}aw, Poland\\ {\small{e-mail:
derzhko@ift.uni.wroc.pl, jjed@ift.uni.wroc.pl}}}

\newcommand{\ie}{i.e.}
\newcommand{\nn}{n.n.}

\begin{document}
\maketitle
\begin{abstract}
Recently, we demonstrated rigorously the stability of charge-stripe
phases in quantum-particle systems that are described by extended
Falicov--Kimball Hamiltonians, with the quantum hopping particles
being either spinless fermions or hardcore bosons. In this paper, by
means of the same methods, we show that any anisotropy of
nearest-neighbor hopping eliminates the $\pi/2$-rotation degeneracy of
the so called dimeric and axial-stripe phases and orients them in the
direction of a weaker hopping.
Moreover, due to the same anisotropy the obtained
phase diagrams of fermions  show a tendency to become similar to
those of hardcore bosons.

\end{abstract}

\section{Introduction}

In our recent paper \cite{DJ2} we addressed the problem of
formation of static charge-stripe phases in systems of correlated
quantum particles. Since an experimental evidence for the existence
of striped phases in materials exhibiting high-temperature
superconductivity has been gathered \cite{tranquada1,tranquada2},
one observes vigorous discussions of the problem of the stability of
such phases in simple models of correlated electrons. Mainly, some
Hubbard-like or $t-J$-like models are considered in this context
(see a review paper by Ole{\'s} \cite{oles1}). Unfortunately, none
of the methods applied for investigating these models enables one to
control tiny energy differences between competing phases: the phase
energies are calculated by means of uncontrolled approximations,
moreover the results are biased by finite-size and boundary effects.
In this situation, to get a deeper insight into the stability problem of
striped phases, some simpler analogue problems
\cite{zhang1,zhang2,lemanski1,lemanski2} have been studied.

Adopting such an approach, we have proposed in Ref. \cite{DJ2} a
kind of an extended Falicov--Kimball model, whose ground-state phase
diagram at the half-filling and at a sufficiently strong coupling is
amenable to rigorous investigations -- a rather exceptional
situation in the field of correlated quantum-particle systems. From
a technical point of view, the standard spinless Falicov--Kimball
model (such as that studied in \cite{brandt,kennedy1}) has been
modified in order to allow for segregated phases, in the regime
specified above. That is, the Hamiltonian has been augmented by a
direct, Ising-like, nearest-neighbor (n.n.) attractive interaction
between the immobile particles. However from the point of view of
crystallization problem, where the immobile particles are
interpreted as ions \cite{kennedy1}, the added attractive
interaction is a physically motivated supplement of the hardcore
repulsion, which is already incorporated in the spinless
Falicov--Kimball model.

We have succeeded in proving that charge-stripe phases are stable in
some domains of phase diagrams. By varying slightly the interaction
parameter of the Ising-like n.n. interaction, the system is driven
from a crystalline (chessboard) phase to a segregated phase, via
quasi-one-dimensional striped phases.

In our studies in Ref. \cite{DJ2} and in this paper, we have
included also a subsidiary direct interaction between the immobile
particles, an Ising-like next nearest-neighbor (n.n.n.) interaction,
much weaker than the n.n. interaction. This interaction can
reinforce or frustrate the n.n. interaction, depending on the sign
of its interaction constant. We find it useful in discussing
similarities and differences between the cases of spinless fermions
and hardcore bosons.

The obtained results on the stability of striped phases open new
possibilities of investigating various characteristics of these
phases. Since, it is the role of a hopping anisotropy that has
received a considerable attention in the recent literature of the
subject \cite{oles1,normand1,normand2,oles2}, we have turned to this
problem, and in this paper we discuss the role of a weak anisotropy
of n.n. hopping.

The paper is organized as follows. In next section we define the
model studied in this paper, describe symmetries of its
grand-canonical phase diagrams, and introduce strong-coupling
expansions of the ground-state energy (effective Hamiltonians) and
the corresponding phase diagrams. Then, in Section 3, we carry out a
detailed analysis of phase diagrams due to truncated effective
Hamiltonians of orders not exceeding four, in presence of a weak
anisotropy of n.n. hopping. In Section 4, we draw conclusions and
provide a summary. Various technical details a placed in Appendices
A, B, and C.

\section{The model and its ground-state phase diagram --- basic properties}

\subsection{The model}

The model to be studied is a simplified version of the one band,
spin $1/2$ Hubbard model, known as the static approximation (one
sort of electrons hops while the other sort is immobile), augmented
by a direct Ising-like interaction $V$ between the immobile
particles. Only a n.n. hopping is taken into account and we allow
for its anisotropy. The total Hamiltonian of the system reads:

\begin{eqnarray}
H_{0}  =  H_{FK} + V,
\label{H0} \\
 H_{FK}=-t_{h} \sum\limits_{\langle x,y \rangle_{1,h}} \, \left(
c^{+}_{x}c_{y} + c^{+}_{y}c_{x} \right) -t_{v} \sum\limits_{\langle
x,y \rangle_{1,v}} \, \left( c^{+}_{x}c_{y} + c^{+}_{y}c_{x}
\right)+ U\sum\limits_{x}\left(c^{+}_{x}c_{x} - \frac{1}{2} \right)
s_{x},
\label{HFK}\\
V=\frac{W}{8} \sum\limits_{\langle x,y \rangle_{1}} s_{x}s_{y}
-\frac{\tilde{\varepsilon}}{16} \sum\limits_{\langle x,y
\rangle_{2}} s_{x}s_{y}.
\label{V}
\end{eqnarray}

In the above formulae, the underlying lattice is a square lattice,
denoted $\Lambda$, consisting of sites $x, y, \ldots$, whose number
is $|\Lambda|$, having the shape of a $\sqrt{|\Lambda|} \times
\sqrt{|\Lambda|}$ torus. In (\ref{HFK},\ref{V}) and below, the sums
$\sum_{\langle x,y \rangle_{i,h}}$, or $\sum_{\langle x,y
\rangle_{i,v}}$, $i=1,2,3$, stand for the summation over all the
$i$-th order \nn{} pairs of lattice sites in $\Lambda$, directed
horizontally (h) or vertically (v), with each pair counted once.

The subsystem of mobile spinless particles (here-after called
{\em the electrons})
is described in terms of creation and annihilation operators of an
electron at site $x$, $c^{+}_{x}$, $c_{x}$, respectively, satisfying
the canonical anticommutation relations (spinless fermions) or the
commutation relations of spin $1/2$ operators $S^{+}_{x}, S^{-}_{x},
S^{z}_{x}$ (hardcore bosons). The total electron-number operator is
$N_{e} = \sum_{x} c^{+}_{x} c_{x}$, and (with a little abuse of
notation) the corresponding electron density is $\rho_{e} = N_{e}
/|\Lambda|$. There is no direct interaction between the mobile
particles. Their energy is due to the anisotropic hopping, with
$t_{h}$ ($t_{v}$) being the \nn{} horizontal (vertical) hopping
intensity, and due to the interaction with the localized particles,
whose strength is controlled by the coupling constant $U$.

The Hamiltonian $H_{FK}$ is well-known as the Hamiltonian of the
spinless Falicov--Kimball model, a simplified version of the
Hamiltonian put forward in \cite{FK}. A review of rigorous results
concerning this model and an extensive list of relevant references
can be found in \cite{GM,JL,GU}).

The subsystem of localized particles (here-after called
{\em the ions}),
is described by a collection of pseudo-spins $\left\{ s_{x}
\right\}_{x \in \Lambda}$, with $s_{x} = 1, -1$ ($s_{x}=1$ if the
site $x$ is occupied by an ion and $s_{x}=-1$ if it is empty),
called the {\em ion configurations}. The total number of ions is
$N_{i} = \sum_{x} ( s_{x} + 1 )/2$ and the ion density is $\rho_{i}
= N_{i}/|\Lambda|$. In our model the ions interact directly by means
of the isotropic Ising-like interaction $V$.

Clearly, in the composite system, whose Hamiltonian is given by
(\ref{H0}) with an arbitrary electron-ion  coupling $U$,
the particle-number operators $N_{e}$, $N_{i}$, and pseudo-spins
$s_{x}$, are conserved. Therefore, the description of the classical
subsystem in terms of the ion configurations $S =\left\{ s_{x}
\right\}_{x \in \Lambda}$ remains valid. Whenever periodic
configurations of pseudo-spins are considered, it is assumed that
$\Lambda$ is sufficiently large, so that it accommodates an integer
number of elementary cells.

\subsection{The ground-state phase diagram in the grand-canonical
ensemble}

In what follows, we shall study the ground-state phase diagram of
the system defined by (\ref{H0}) in the grand-canonical ensemble.
That is, let
\begin{equation}
H \left( \mu_{e}, \mu_{i} \right) = H_{0} - \mu_{e}N_{e} -
\mu_{i}N_{i}, \label{Hmu}
\end{equation}
where $\mu_{e}$, $\mu_{i}$ are the chemical potentials of the
electrons and ions, respectively, and let $E_{S}\left(
\mu_{e}, \mu_{i} \right)$ be the ground-state energy of $H \left(
\mu_{e}, \mu_{i} \right)$, for a given configuration $S$ of the
ions. Then, the ground-state energy of $H \left( \mu_{e}, \mu_{i}
\right)$, $E_{G}\left( \mu_{e}, \mu_{i} \right)$,  is defined as
$E_{G}\left( \mu_{e}, \mu_{i} \right) = \min \left\{ E_{S} \left(
\mu_{e}, \mu_{i} \right): S \right\}$. The minimum is attained at
the set $G$ of the ground-state configurations of ions. We shall
determine the subsets of the space of the Hamiltonian's energy parameters,
where $G$ consists of periodic configurations of ions,
uniformly in the size of the underlying square lattice.

In studies of grand-canonical phase diagrams an important role is
played by unitary transformations ({\em hole--particle
transformations}) that exchange particles and holes, $c^{+}_{x}c_{x}
\rightarrow 1 - c^{+}_{x}c_{x}$, $s_{x} \rightarrow -s_{x}$, and for
some $\left( \mu^{0}_{e}, \mu^{0}_{i} \right)$ leave the Hamiltonian
$H \left( \mu_{e}, \mu_{i} \right)$ invariant. For the mobile
particles such a role is played by the transformations: $c_{x}^{+}
\rightarrow \epsilon_{x} c_{x}$, where $\epsilon_{x} = 1$ for bosons
while for fermions $\epsilon_{x} = 1$ at the even sublattice of
$\Lambda$ and $\epsilon_{x} = -1$ at the odd one. Clearly, since
$H_{0}$ is invariant under the joint hole--particle transformation
of mobile and localized particles, $H \left( \mu_{e}, \mu_{i}
\right)$ is hole--particle invariant at the point $(0,0)$. At the
hole--particle symmetry point, the system under consideration has
very special properties, which simplify studies of its phase diagram
\cite{kennedy1}. Moreover, by means of the defined above
hole-particle transformations one can determine a number of
symmetries of the grand-canonical phase diagram \cite{GJL}. The
peculiarity of the model is that the case of attraction ($U<0$) and
the case of repulsion ($U>0$) are related by a unitary
transformation (the hole-particle transformation for ions): if $S
=\left\{ s_{x} \right\}_{x \in \Lambda}$ is a ground-state
configuration at $\left( \mu_{e}, \mu_{i} \right)$ for $U>0$, then
$-S =\left\{- s_{x} \right\}_{x \in \Lambda}$ is the ground-state
configuration at $\left( \mu_{e}, -\mu_{i} \right)$ for $U<0$.
Consequently, without any loss of generality one can fix the sign of
the coupling constant $U$. Moreover (with the sign of $U$ fixed),
there is an {\em inversion symmetry\/} of the grand-canonical phase
diagram, that is, if $S$ is a ground-state configuration at $\left(
\mu_{e}, \mu_{i} \right)$, then $-S$ is the ground-state
configuration at $\left( -\mu_{e}, -\mu_{i} \right)$. Therefore, it
is enough to determine the phase diagram in a half-plane specified
by fixing the sign of one of the chemical potentials.

Our aim in this paper is to investigate the ground-state phase
diagrams of our systems, for general values of the energy
parameters that appear in $H \left( \mu_{e}, \mu_{i} \right)$.
According to the state of art, this is feasible only in the {\em
strong-coupling regime}, \ie{} when $|t/U|$ is sufficiently small.
Therefore, from now on we shall consider exclusively the case of a
large positive coupling $U$, and we express all the parameters of
$H \left( \mu_{e}, \mu_{i} \right)$ in the units of $U$,
preserving the previous notation.

\subsection{The ground-state energy and phase diagram
in the strong-coupling regime}

In the strong-coupling regime, the ground-state energy $E_{S} \left(
\mu_{e}, \mu_{i} \right)$ can be expanded in a power series in $t$.
One of the ways to achieve this, for fermions and for hardcore
bosons, is a method of unitarily equivalent interactions \cite{DFF}.
The result, with the expansion terms up to order four shown
explicitly, reads:

\begin{eqnarray}
\label{Expfmn}
E^{f}_{S}\left(\mu_e, \mu_i \right) &=&
\left(E^{f}_{S}\right)^{(4)}\left(\mu_e, \mu_i \right) + \left(R_{S}^{f}\right)^{(4)},
\nonumber \\
\left(E^{f}_{S}\right)^{(4)}\left(\mu_e, \mu_i \right) &=&
-\frac{1}{2}\left(\mu_i -\mu_e \right)
\sum\limits_{x} \left( s_{x} + 1 \right)+
\left[ \frac{t^{2}}{4}- \frac{3t^{4}}{16}-\frac{3}{8}\gamma
t^{4} + \frac{W}{8} \right] \sum\limits_{\langle x,y \rangle_{1,h}}
s_{x}s_{y} \nonumber \\
&+& \left[ \gamma \frac{t^{2}}{4} -\frac{3}{8}\gamma t^{4}
- \gamma^{2}\frac{3t^{4}}{16} + \frac{W}{8} \right]
\sum\limits_{\langle x,y \rangle_{1,v}} s_{x}s_{y}+
\nonumber \\
&+& \left[ \gamma \frac{3t^{4}}{16}-\frac{\tilde{\varepsilon}}{16}
\right] \sum\limits_{\langle x,y \rangle_{2}} s_{x}s_{y} +
\frac{t^{4}}{8} \sum\limits_{\langle x,y \rangle_{3,h}}
s_{x}s_{y}+
\nonumber
\\
&+& \gamma^{2} \frac{t^{4}}{8} \sum\limits_{\langle x,y
\rangle_{3,v}} s_{x}s_{y} + \gamma \frac{t^{4}}{16} \sum\limits_{P}
\left(1+5s_{P}\right),
\end{eqnarray}
in the case of hopping fermions, and
\begin{eqnarray}
\label{Exphcb}
E^{b}_{S}\left(\mu_e, \mu_i \right) &=&
\left(E^{b}_{S}\right)^{(4)}\left(\mu_e, \mu_i \right) + \left(R_{S}^{b}\right)^{(4)},
\nonumber \\
\left(E^{b}_{S}\right)^{(4)}\left(\mu_e, \mu_i \right) &=&
-\frac{1}{2}\left(\mu_i -\mu_e \right)
\sum\limits_{x} \left( s_{x} + 1 \right)+
\left[ \frac{t^{2}}{4}- \frac{3t^{4}}{16} -\frac{1}{8}\gamma
t^{4} + \frac{W}{8} \right] \sum\limits_{\langle x,y \rangle_{1,h}}
s_{x}s_{y} + \nonumber \\
&+&  \left[ \gamma \frac{t^{2}}{4} -\frac{1}{8}\gamma t^{4}
- \gamma^{2}\frac{3t^{4}}{16} + \frac{W}{8} \right]
\sum\limits_{\langle x,y \rangle_{1,v}} s_{x}s_{y}+
\nonumber \\
&+& \left[ \gamma \frac{5t^{4}}{16}-\frac{\tilde{\varepsilon}}{16}
\right] \sum\limits_{\langle x,y \rangle_{2}} s_{x}s_{y} +
\frac{t^{4}}{8} \sum\limits_{\langle x,y \rangle_{3,h}}
s_{x}s_{y} +
\nonumber
\\
&+& \gamma^{2} \frac{t^{4}}{8} \sum\limits_{\langle x,y
\rangle_{3,v}} s_{x}s_{y} - \gamma \frac{t^{4}}{16} \sum\limits_{P}
\left(5+s_{P}\right),
\end{eqnarray}
in the case of hopping hardcore bosons. The both of the above
expressions are given  up to a term independent of the ion
configurations and the chemical potentials. In (\ref{Expfmn}) and
(\ref{Exphcb}), we have set $t_{h} \equiv t$, and $t_{v} =
\sqrt{\gamma}t$, with  $0 \leq \gamma \leq 1$, $P$ denotes a
$(2 \times 2)$-sites plaquette of the square lattice $\Lambda$,
$s_{P}$ stands for the product of pseudo-spins assigned to the corners of $P$.
The remainders, $\left(R_{S}^{f}\right)^{(4)}$ and $\left(R_{S}^{b}\right)^{(4)}$,
which are independent of the chemical potentials and the parameters $W$ and
$\tilde{\varepsilon}$, but dependent on $\gamma$,
collect those terms of the expansions that
are proportional to $t^{2m}$, with $m=3,4,\ldots$. It can be proved
that the above expansions  are absolutely convergent, uniformly in
$\Lambda$, provided that $t < 1/16$ and $\left| \mu_{e} \right| < 1
- 16t$ \cite{GMMU}. Moreover, under these conditions the
ground-state densities of particles satisfy the half-filling relation:
$\rho_{e} + \rho_{i} =1$.

We note that on taking into account the inversion symmetry of the
phase diagram in the $\left( \mu_{e}, \mu_{i} \right)$-plane and the
fact that the ground-state energies depend only on the difference of
the chemical potentials, in order to determine the phase diagram in
the stripe $\left| \mu_{e} \right| < 1 - 16t$ it is enough to
consider the phase diagram at the half-line $\mu_{e}=0$, $\mu_{i} <
0$ (or $\mu_{i} >0$). At this half-line we set $\mu_{i} \equiv \mu$.

\section{Phase diagrams according to truncated expansions}

Due to the convergence of the expansions (\ref{Expfmn}) and
(\ref{Exphcb}), it is possible to establish rigorously a part of the
phase diagram, that is the ground-state configurations of ions are
determined in the space of the Hamiltonian's energy parameters
everywhere, except some small regions. This is accomplished by
determining the ground-state phase diagram of the expansion
truncated at the order $k$ ({\em the k-th order phase diagram}),
that is according to the $k$-th order
effective Hamiltonians $(E^{f}_{S})^{(k)} \left( 0, \mu
\right)$ and $(E^{b}_{S})^{(k)} \left( 0, \mu \right)$. The
remainder of the expansion, that consists of a term of the next
order, $k^{\prime}>k$, and other higher-order terms, cannot modify the
obtained phase diagram, except some narrow regions along the phase
boundaries whose width is of the order $k^{\prime}$. The procedure
of constructing the phase diagram is recursive: the phase diagram of
the effective interaction truncated at the order $k^{\prime}$ is
constructed on the basis of the phase diagram obtained at the
preceding order $k$: the conditions imposed on the ground-state
configurations by the $k$-th order effective Hamiltonian have to be
obeyed by the ground-state configurations of the $k^{\prime}$-th
order effective Hamiltonian. In other words, the $k^{\prime}$-th
order terms of the expansion cannot change the hierarchy of
configuration's energies established by the $k$-th order effective
Hamiltonian; they can only split the energies of configurations in
cases of degeneracy.
At each step of the construction, we use the $m$-potential method introduced
in \cite{Slawny}, with technical developments described in
\cite{GJL,kennedy2,GMMU}. For details specific to this paper see
also Ref.\cite{DJ2} and  Appendix A.

In Ref.\cite{DJ2} we have obtained the ground-state phase diagrams,
according to the fourth-order isotropic effective Hamiltonians
$\left. (E^{f}_{S})^{(4)} \left( 0, \mu \right) \right|_{\gamma=1}$
and $\left.(E^{b}_{S})^{(4)} \left( 0, \mu \right) \right|_{\gamma=1}$,
for hole-particle symmetric systems ($\mu =0$) with a weak (of fourth order)
n.n.n. subsidiary interaction, and for unsymmetrical systems
without the subsidiary interaction ($\tilde{\varepsilon}=0$), see the top
phase diagrams in Figs.~\ref{sppd},\ref{epspdf},\ref{epspdb}.
Second-order phase diagrams of Ref.\cite{DJ2} consist exclusively of phases
whose configurations are invariant with respect
to $\pi/2$-rotations, with {\em macroscopic degeneracies}
(i.e. the number of configurations grows exponentially with the number of sites)
at the boundaries of phase domains.
Stripe phases, whose configurations are not invariant with respect
to $\pi/2$-rotations, appear on perturbing the second-order
phase diagrams by the fourth-order isotropic interactions.
Here, we would like to observe the influence of a weak anisotropy of n.n. hopping
on these stripe phases.
Since we are working with truncated effective Hamiltonians,
we have to assign an order to the deviation of the anisotropy parameter $\gamma$
from the value $1$ (corresponding to the isotropic case). Therefore, we
introduce an anisotropy order, $a$, and a new anisotropy parameter,
$\beta_{a}$:
\begin{eqnarray}
\gamma=1-\beta_{a}t^{a}.
\end{eqnarray}
The orders of the deviations can be neither too small,
not to modify the second-order effective Hamiltonians, nor too high, to effect
the considered effective Hamiltonian of the highest order.
Since here, the highest order of the effective Hamiltonians is $k=4$,
the weakest admissible deviation from the isotropic case
corresponds to the highest anisotropy order $a=2$.
Then, we can consider an intermediate deviation, i.e. $0<a<2$.
The strongest deviation, i.e. $a=0$, is not admissible,
since it modifies the second-order effective Hamiltonians.

\subsection{The smallest deviation from the isotropic case}

In the sequel, we drop the arguments of ground-state energies, that is
we set $\left( E^{f}_{S} \right)^{(4)}\left(0, \mu \right) \equiv
\left( E^{f}_{S} \right)^{(4)}$, etc.
In the case of the smallest deviation from the isotropic case
the fourth-order effective Hamiltonian for fermions reads:
\begin{eqnarray}
\label{E4sd}
\left( E^{f}_{S} \right)^{(4)} &=&
\left.\left( E^{f}_{S} \right)^{(4)}\right|_{\gamma=1}
- \beta_{2} \frac{t^{4}}{4}
\sum\limits_{\langle x,y \rangle_{1,v}} s_{x}s_{y},
\end{eqnarray}
while for hardcore bosons only the first term, representing the
isotropic fourth-order effective Hamiltonian, has to be changed properly.

\begin{figure}[p]
\centering \includegraphics[width=\textwidth]{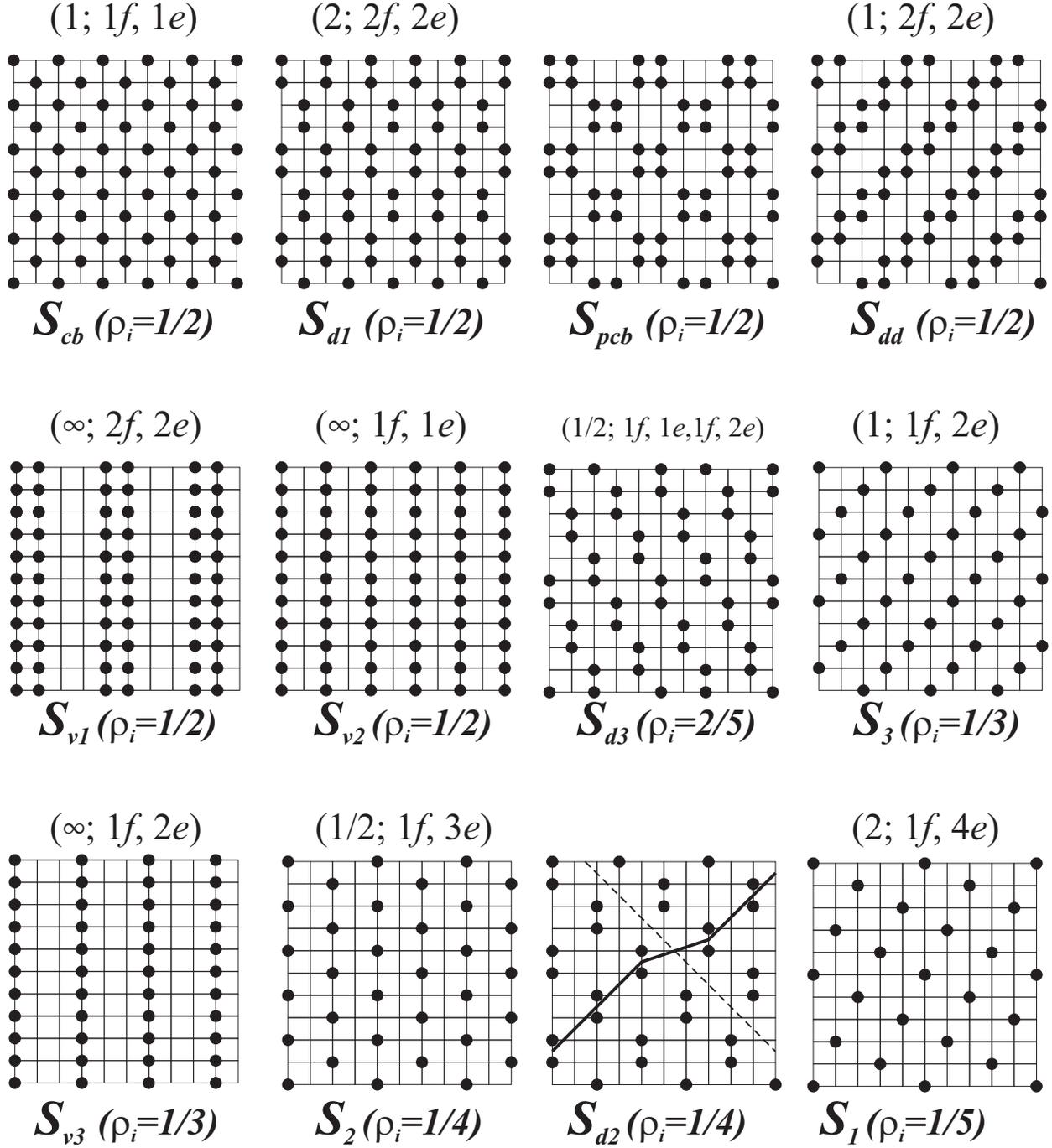}
\caption{\scriptsize{The representative configurations of the phases
that appear in the phase diagrams of $\left. H_0\right|_{\gamma=1}$.
The remaining configurations can be
obtained by applying the spatial symmetries of $\left. H_0\right|_{\gamma=1}$.
As a representative configuration of the set ${\mathcal{S}}_{d2}$, whose
degeneracy grows like $\exp{(const \sqrt{\Lambda})}$,
we show a configuration with one defect
line (the dashed line); the continuous line is a guide for the eye.
Explanations of the symbols on the top of the representative
configurations and other comments are given in Section 3.
}}
\label{phases1}
\end{figure}

Apparently, the effective Hamiltonians of order zero and two are
isotropic, and there is no difference between the cases of
hopping fermions and hopping bosons.
The corresponding phase diagrams, determined in Ref.\cite{DJ2},
are built of three {\em phases}, i.e. sets of configurations with
the same energy in some open regions of phase diagrams, called
{\em phase domains},
${\mathcal{S}}_{+}$, ${\mathcal{S}}_{-}$, and ${\mathcal{S}}_{cb}$.
The first two phases consist of single translation invariant
configurations, one completely filled with ions and one completely
empty, respectively; a representative configuration of the
chessboard phase, ${\mathcal{S}}_{cb}$, is shown in Fig.~\ref{phases1}.
The three named phases coexist at the open half-line
$W=-2t^{2}$, $\mu =0$, $\tilde{\varepsilon}>0$.
However, if $\tilde{\varepsilon}$ attains zero, the finite
(independent of the size of $\Lambda$) degeneracy of the open half-line
changes into a macroscopic one.
It is the point $W=-2t^{2}$, $\mu =0$, $\tilde{\varepsilon}=0$
around which the fourth-order diagrams,
containing various stripe phases, have been constructed in Ref.\cite{DJ2}.
Now, to observe the effect of the fourth-order anisotropy term of
(\ref{E4sd}) the construction of the fourth-order diagrams has to be
carried out again.
As in the isotropic case, this is facilitated by introducing new variables,
$\omega$, $\delta$, and $\varepsilon$:
\begin{equation}
W=-2t^{2}+t^{4} \omega, \qquad
\mu=t^{4}\delta, \qquad
\tilde{\varepsilon}=t^{4}\varepsilon,
\end{equation}
and rewriting the fourth order effective Hamiltonian in the form,
\begin{eqnarray}
\left(E^{f}_{S} \right)^{(4)} =
\frac{t^{4}}{2} \sum\limits_{T} \left( H^{f}_{T}
\right)^{(4)},
\end{eqnarray}
where
\begin{eqnarray}
\left( H^{f}_{T} \right)^{(4)} =
\left.\left( H^{f}_{T} \right)^{(4)} \right|_{\gamma=1} -
\frac{\beta_{2}}{12} \sideset{}{''}\sum\limits_{\langle x,y
\rangle_{1,v}} s_{x}s_{y},
\label{HTf4}
\end{eqnarray}
with analogous expressions in the bosonic case, and with the isotropic
potentials $\left.\left( H^{f}_{T} \right)^{(4)} \right|_{\gamma=1}$,
$\left.\left( H^{b}_{T} \right)^{(4)} \right|_{\gamma=1}$
given by
\begin{eqnarray}
\left.\left( H^{f}_{T} \right)^{(4)} \right|_{\gamma=1}&=&
-\delta \left( s_{5} +1
\right) +\frac{1}{24} \left( \omega-\frac{9}{2} \right)
\sideset{}{''}\sum\limits_{\langle x,y \rangle_{1}} s_{x}s_{y}
 + \frac{1}{32} \left(3-\varepsilon \right)
\sideset{}{''}\sum\limits_{\langle x,y \rangle_{2}} s_{x}s_{y} +
\nonumber \\
&+& \frac{1}{12} \sideset{}{''}\sum\limits_{\langle x,y \rangle_{3}}
s_{x}s_{y} + \frac{1}{32} \sideset{}{''}\sum\limits_{P} \left(
5s_{P}+1 \right),
\label{HTf41}
\end{eqnarray}
\begin{eqnarray}
\left.\left( H^{b}_{T} \right)^{(4)} \right|_{\gamma=1}&=&
-\delta \left( s_{5} +1
\right) +\frac{1}{24} \left( \omega-\frac{5}{2} \right)
\sideset{}{''}\sum\limits_{\langle x,y \rangle_{1}} s_{x}s_{y}
 + \frac{1}{32} \left(5-\varepsilon \right)
\sideset{}{''}\sum\limits_{\langle x,y \rangle_{2}} s_{x}s_{y} +
\nonumber \\
&+& \frac{1}{12} \sideset{}{''}\sum\limits_{\langle x,y \rangle_{3}}
s_{x}s_{y} - \frac{1}{32} \sideset{}{''}\sum\limits_{P} \left(
5+s_{P} \right).
\label{HTb41}
\end{eqnarray}

In the above formulae $T$ stands for a $(3 \times 3)$-sites
plaquette of a square lattice, later on called {\em the
$T$-plaquette}, and $s_5$ stands for the central site of a
$T$-plaquette. The summation in double-primed sums runs over a
$T$-plaquette.

We would like to get an idea of the phase diagram in the space of the
four energy parameters $(\omega , \varepsilon , \beta_{2}, \delta )$,
that appear in the Hamiltonian. Due to the fact that the domains
occupied by the phases are polyhedral sets (see Appendix A), this goal
can be achieved by studying phase diagrams in two-dimensional hyperplanes.
Of particular interest are those hyperplanes that result from intersecting
the four-dimensional space by hyperplanes $\delta=0$ and $\beta_2 = const$,
and by hyperplanes $\varepsilon=0$ and $\beta_2 = const$.
A collection of such sections for suitable values of  $\beta_2$,
forming a finite increasing from zero sequence,
enable us to observe how the anisotropy effects our system
if it is hole-particle symmetric and if it is not, respectively.

\begin{figure}[p]
\begin{center}
\includegraphics[totalheight=0.3\textwidth,origin=c]{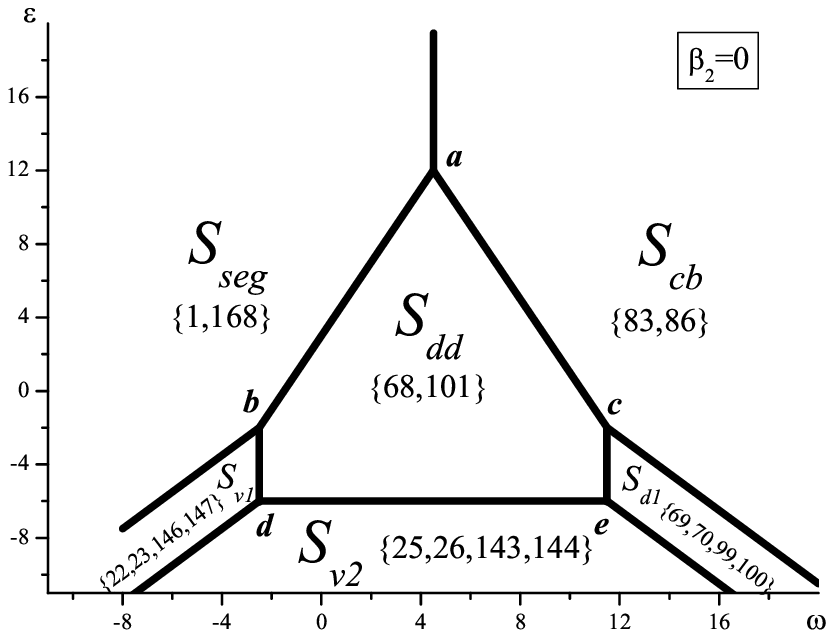}
\includegraphics[totalheight=0.3\textwidth,origin=c]{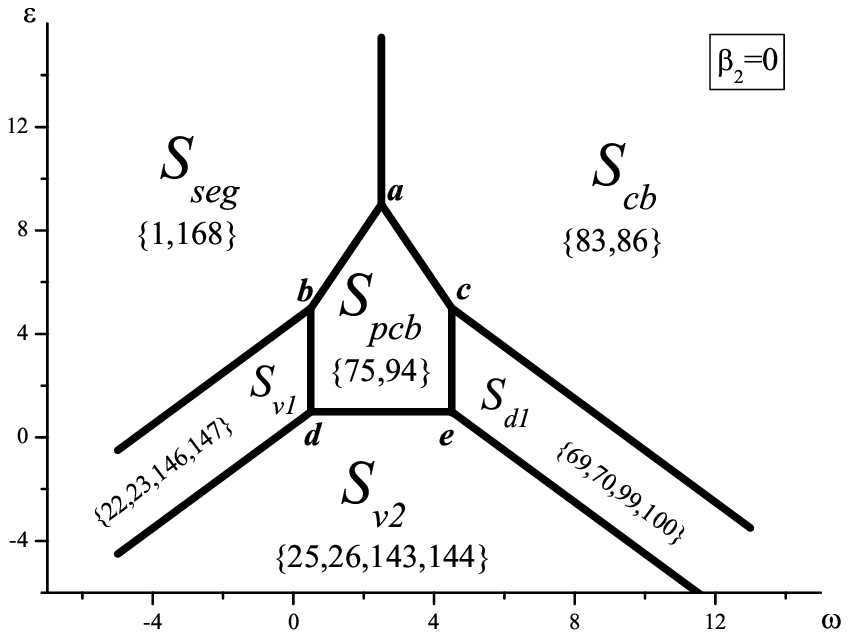}\\
\includegraphics[totalheight=0.3\textwidth,origin=c]{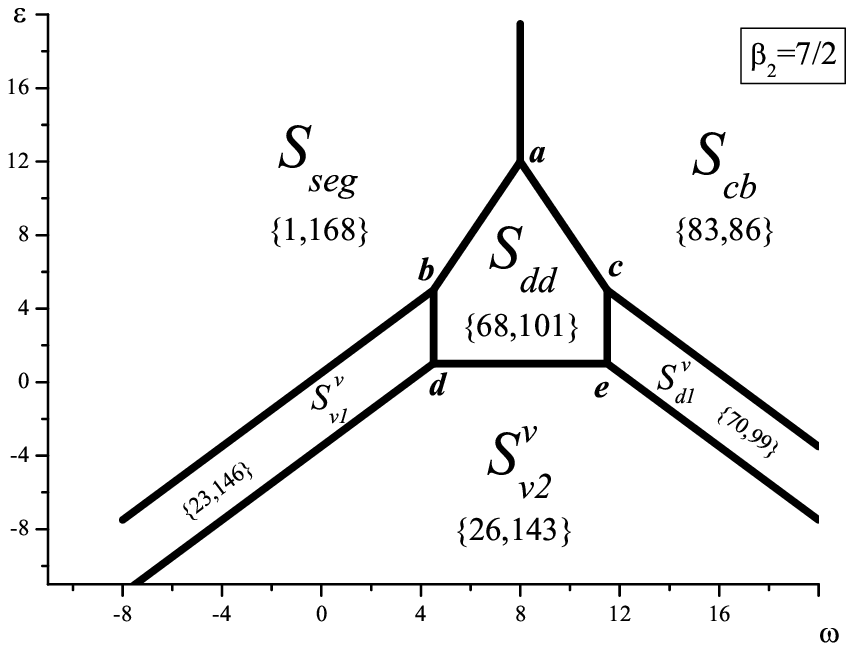}
\includegraphics[totalheight=0.3\textwidth,origin=c]{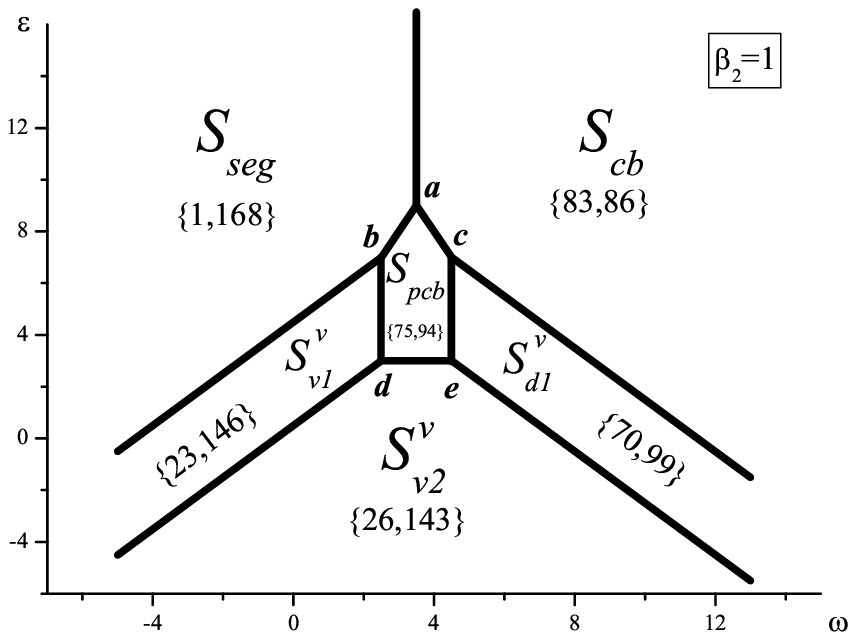}\\
\includegraphics[totalheight=0.3\textwidth,origin=c]{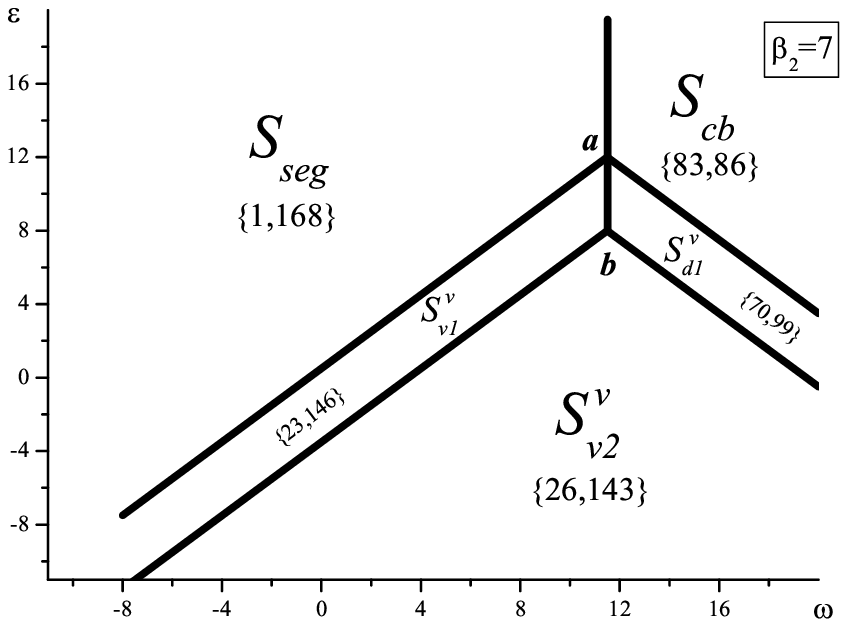}
\includegraphics[totalheight=0.3\textwidth,origin=c]{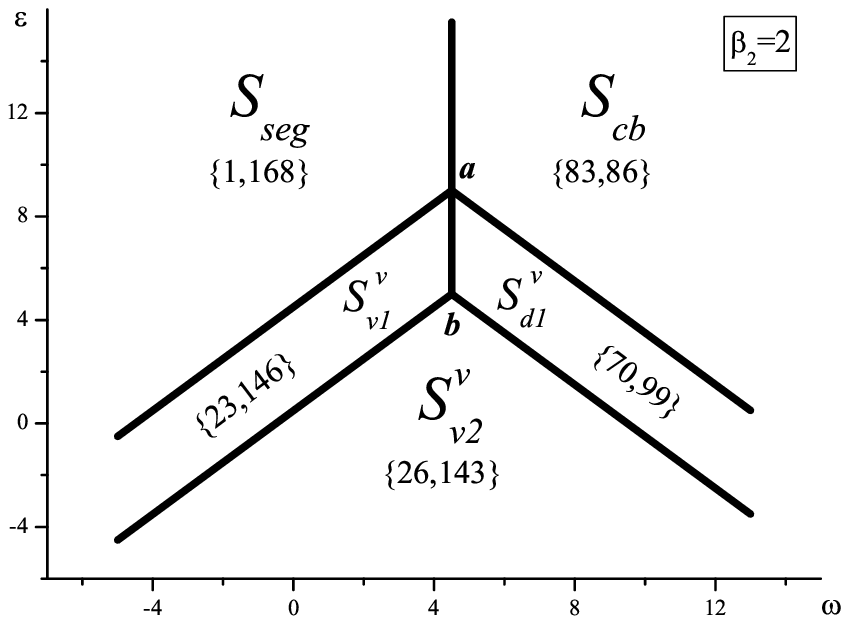}\\
\includegraphics[totalheight=0.3\textwidth,origin=c]{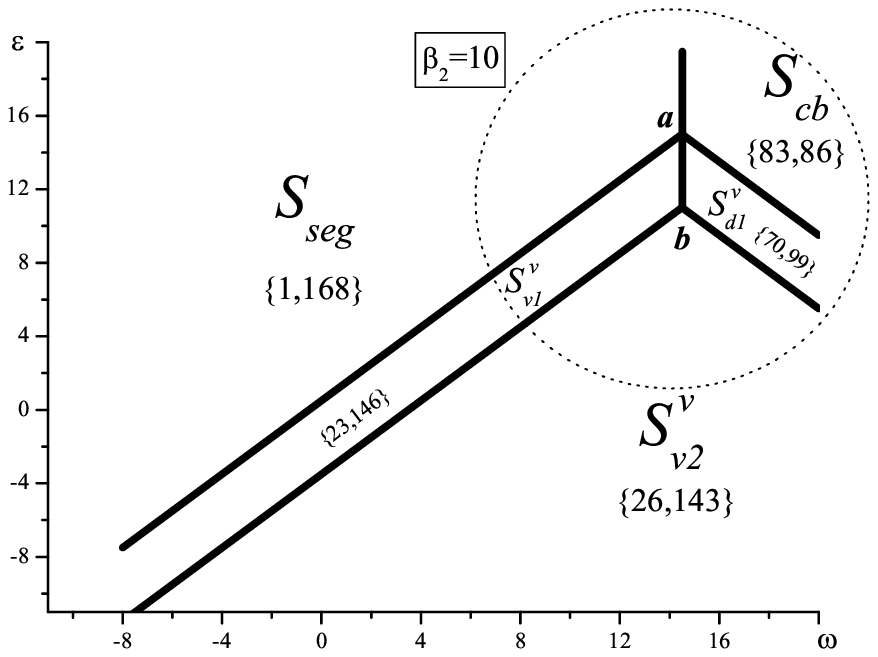}
\includegraphics[totalheight=0.3\textwidth,origin=c]{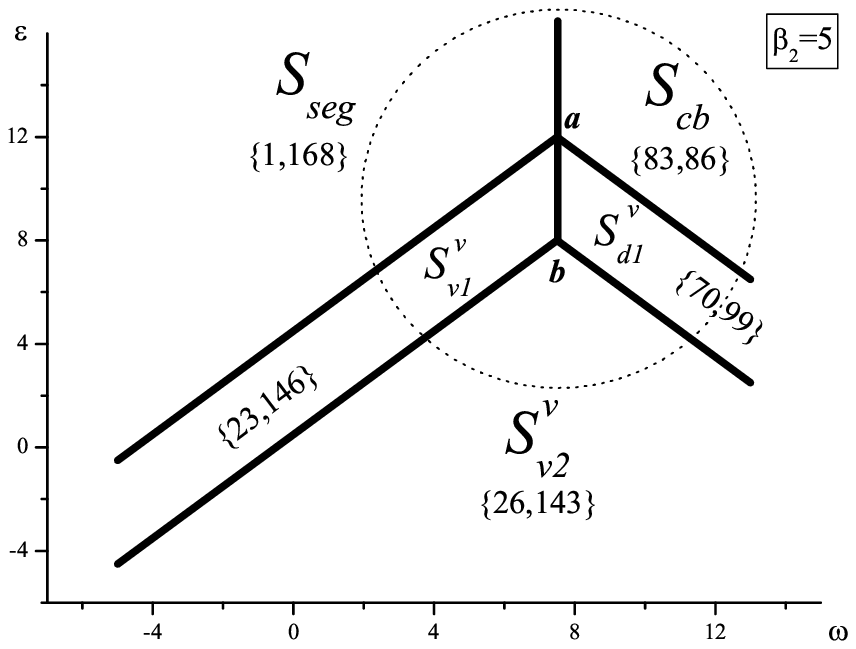}
\end{center}
\caption{\scriptsize{The case of the weakest anisotropy and of
hole-particle symmetric systems ($\mu =0$). Phase diagrams of
$\left(H_{T}^{f}\right)^{(4)}$ (given by (\ref{HTf4})) --- left
column, and $\left(H_{T}^{b}\right)^{(4)}$ --- right column, for an
increasing sequence of values of $\beta_2$.
The representative ion
configurations of the displayed phases are shown in
Fig.~\ref{phases1} (for more comments see Section 3).
For fermions,
the critical value is $\beta_2=7$, while for bosons
it is $\beta_2 =2$.
The equations defining the boundary lines of
the phase domains are given in Tab.~\ref{tb1} and Tab.~\ref{tb2} of
Appendix B, while the corresponding zero-potential coefficients
$\{\alpha_{i} \}$ --- in Tab.~\ref{tb8} -- Tab.~\ref{tb15} of Appendix
C. In the bottom diagrams, the regions surrounded by dotted circles,
are reconsidered in the case of an intermediate anisotropy.
}}
\label{sppd}
\end{figure}

\begin{figure}[p]
\begin{center}
\includegraphics[totalheight=0.3\textwidth,origin=c]{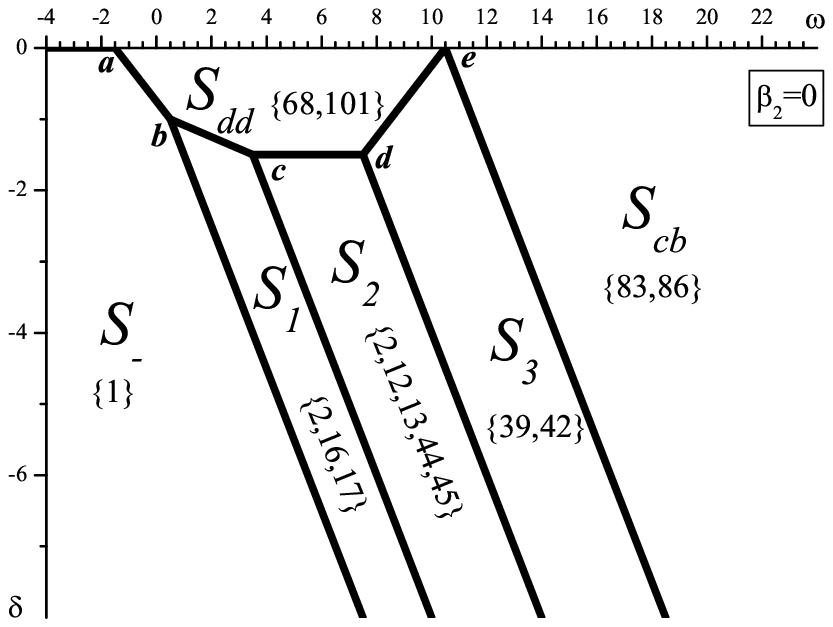}
\includegraphics[totalheight=0.3\textwidth,origin=c]{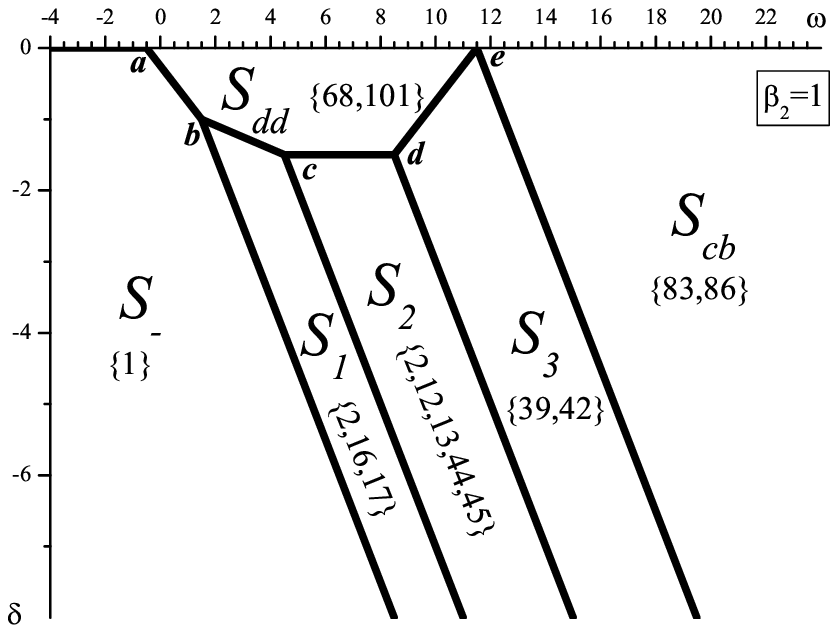}\\
\includegraphics[totalheight=0.3\textwidth,origin=c]{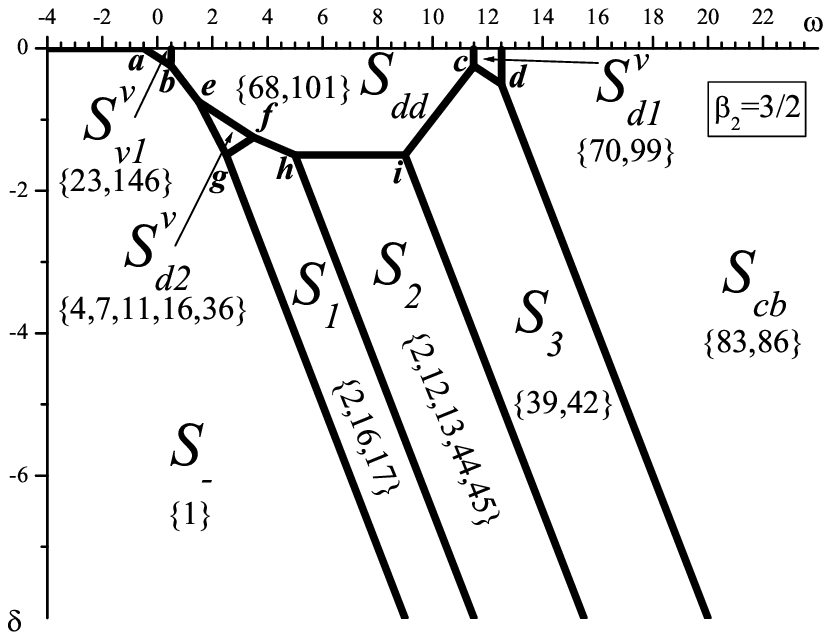}
\includegraphics[totalheight=0.3\textwidth,origin=c]{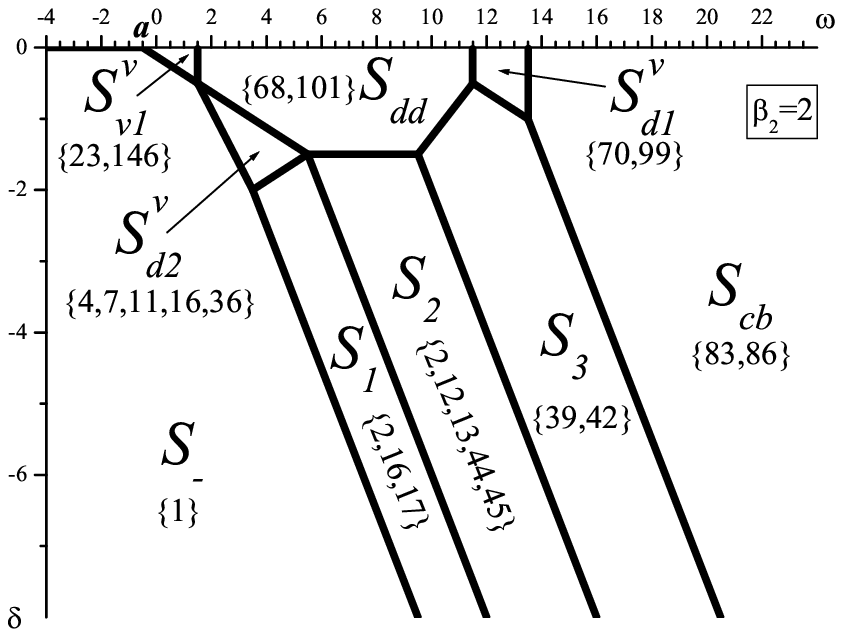}\\
\includegraphics[totalheight=0.3\textwidth,origin=c]{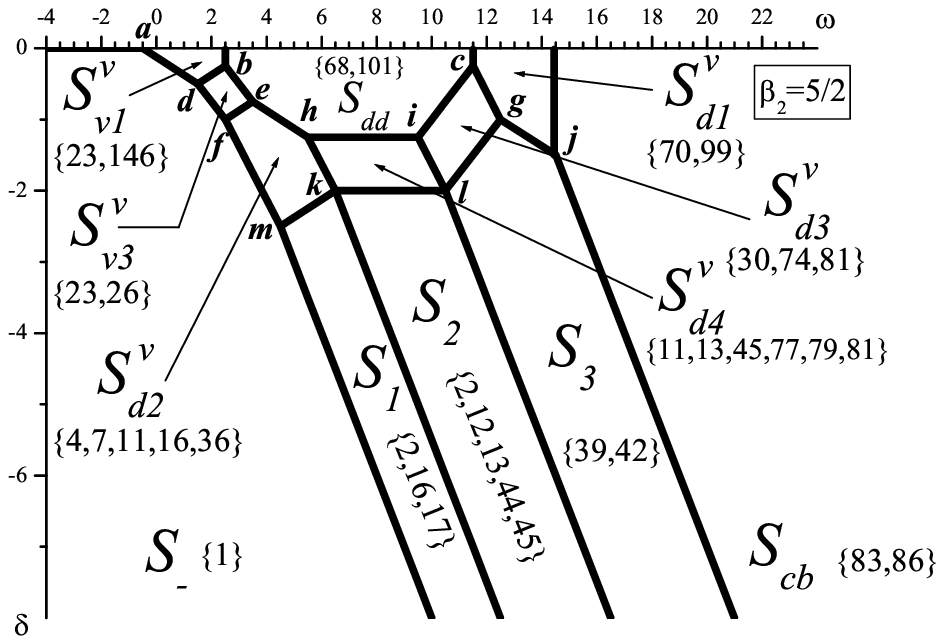}
\includegraphics[totalheight=0.3\textwidth,origin=c]{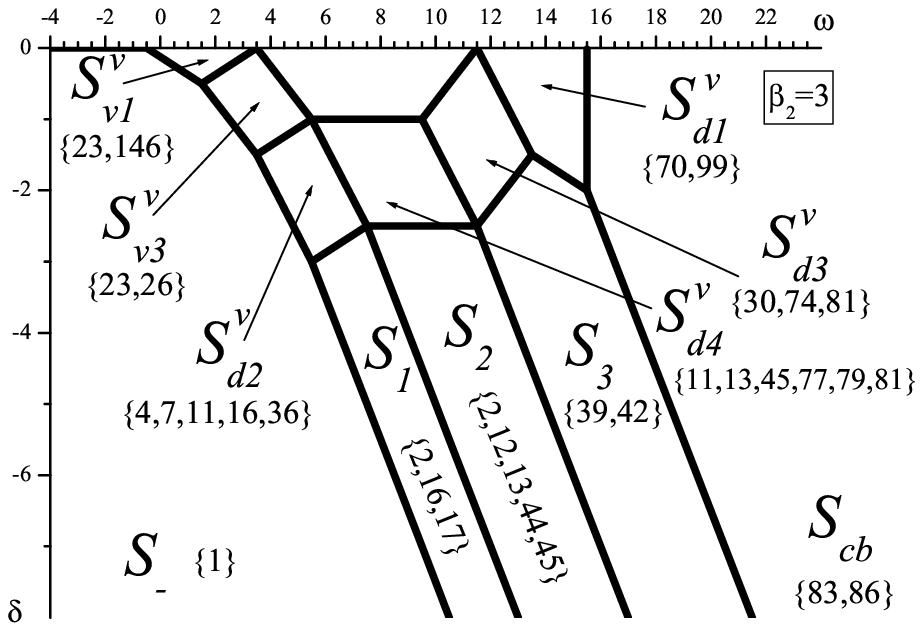}\\
\includegraphics[totalheight=0.3\textwidth,origin=c]{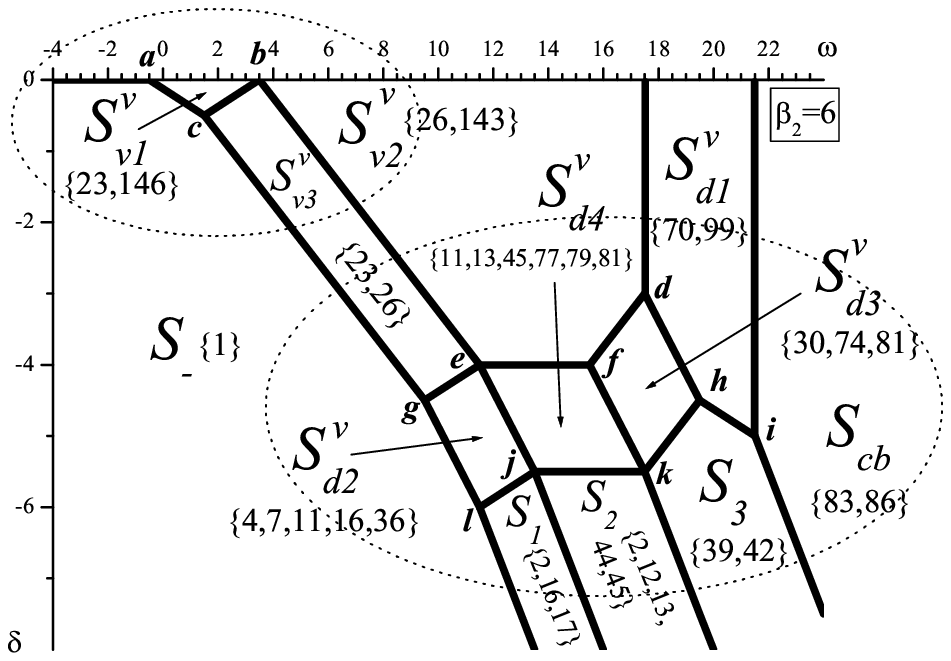}
\end{center}
\caption{\scriptsize{The case of the weakest anisotropy, off the
hole-particle symmetry, with $\varepsilon=0$. The phase diagram of
$\left(H_{T}^{f}\right)^{(4)}$ (given by (\ref{HTf4})) for an
increasing sequence of values of $\beta_2$.
The representative ion configurations of the displayed phases are shown
in Fig.~\ref{phases1} (for more comments see Section 3).
The critical values
are: $\beta_2 =1$, $\beta_2 =2$, and $\beta_2 =3$. In the blank region
of the phase diagram for the critical $\beta_2 =1$,
the following $T$-plaquette configurations have the minimal energy:
26,52,53,68,81,88,101,116,117,143 (see Fig.~\ref{bc168}).
By means of these $T$-plaquette configurations one can construct
${\mathcal{S}}^{v}_{v2}$, ${\mathcal{S}}_{dd}$, and many other configurations.
The equations defining the boundary lines of
the phase domains are given in Tab.~\ref{tb3} of Appendix B, while
the corresponding zero-potential coefficients $\{ \alpha_{i} \}$ ---
in Tab.~\ref{tb16} -- Tab.~\ref{tb22} of Appendix C.
In the bottom diagram, the regions surrounded by dotted ellipses are
reconsidered in the case of an intermediate anisotropy.
}}
\label{epspdf}
\end{figure}

\begin{figure}[htp]
\includegraphics[width=0.43\textwidth]{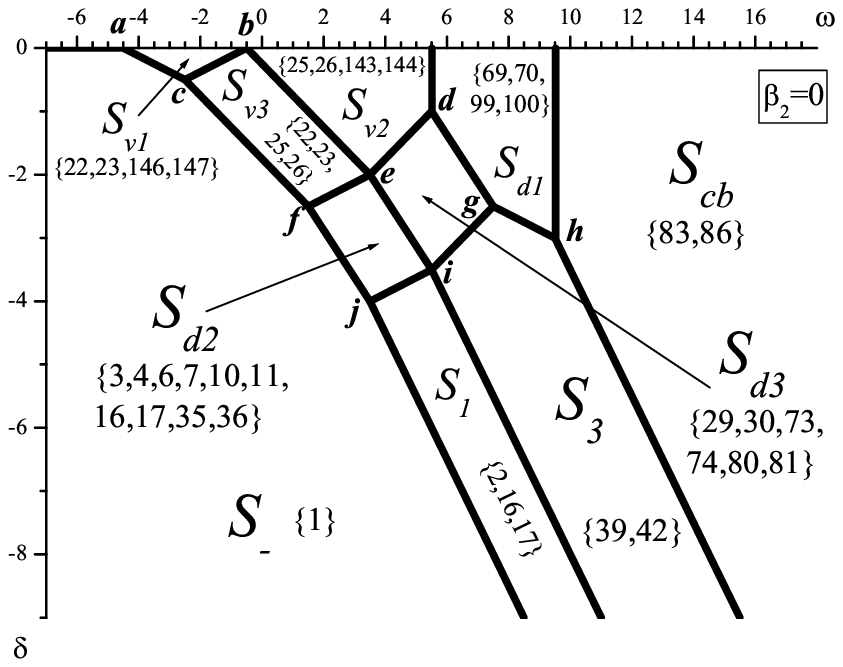}
\hfill
\includegraphics[width=0.47\textwidth]{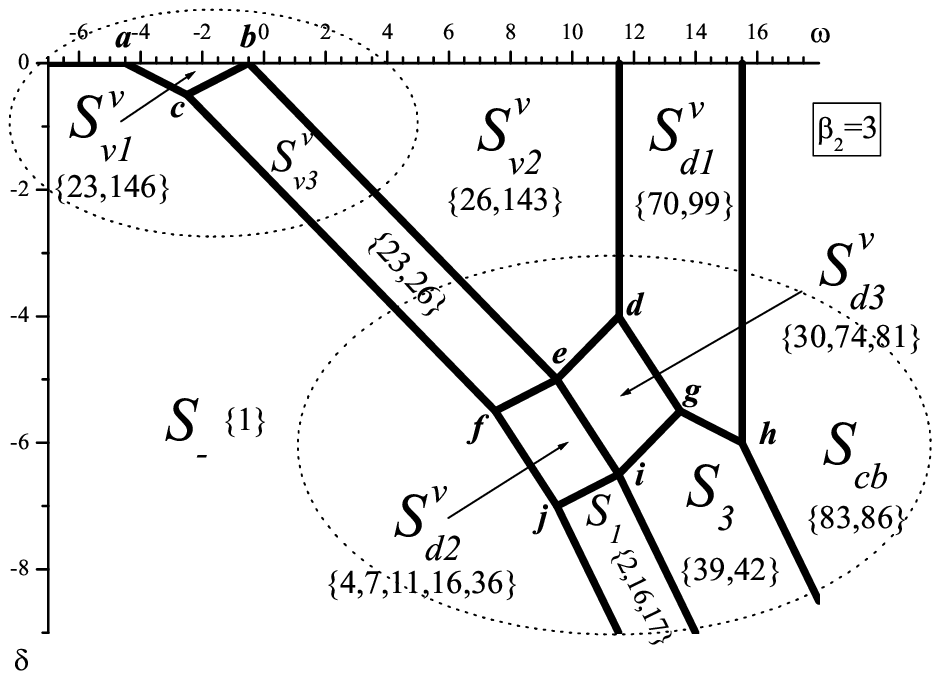}
\caption{\scriptsize{The case of the weakest anisotropy, off the
hole-particle symmetry, with $\varepsilon=0$. The phase diagram of
$\left(H_{T}^{b}\right)^{(4)}$ (given by (\ref{HTf4})); the
isotropic diagram ($\beta_2=0$) and the anisotropic diagram
($\beta_2=3$).
The representative ion configurations of the
displayed phases are shown in Fig.~\ref{phases1} (for more comments see Section 3).
Here, no critical values have been detected.
The equations
defining the boundary lines of the phase domains are given in
Tab.~\ref{tb4} of Appendix B, while the corresponding zero-potential
coefficients $\{ \alpha_{i} \}$ --- in Tab.~\ref{tb23} and
Tab.~\ref{tb24} of Appendix C.
In the right diagram, the regions surrounded by dotted ellipses are
reconsidered in the case of an intermediate anisotropy.
}}
\label{epspdb}
\end{figure}

Specifically, in Fig.~\ref{sppd} we show phase diagrams
in the hole-particle-symmetric case,
while in the absence of the hole-particle symmetry, the phase diagrams are
shown in Fig.~\ref{epspdf} -- for fermions, and in Fig.~\ref{epspdb}
-- for bosons.

The phases that appear in these phase diagrams can conveniently be described
in terms of the phases found in ground-state phase diagrams of the isotropic systems
\cite{DJ2}, with Hamiltonian $\left. H_{0}\right|_{\gamma=1}$, and presented in
Fig.~\ref{phases1}. The phases of anisotropic systems, considered here, either coincide
with or are simple modifications of the isotropic phases.
In fact, in Fig.~\ref{phases1} we display only representative configurations of the
phases of isotropic systems. The remaining configurations can be obtained
by means of the spatial symmetry operations of $\left. H_{0}\right|_{\gamma=1}$,
like translations and rotations by $\pi /2$. The numbers in curly brackets,
placed by the symbols of phases, stand for the numbers of the $T$-plaquette
configurations (according to Fig.~\ref{bc168}) that are obtained by restricting
the configurations of a phase to a $T$-plaquette.
By the same symbols as the phases we denote also their domains.

Among the phases of isotropic systems, we can distinguish a class of
{\em dimeric phases}, ${\mathcal{S}}_{d1}$, \ldots, ${\mathcal{S}}_{d4}$,
and a class of {\em axial-stripe phases}, ${\mathcal{S}}_{v1}$, ${\mathcal{S}}_{v2}$,
and ${\mathcal{S}}_{v3}$.
The configurations of dimeric phases consist of isolated pairs of n.n. occupied sites
(in the sequel called {\em the dimers}).
In the configurations of axial-stripe phases, the ions fill completely some,
parallel to one of the axes, lattice lines, so that a periodic pattern of {\em stripes}
is formed. Out of the dimeric or axial-stripe
configurations of Fig.~\ref{phases1}, only those with dimers or stripes,
respectively, oriented vertically appear in the phase diagrams of anisotropic systems.
Such a restricted phases are marked in anisotropic phase diagrams by
the additional superscript, $v$.

All the phases presented in Fig.~\ref{phases1},
except ${\mathcal{S}}_{d2}$ and ${\mathcal{S}}_{d4}$, contain exclusively periodic
configurations, related by the
spatial symmetries of $\left. H_{0}\right|_{\gamma=1}$,
hence of the degeneracy independent of the size of the underlying lattice.
The set  ${\mathcal{S}}_{pcb}$ (of plaquette-chessboard
configurations) contains configurations built of elementary
plaquettes  with occupied sites, forming a square lattice according
to the same rules as filled sites form a square lattice in the
chessboard configurations of ${\mathcal{S}}_{cb}$. The
remaining phases consist of configurations that
have a {\em quasi-one-dimensional} structure. That is, they are built
of sequences of completely filled with ions lattice lines of given slope.
Such a configuration can be specified by giving the slope of the
filled lattice lines of a representative configuration and than
the succession of filled (f) and empty (e) consecutive lattice lines
in a period. For instance, the representative configuration of
${\mathcal{S}}_{d1}$ (see Fig.~\ref{phases1}) is built of filled
lines with the slope $2$ and, in the period, two consecutive
filled lines are followed by two empty lines, which is denoted
$(2;2f,2e)$. This kind of description of the remaining
quasi-one-dimensional configurations is given in Fig.~\ref{phases1}.
The phase ${\mathcal{S}}_{dd}$ is an example of a {\em diagonal-stripe phase}.

\begin{figure}[ht]
\begin{center}
\centering \includegraphics[width=0.5\textwidth]{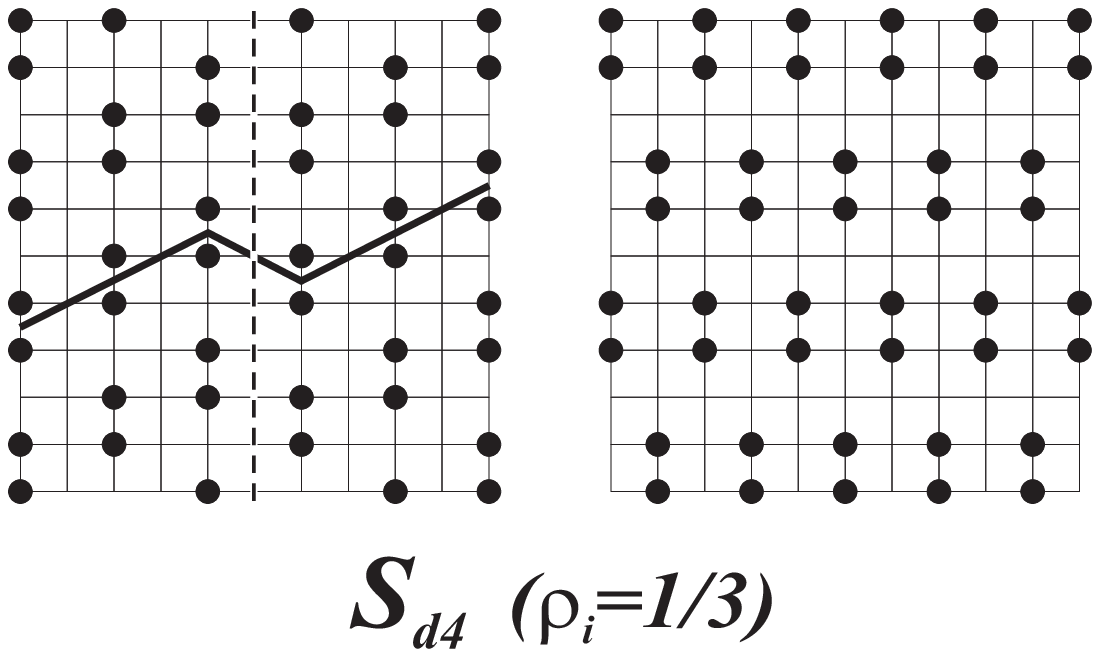}
\caption{\scriptsize{The representative configurations of the set
${\mathcal{S}}^{v}_{d4}$ whose degeneracy grows like
$\exp{(const \sqrt{\Lambda})}$. The left
configuration is an example of configurations with defect lines.
Here, there is one defect line (the dashed line, the continuous line
is a guide for the eye): the vertical lattice line separating
a periodic configuration of vertical dimers from its vertical
translate by one lattice constant. The right configuration is a
periodic configuration of dimers. For more comments see Section 3.
}}
\label{phases2}
\end{center}
\end{figure}

Only in the domains ${\mathcal{S}}_{d2}$ and ${\mathcal{S}}_{d4}$,
which appear in the phase diagrams shown in Figs.~\ref{epspdf},~\ref{epspdb},
that is off the hole-particle symmetry case, the situation is more complex,
their degeneracy grows indefinitely with the size of the lattice.
In ${\mathcal{S}}_{d2}$ one can distinguish two classes,
${\mathcal{S}}_{d2a}$ and ${\mathcal{S}}_{d2b}$, of periodic
configurations with parallelogram elementary cells. A
configuration in ${\mathcal{S}}_{d2a}$ consists of vertical
(horizontal) dimers of occupied sites that form a square lattice,
where the sides of the elementary squares have the length
$2\sqrt{2}$ and the slope $\pm 1$. In a configuration of
${\mathcal{S}}_{d2b}$, the elementary parallelograms formed by
dimers have the sides of the length $2\sqrt{2}$ and the slope $\pm
1$, and the sides of the length $\sqrt{10}$ and the slope $\pm
1/3$. Two configurations, one from ${\mathcal{S}}_{d2a}$ and one
from ${\mathcal{S}}_{d2b}$, having the same kind of dimers
(vertical or horizontal), can be merged together along a "defect
line" of the slope $\pm 1$ (the dashed line), as
shown in Fig.~\ref{phases1}, without increasing the energy. By
introducing more defect lines one can construct many ground-state
configurations whose number scales with the size of the lattice as
$\exp{(const \sqrt{\Lambda})}$. The same kind of degeneracy is in
${\mathcal{S}}_{d4}$. This phase consists of three classes of
periodic configurations of dimers.
In  the class ${\mathcal{S}}_{d4a}$, the elementary cell can be chosen
as a parallelogram whose two sides of length $3$ are parallel to dimers
(which are vertical or horizontal). If the dimers are oriented vertically,
then the other two sides have the slope $1/2$ and the length
$\sqrt{5}$. By reflecting an elementary cell of ${\mathcal{S}}_{d4a}$ in
a lattice line passing through its side that is parallel to dimers,
we obtain an elementary cell of the class ${\mathcal{S}}_{d4b}$. In the
third class, ${\mathcal{S}}_{d4c}$, an elementary cell can be chosen as
a rhomb formed by the centers of dimers, with the sides of length
$\sqrt{10}$. Two configurations, one from ${\mathcal{S}}_{d4a}$ and one
from ${\mathcal{S}}_{d4b}$, having the same kind of dimers
(vertical or horizontal), can be merged together along a "defect
line" parallel to dimers (dashed line in Fig.~\ref{phases2})
without increasing the energy.
In this way, a numerous family of configurations can be constructed,
with the number of configurations growing like
$\exp{(const \sqrt{\Lambda})}$.

It follows from the polyhedral shape of the phase domains, that the set of
values of $\beta_2$ is partitioned into open intervals, where
the boundary lines of phase domains do not change their direction,
only their distance to the origin varies in an affine way.
The boundary points of these open intervals are special values of the
anisotropy for the phase diagrams,
and in the sequel we call them {\em the critical values}.
As a critical value of anisotropy is approached,
some boundary lines merge into a line or a point, which results in
disappearance of some phase domains. And vice versa, some points and lines break off,
creating new phase domains.

In particular, in the hole-particle symmetric case
it can be inferred from (Fig.~\ref{sppd}) that at least up to $\beta_2=10$
-- for hopping fermions, and at least up to $\beta_2=5$ -- for hardcore bosons,
there is only one critical value of $\beta_2$. For fermions, it amounts to $\beta_2 =7$,
where the phase ${\mathcal{S}}_{dd}$ disappears, while for hardcore bosons it is
$\beta_2 =2$, where the phase ${\mathcal{S}}_{pcb}$ disappears.

Off the hole-particle symmetry,
for bosons, there are no critical values of $\beta_{2}$,
at least up to $\beta_{2}=3$.
On the other hand, for fermions and for $\beta_{2}\leq 6$ there are
as many as three critical values of $\beta_{2}$. In increasing order,
the first is $\beta_{2}=1$, where the  phases ${\mathcal{S}}^{v}_{v1}$,
${\mathcal{S}}^{v}_{d1}$, and ${\mathcal{S}}^{v}_{d2}$ appear.
The second is $\beta_{2}=2$, where the phases ${\mathcal{S}}^{v}_{d3}$,
${\mathcal{S}}^{v}_{d4}$, and ${\mathcal{S}}^{v}_{v3}$ appear.
And the last one is $\beta_{2}=3$, where ${\mathcal{S}}_{dd}$ is
replaced by ${\mathcal{S}}^{v}_{v2}$.

If there are no more critical values of the anisotropy parameter
$\beta_2$, the shape of the fourth-order phase diagrams for
any value $\beta_2$ larger than the greatest considered in
Figs.~\ref{sppd},~\ref{epspdf},~\ref{epspdb}
remains the same. On increasing the anisotropy parameter,
the diagrams undergo only some translations.
To verify whether for the values of $\beta_2$ larger
than those considered in this section new critical values do appear,
we proceed to investigating stronger, i.e. intermediate, deviations
from the isotropic case.

\subsection{The intermediate deviation from the isotropic case}

\begin{figure}[th]
\includegraphics[width=0.42\textwidth]{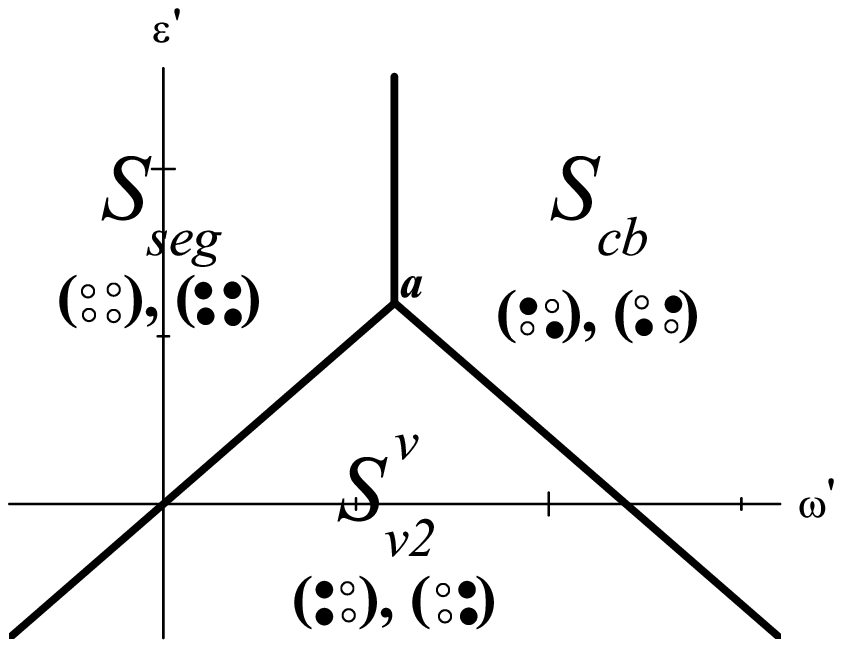}
\hfill
\includegraphics[width=0.42\textwidth]{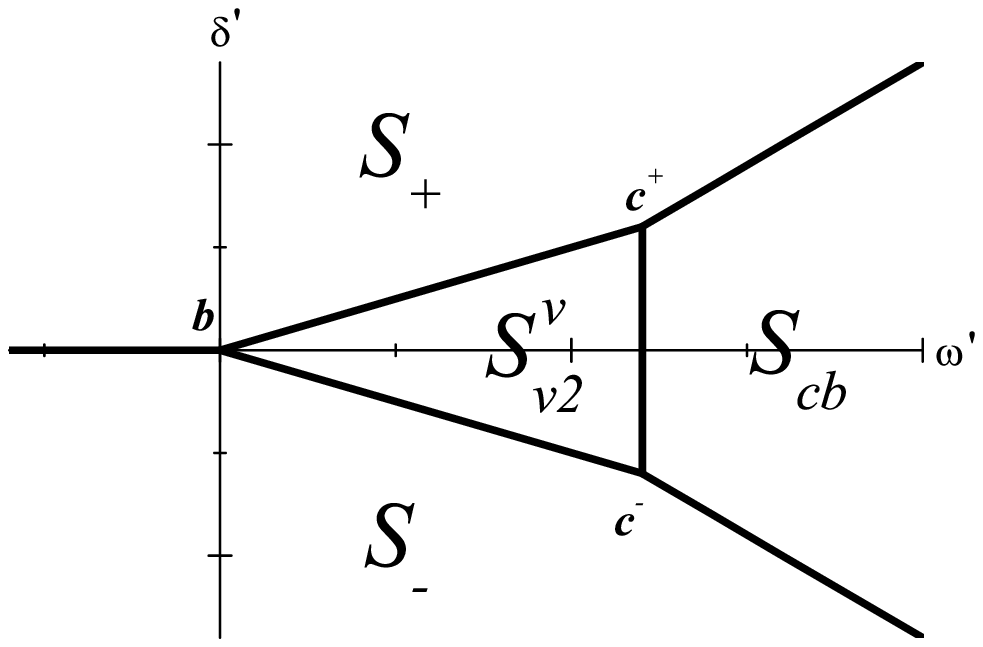}
\\
\parbox[t]{0.47\textwidth}{\caption{{\scriptsize{
The case of an intermediate anisotropy ($0<a<2$) and of
hole-particle symmetric systems ($\mu =0$).
The phase diagram (common for hopping fermion and hardcore boson systems)
of $H_{P}^{(2+a)}$. The coordinates of point ${\bf a}$ are
$\omega^{\prime}=\beta_{a}$, $\varepsilon^{\prime}=\beta_{a}$.
The boundary lines of ${\mathcal{S}}^{v}_{v2}$, from left to right, are:
$\varepsilon^{\prime}=\omega^{\prime}$ and $\varepsilon^{\prime}
=-\omega^{\prime}+2\beta_{a}$. The boundary
between ${\mathcal{S}}_{seg}$ and ${\mathcal{S}}_{cb}$ is
$\omega^{\prime}=\beta_{a}$.}}}
\label{pd3-2}}
\hfill
\parbox[t]{0.47\textwidth}{\caption{{\scriptsize{
The case of an intermediate anisotropy ($0<a<2$),
off the hole-particle symmetry, with $\varepsilon^{\prime}=0$.
The phase diagram (common for hopping fermion and hardcore boson systems)
of $H_{P}^{(2+a)}$. In the $(\omega^{\prime},\delta^{\prime})$-plane,
${\bf b}=(0,0)$, ${\bf c^{+}}=(2\beta_{a},\beta_{a})$,
${\bf c^{-}}=(2\beta_{a},-\beta_{a})$.
The boundary lines of ${\mathcal{S}}_{-}$, from left to right, are:
$\delta^{\prime}=0$,
$\delta^{\prime}=-\frac{\omega^{\prime}}{2}$
and $\delta^{\prime}=-\omega^{\prime}+\beta_{a}$.
The boundary lines of ${\mathcal{S}}_{+}$ are obtained by
changing $\delta^{\prime} \rightarrow -\delta^{\prime}$.
}}}
\label{pd3-1}}
\end{figure}

Now, the fourth-order
effective Hamiltonian for fermions assumes the form:
\begin{eqnarray}
\left( E^{f}_{S} \right)^{(4)} &=&
\left.\left( E^{f}_{S} \right)^{(4)}\right|_{\gamma=1}
- \beta_{2} \frac{t^{2+a}}{4}
\sum\limits_{\langle x,y \rangle_{1,v}} s_{x}s_{y},
\end{eqnarray}
and for hardcore bosons an analogous formula holds true.
Obviously, the phase diagrams in the
zeroth and second orders remain the same as in the isotropic case,
described above.
Therefore, we proceed to constructing the phase diagram in next
order, which is ($2+a$)-order with $0<a<2$, and the corresponding
effective Hamiltonian reads:
\begin{equation}
E^{(2+a)}_{S} = E^{(2)}_{S} - \beta_{2} \frac{t^{2+a}}{4}
\sum\limits_{\langle x,y \rangle_{1,v}} s_{x}s_{y},
\end{equation}
where $ E^{(2)}_{S}$ stands for the, common for fermions and bosons,
second-order effective Hamiltonian. For the reasons given in the previous
subsection, we consider a neighborhood of the point
$W=-2t^{2}$, $\mu =0$, $\tilde{\varepsilon}=0$,
where the energies of all the configurations are equal Ref.\cite{DJ2}.
In this neighborhood it is convenient to introduce new variables,
$\delta^{\prime}$, $\varepsilon^{\prime}$, and $\omega^{\prime}$,
\begin{eqnarray}
\mu=t^{2+a} \delta^{\prime} , \qquad
\tilde{\varepsilon}=t^{2+a} \varepsilon^{\prime} , \qquad
W=-2t^{2}+t^{2+a} \omega^{\prime} ,
\end{eqnarray}
and rewrite the expansion up to the order $2+a$:
\begin{eqnarray}
E^{(2+a)}_{S}  & = &
\frac{t^{2+a}}{2} \left\{ - \delta^{\prime} \sum\limits_{x} \left( s_{x} + 1
\right) + \frac{\omega^{\prime}}{4} \sum\limits_{\langle x,y \rangle_{1}}
s_{x}s_{y} - \frac{\varepsilon^{\prime}}{8}
\sum\limits_{\langle x,y \rangle_{2}}
s_{x}s_{y}- \frac{\beta_{a}}{2} \sum\limits_{\langle x,y
\rangle_{1,v}} s_{x}s_{y} \right\} \nonumber \\
& = & \frac{t^{2+a}}{2} \sum\limits_{P} H_{P}^{(2+a)},
\end{eqnarray}
where
\begin{eqnarray}
H_{P}^{(2+a)} = - \frac{\delta^{\prime}}{4}
\sideset{}{'}\sum\limits_{x} \left( s_{x} + 1 \right) +
\frac{\omega^{\prime}}{8}
\sideset{}{'}\sum\limits_{\langle x,y \rangle_{1}} s_{x}s_{y} -
\frac{\varepsilon^{\prime}}{8}
\sideset{}{'}\sum\limits_{\langle x,y \rangle_{2}} s_{x}s_{y} -
\frac{\beta_{a}}{4}
\sideset{}{'}\sum\limits_{\langle x,y
\rangle_{1,v}} s_{x}s_{y}.
\end{eqnarray}
The summations in the primed sums run over a plaquette $P$.
The plaquette potentials $H_{P}^{(2+a)}$ have to be minimized over
all the plaquette configurations.
As in the previous case, we are interested in phase diagrams
of hole-particle symmetric systems ($\delta^{\prime} =0$) or
unsymmetrical systems with $\varepsilon^{\prime}=0$.
It turns out that for such energy parameters and plaquette configurations
the potentials $H_{P}^{(2+a)}$ are m-potentials.
The resulting phase diagrams are shown in
Fig.~\ref{pd3-2} and Fig.~\ref{pd3-1}.
We note that each of the points
${\bf a}$: $\omega^{\prime}=\beta_{a}$, $\varepsilon^{\prime}=\beta_{a}$,
${\bf b}$: $\omega^{\prime}=0$, $\delta^{\prime}=0$,
${\bf c^{-}}$: $\omega^{\prime}=2\beta_{a}$, $\delta^{\prime}=-\beta_{a}$,
and ${\bf c^{+}}$: $\omega^{\prime}=2\beta_{a}$, $\delta^{\prime}=\beta_{a}$,
is the coexistence point of three periodic phases. That is,
the only plaquette configurations minimizing the potential $H_{P}^{(2+a)}$
at such a point are those obtained by restricting the configurations of
coexisting phases to a  plaquette (these plaquette configurations are shown
in Fig.~\ref{pd3-2})

\begin{figure}[t]
\includegraphics[width=0.46\textwidth]{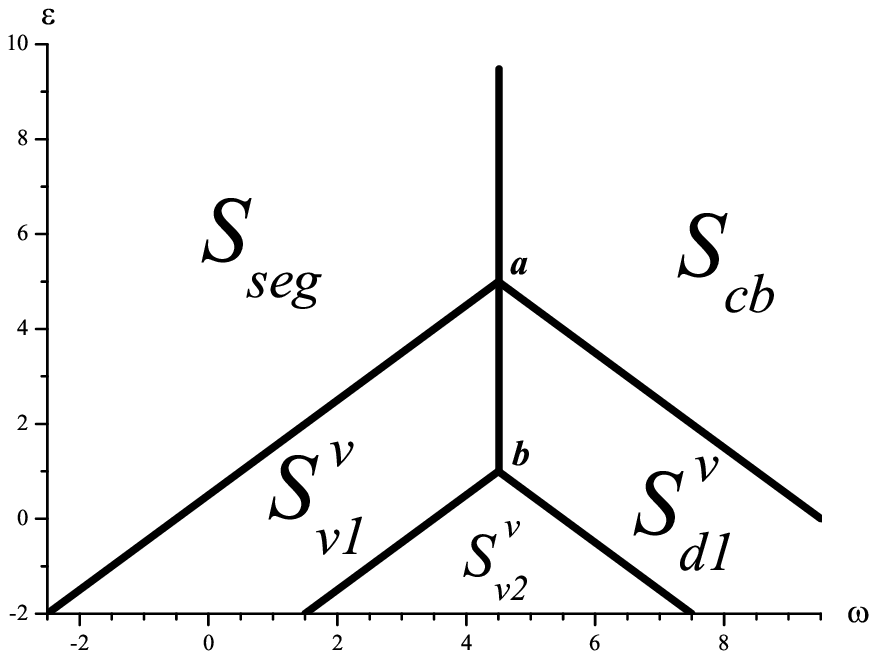}
\hfill
\includegraphics[width=0.46\textwidth]{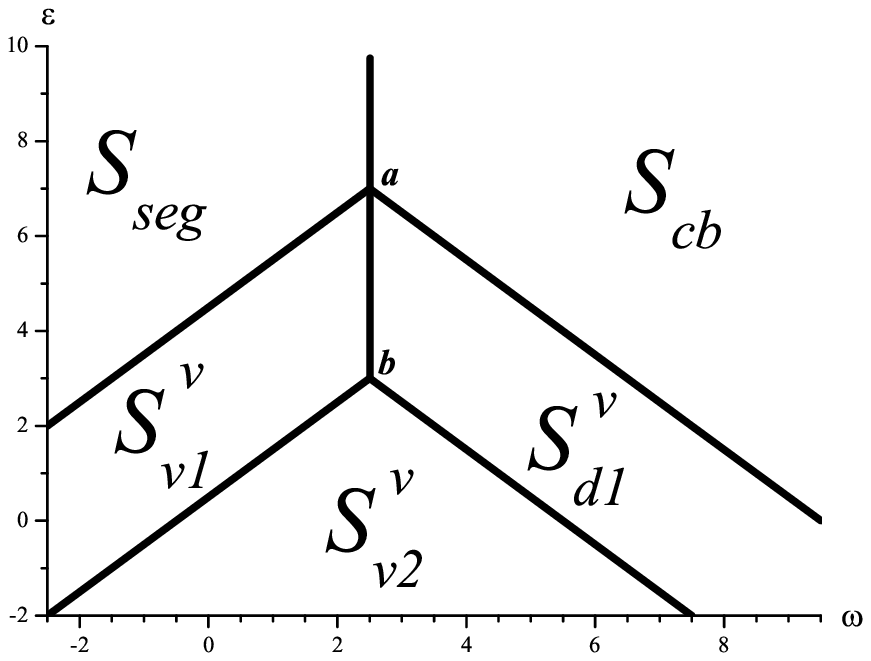}
\\
\parbox[t]{0.46\textwidth}{\caption{{\scriptsize{
The case of an intermediate anisotropy ($0<a<2$) and of
hole-particle symmetric systems ($\mu =0$). The phase diagram of
$\left( E_{S}^{f} \right)^{(4)}$ in a neighborhood of point ${\bf
a}$ ($\omega^{\prime}=\beta_{a}$, $\varepsilon^{\prime}=\beta_{a}$)
(see formula (\ref{Ef4a})). The equations defining the boundary
lines of the phase domains are given in Tab.~\ref{tb5} of Appendix B.
}}}
\label{p1fmnd0}}
\hfill
\parbox[t]{0.46\textwidth}{\caption{{\scriptsize{
The case of an intermediate anisotropy ($0<a<2$) and of
hole-particle symmetric systems ($\mu =0$). The phase diagram of
$\left( E_{S}^{b} \right)^{(4)}$ in a neighborhood of point ${\bf
a}$ ($\omega^{\prime}=\beta_{a}$, $\varepsilon^{\prime}=\beta_{a}$)
(see formula (\ref{Ef4a})). The equations defining the boundary
lines of the phase domains are given in Tab.~\ref{tb5} of Appendix B.
}}}
\label{p1hcbd0}}
\end{figure}

\begin{figure}[ph]
\centering \includegraphics[width=0.9\textwidth]{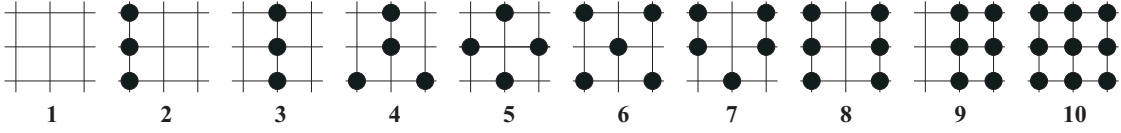}
\caption{\scriptsize{All the admissible $T$-plaquette configurations,
up to rotations by $\pi$,
that are used in constructing the fourth-order phase diagram in a neighborhood
of point {\bf a} of Fig.~\ref{pd3-2},
i.e. the elements of ${\mathcal{S}}^{{\bf a}}_{T}$.
}}
\label{bc10}
\end{figure}

Now, following our recursive procedure of constructing phase diagrams to some order,
we are ready to investigate the effect of fourth-order interactions. As in earlier steps,
it is enough to consider neighborhoods of the coexistence points
${\bf a}$, ${\bf b}$, and ${\bf c^{-}}$ of Figs.~\ref{pd3-2},\ref{pd3-1}
(the diagram in a neighborhood of ${\bf c^{+}}$ can be obtained from that around
${\bf c^{-}}$ by a symmetry operation).
\paragraph{A. A neighborhood of ${\bf a}$}$\\$
Here $\delta^{\prime}=0$, hence the systems are hole-particle
invariant. A convenient change of variables is:
\begin{eqnarray}
\omega^{\prime} = \beta_{a} + t^{2-a}\omega ,  \qquad
\varepsilon^{\prime}=\beta_{a}+ t^{2-a}\varepsilon.
\end{eqnarray}
In these variables, the fourth-order effective Hamiltonian reads:
\begin{eqnarray}
\left( E_{S}^{f} \right)^{(4)} =  \frac{t^{2+a}}{2} \sum\limits_{P}
\left. H^{(2+a)}_{P}\right|_{\substack{\delta^{\prime}=0\\
\varepsilon^{\prime}=\omega^{\prime}=\beta_{a}}}+\frac{t^{4}}{2}\sum\limits_{T}\left.\left( H^{f}_{T}
\right)^{(4)}\right|_{\gamma=1},
\label{Ef4a}
\end{eqnarray}
with a similar formula for hard-core bosons, where
\begin{eqnarray}
\left. H^{(2+a)}_{P}\right|_{\substack{\delta^{\prime}=0\\
\varepsilon^{\prime}=\omega^{\prime}=\beta_{a}}} =
\frac{\beta_{a}}{8}
\left(\sideset{}{'}\sum\limits_{\langle x,y \rangle_{1,h}} s_{x}s_{y} -
\sideset{}{'}\sum\limits_{\langle x,y \rangle_{1,v}}
s_{x}s_{y}-
\sideset{}{'}\sum\limits_{\langle x,y \rangle_{2}} s_{x}s_{y}\right).
\end{eqnarray}
In the previous order, it has been established that the only plaquette configurations
minimizing $\left. H^{(2+a)}_{P}\right|_{\substack{\delta^{\prime}=0\\
\varepsilon^{\prime}=\omega^{\prime}=\beta_{a}}}$ are the ones obtained by
restricting to a plaquette $P$ the periodic configurations
${\mathcal{S}}_{+}$, ${\mathcal{S}}_{-}$, ${\mathcal{S}}_{cb}$,
and ${\mathcal{S}}^{v}_{v2}$. Let us denote this set of plaquette configurations
by ${\mathcal{S}}^{{\bf a}}_{P}$. Consequently, the minimization of
the fourth-order potentials
$\left.\left( H^{f}_{T} \right)^{(4)}\right|_{\gamma=1}$,
$\left.\left( H^{b}_{T} \right)^{(4)}\right|_{\gamma=1}$,
should be carried out only over the set ${\mathcal{S}}^{{\bf a}}_{T}$
of those $T$-plaquette configurations
whose restriction to a plaquette $P$ belongs to
${\mathcal{S}}^{{\bf a}}_{P}$.
The configurations of the set ${\mathcal{S}}^{{\bf a}}_{T}$
are displayed in Fig.~\ref{bc10}.
It appears that on the set ${\mathcal{S}}^{{\bf a}}_{T}$,
the  potentials
$\left.\left( H^{f}_{T} \right)^{(4)}\right|_{\gamma=1}$,
$\left.\left( H^{b}_{T} \right)^{(4)}\right|_{\gamma=1}$,
are $m$-potentials. The obtained phase diagrams, independent of the
anisotropy parameter $\beta_{a}$, are shown in
Fig.~\ref{p1fmnd0} and Fig.~\ref{p1hcbd0}.

\begin{figure}[p]
\includegraphics[width=0.45\textwidth]{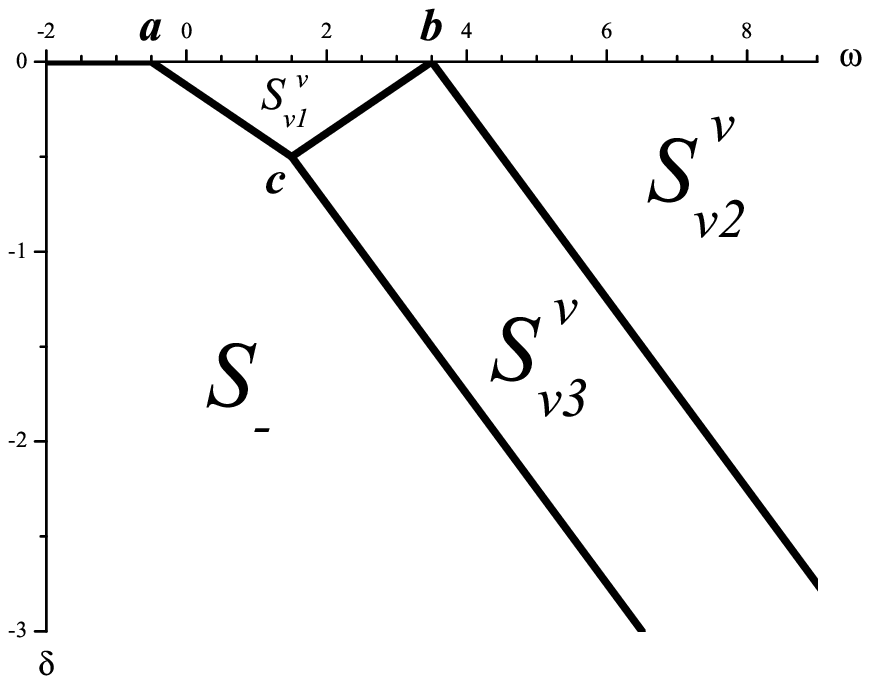}
\hfill
\includegraphics[width=0.45\textwidth]{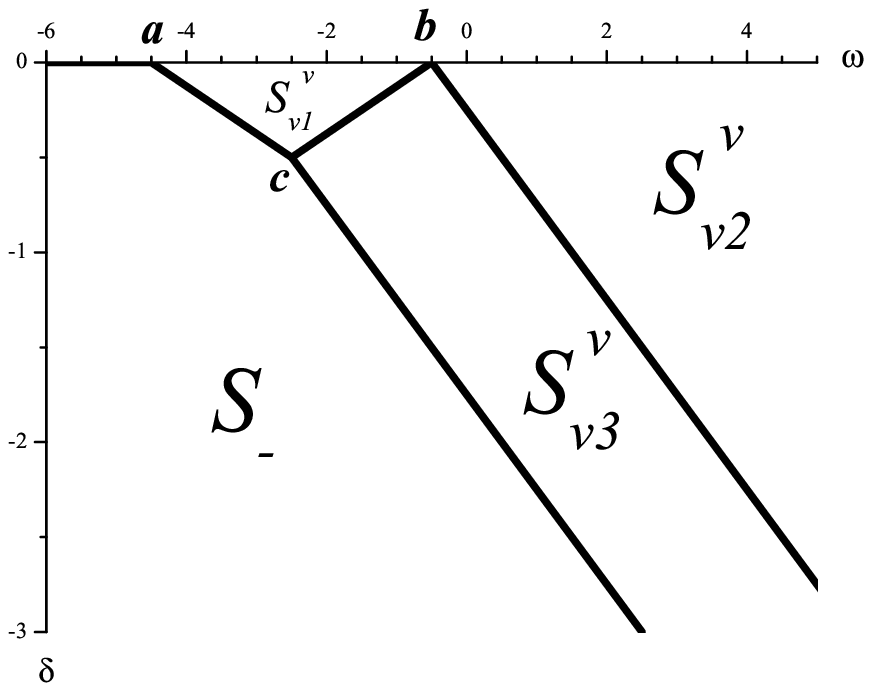}
\\
\parbox[t]{0.46\textwidth}{\caption{{\scriptsize{
The case of an intermediate anisotropy ($0<a<2$), off the
hole-particle symmetry, with $\varepsilon^{\prime}=0$. The phase
diagram of $\left( E_{S}^{f} \right)^{(4)}$ in a neighborhood of
point ${\bf b}$ ($\omega^{\prime}=0$, $\delta^{\prime}=0$), see
formula (\ref{Ef4b}). The equations defining the boundary lines of
the phase domains are given in Tab.~\ref{tb6} of Appendix B, while
the corresponding zero-potential coefficients $\{ \alpha_{i} \}$ in
Tab.~\ref{tb25} of Appendix C.
}}}
\label{p1fmne0}}
\hfill
\parbox[t]{0.46\textwidth}{\caption{{\scriptsize{
The case of an intermediate anisotropy ($0<a<2$), off the
hole-particle symmetry, with $\varepsilon^{\prime}=0$. The phase
diagram of $\left( E_{S}^{b} \right)^{(4)}$ in a neighborhood of
point ${\bf b}$ ($\omega^{\prime}=0$, $\delta^{\prime}=0)$), see
formula (\ref{Ef4b}). The equations defining the boundary lines of
the phase domains are given in Tab.~\ref{tb6} of Appendix B, while
the corresponding zero-potential coefficients $\{ \alpha_{i} \}$ in
Tab.~\ref{tb26} of Appendix C.
}}}
\label{p1hcbe0}}
\end{figure}

\begin{figure}[hp]
\includegraphics[width=0.45\textwidth]{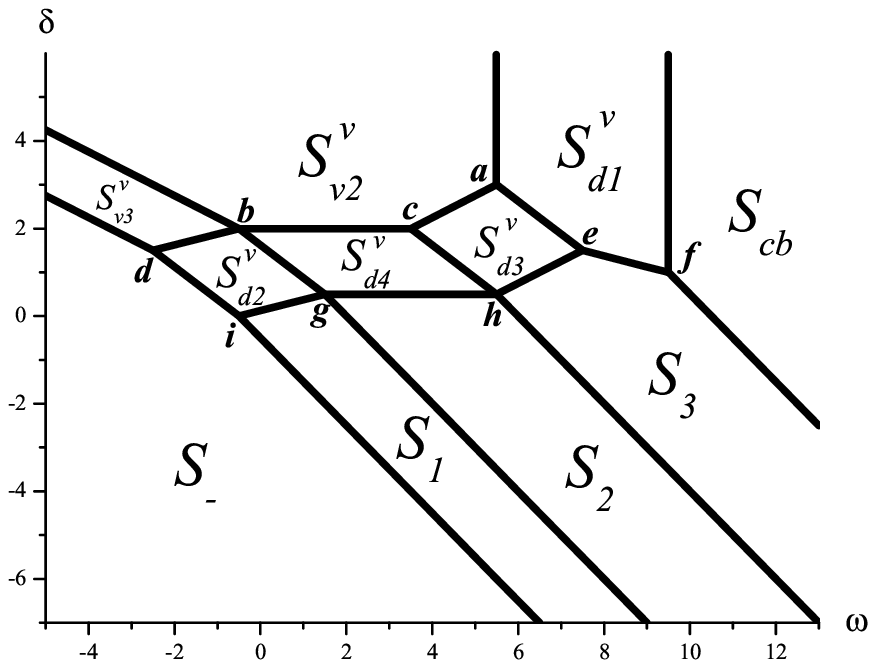}
\hfill
\includegraphics[width=0.45\textwidth]{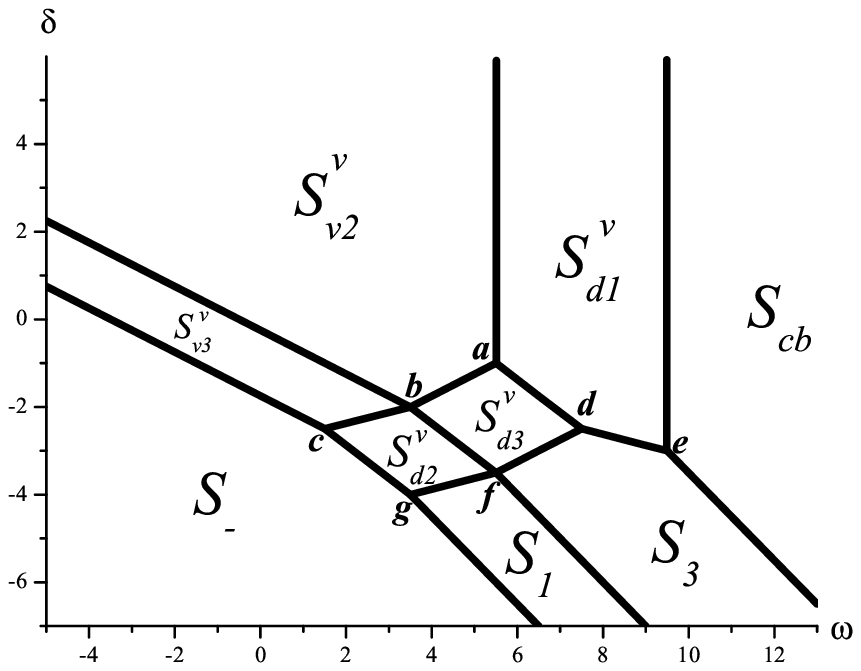}
\\
\parbox[t]{0.46\textwidth}{\caption{{\scriptsize{
The case of an intermediate anisotropy ($0<a<2$), off the
hole-particle symmetry, with $\varepsilon^{\prime}=0$. The phase
diagram of $\left( E_{S}^{f} \right)^{(4)}$ in a neighborhood of
point ${\bf c^{-}}$ ($\omega^{\prime}=2\beta_{a}$,
$\delta^{\prime}=-\beta_{a}$), see formula (\ref{Ef4c}). The
equations defining the boundary lines of the phase domains are given
in Tab.~\ref{tb7} of Appendix B, while the corresponding
zero-potential coefficients $\{ \alpha_{i} \}$ in Tab.~\ref{tb27} of
Appendix C.
}}}
\label{p2fmne0}}
\hfill
\parbox[t]{0.46\textwidth}{\caption{{\scriptsize{
The case of an intermediate anisotropy ($0<a<2$), off the
hole-particle symmetry, with $\varepsilon^{\prime}=0$. The phase
diagram of $\left( E_{S}^{b} \right)^{(4)}$ in a neighborhood of
point ${\bf c^{-}}$ ($\omega^{\prime}=2\beta_{a}$,
$\delta^{\prime}=-\beta_{a}$), see formula (\ref{Ef4c}). The
equations defining the boundary lines of the phase domains are given
in Tab.~\ref{tb7} of Appendix B, while the corresponding
zero-potential coefficients $\{ \alpha_{i} \}$ in Tab.~\ref{tb28} of
Appendix C.
}}}
\label{p2hcbe0}}
\end{figure}

\begin{figure}[ph]
\centering \includegraphics[width=0.55\textwidth]{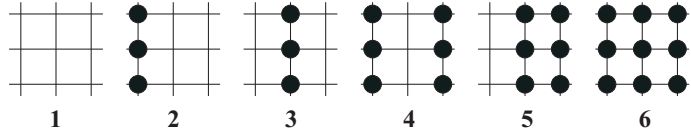}
\caption{\scriptsize{All the admissible $T$-plaquette configurations
that are used in constructing the fourth-order phase diagram,
up to rotations by $\pi$,
in a neighborhood  of point {\bf b} of Fig.~\ref{pd3-1},
i.e. the elements of ${\mathcal{S}}^{{\bf b}}_{T}$
(the vertical n.n. pairs that are occupied by one ion are forbidden).
}}
\label{bc6}
\end{figure}

\begin{figure}[hpt]
\centering \includegraphics[width=0.9\textwidth]{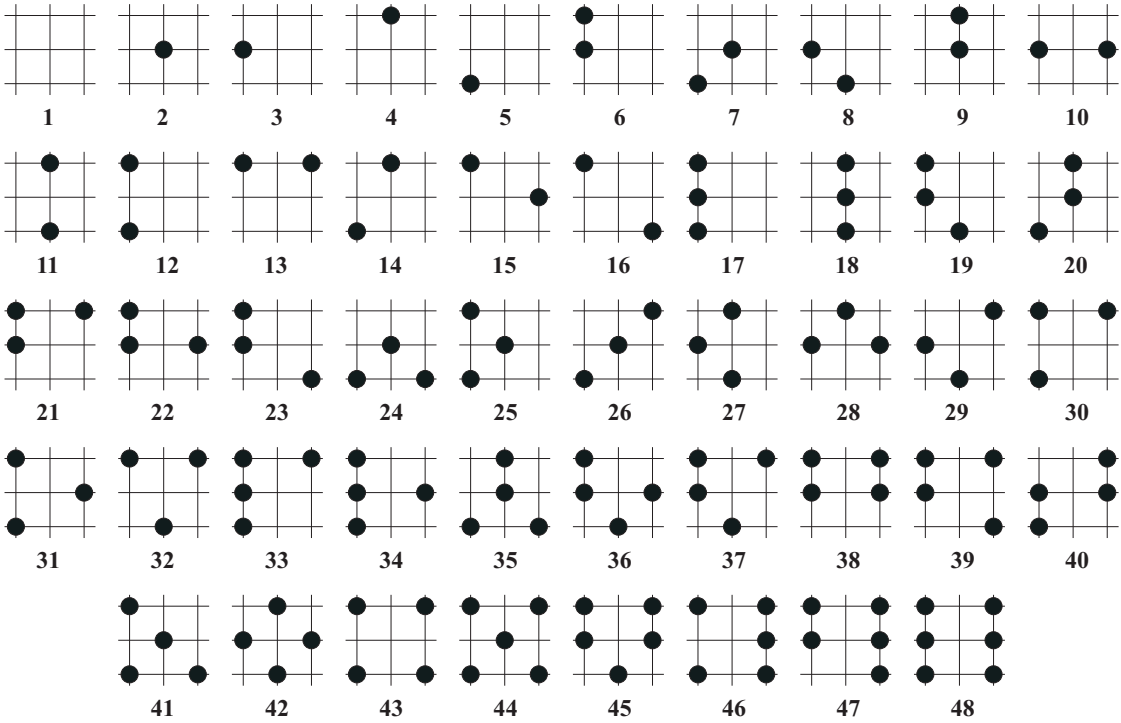}
\caption{\scriptsize{All the admissible $T$-plaquette configurations,
up to rotations by $\pi$ and reflections
in lattice lines parallel to the axes,
that are used in constructing the fourth-order phase diagram
in a neighborhood  of point ${\bf c}^{-}$ of Fig.~\ref{pd3-1},
i.e. the elements of ${\mathcal{S}}^{{\bf c}^{-}}_{T}$
(the horizontal n.n. pairs that are occupied by two ions are forbidden).
}}
\label{bc48}
\end{figure}

\paragraph{B. A neighborhood of ${\bf b}$}$\\$
Here the system is not hole-particle symmetric and
$\varepsilon^{\prime}=0$. A convenient change of variables is:
\begin{eqnarray}
\omega^{\prime} = t^{2-a}\omega ,  \qquad
\delta^{\prime} = t^{2-a} \delta .
\end{eqnarray}
Then, the fourth-order effective Hamiltonian takes the form,
\begin{eqnarray}
\left( E_{S}^{f} \right)^{(4)} =
\frac{t^{2+a}}{2} \sum\limits_{P}
\left. H^{(2+a)}_{P}\right|_{\substack{\delta^{\prime}=0\\
\varepsilon^{\prime}=\omega^{\prime}=0}}
+\frac{t^{4}}{2}\sum\limits_{T}\left.\left(
H^{f}_{T} \right)^{(4)}\right|_{\gamma=1},
\label{Ef4b}
\end{eqnarray}
where
\begin{equation}
\left. H^{(2+a)}_{P}\right|_{\substack{\delta^{\prime}=0\\
\varepsilon^{\prime}=\omega^{\prime}=0}}=
-  \frac{\beta_{a}}{4} \sideset{}{'}\sum\limits_{\langle x,y
\rangle_{1,v}} s_{x}s_{y}
\end{equation}
with a  similar formula for hard-core bosons.
The minimum of $\left. H^{(2+a)}_{P}\right|_{\substack{\delta^{\prime}=0\\
\varepsilon^{\prime}=\omega^{\prime}=0}}$ is attained at the
configurations belonging to ${\mathcal{S}}^{{\bf b}}_{P}$, i.e. the
plaquette configurations obtained by restricting the periodic
configurations ${\mathcal{S}}_{+}$, ${\mathcal{S}}_{-}$, and
${\mathcal{S}}^{v}_{v2}$ to a plaquette $P$. Let
${\mathcal{S}}^{{\bf b}}_{T}$ be the corresponding set of
$T$-plaquette configurations (there are no vertical pairs of n.n.
sites occupied by one ion). This set is shown in Fig.~\ref{bc6}.
Here the
potentials $\left.\left( H^{f}_{T} \right)^{(4)}\right|_{\gamma=1}$
are not the $m$-potentials. The obtained phase diagrams  are shown
in Figs.~\ref{p1fmne0},~\ref{p1hcbe0}.
\paragraph{C. A neighborhood of ${\bf c}^{-}$}$\\$
Here the system is not hole-particle symmetric and
$\varepsilon^{\prime}=0$. A convenient change of variables is:
\begin{eqnarray}
\omega^{\prime} = 2\beta_{a}+t^{2-a}\omega ,  \qquad
\delta^{\prime} = - \beta_{a} + t^{2-a} \delta .
\end{eqnarray}
The fourth-order effective Hamiltonian reads:
\begin{eqnarray}
\left( E_{S}^{f} \right)^{(4)} =
\frac{t^{2+a}}{2} \sum\limits_{P}
\left. H^{(2+a)}_{P}\right|_{\substack{\varepsilon^{\prime}=0\\
\delta^{\prime}=-\beta_a, \omega^{\prime}=2\beta_a}}
+\frac{t^{4}}{2}\sum\limits_{T} \left.\left(
H^{f}_{T} \right)^{(4)}\right|_{\gamma=1}
\label{Ef4c}
\end{eqnarray}
with a similar formula for hard-core bosons, where
\begin{equation}
\left. H^{(2+a)}_{P}\right|_{\substack{\varepsilon^{\prime}=0\\
\delta^{\prime}=-\beta_a, \omega^{\prime}=2\beta_a}}=
\frac{\beta_{a}}{4} \sideset{}{'}\sum\limits_{\langle x,y
\rangle_{1,h}} \left( s_{x}+ s_{y} + s_{x}s_{y}+ 1 \right)
\end{equation}
The potentials $\left. H^{(2+a)}_{P}\right|_{\substack{\varepsilon^{\prime}=0\\
\delta^{\prime}=-\beta_a, \omega^{\prime}=2\beta_a}}$ are minimized by restrictions
to a plaquette $P$ of periodic configurations
${\mathcal{S}}_{-}$, ${\mathcal{S}}_{cb}$, and ${\mathcal{S}}^{v}_{v2}$,
that constitute the set ${\mathcal{S}}^{c^{-}}_{P}$.
The corresponding set ${\mathcal{S}}^{c^{-}}_{T}$ of $T$-plaquette
configurations consists of configurations where no horizontal pair
of n.n. sites is occupied by two ions (see Fig.~\ref{bc48}).
Here the potentials
$\left.\left( H^{f}_{T} \right)^{(4)}\right|_{\gamma=1}$ are not the
$m$-potentials. The corresponding phase diagrams are shown in
Fig.~\ref{p2fmne0} and Fig.~\ref{p2hcbe0}.

\section{Discussion of the phase diagrams and conclusions}

In  subsection 3.2, we have obtained phase diagrams
according to the truncated effective Hamiltonians of fourth order,
with an intermediate anisotropy of hopping. These diagrams have been
constructed to see what happens to the fourth-order phase diagrams
with the weakest anisotropy if the anisotropy parameter $\beta_2$
grows beyond the values taken into account in subsection 3.1.
Let us consider firstly the diagrams in a neighborhood of point
${\bf a}$. Apparently,
up to a rescaling the phase diagram in Fig.~\ref{p1fmnd0}
looks the same as the part in the dotted circle of the
fermionic phase diagram in Fig.~\ref{sppd}.
Similarly, up to a rescaling the phase diagram in Fig.~\ref{p1hcbd0}
looks the same as the the part in the dotted circle of the bosonic phase diagram
in Fig.~\ref{sppd}.
The same remarks apply to the phase diagrams in neighborhoods of points
${\bf b}$ and ${\bf c^{-}}$.
The phase diagram in Fig.~\ref{p1fmne0} reproduces
the part in the upper dotted ellipse in Fig.~\ref{epspdf},
while the diagram in Fig.~\ref{p1hcbe0}
-- the part in the upper dotted ellipse in Fig.~\ref{epspdb}.
Then, the phase diagram in Fig.~\ref{p2fmne0} reproduces
the part in the lower dotted ellipse in Fig.~\ref{epspdf},
while the diagram in Fig.~\ref{p2hcbe0}
-- the part in the lower dotted ellipse in Fig.~\ref{epspdb}.
Therefore, we conclude that in the case of the weakest anisotropy
in the fourth-order effective Hamiltonians, we have determined all the
critical values of the anisotropy parameter $\beta_2$.
For the both kinds of hopping particles, if $\beta_2$ exceeds
the greatest critical value, the obtained phase diagram undergoes only
a translation (varying with $\beta_2$).

Now, the basic question to be answered is concerned with the relation
between these fourth-order phase diagrams and the
phase diagrams of quantum systems described by Hamiltonians $H_{0}$.
By adapting the arguments presented in Refs.
\cite{kennedy2,GMMU}, we can demonstrate, see for instance Ref.
\cite{DJ1}, that if the remainders,
$\left(R_{S}^{f}\right)^{(4)}$ and $\left(R_{S}^{b}\right)^{(4)}$,
are taken into account,
then there is a sufficiently small $t_{0}$ such that
for $t<t_{0}$ the phase diagrams of quantum systems look the same as the phase
diagrams according  to the effective Hamiltonians truncated
at the fourth order, except some narrow regions, of width $O(t^2)$ (at the
diagrams displayed above),
located along the phase-domains boundaries, and except the domains
${\mathcal{S}}^{v}_{d2}$ and ${\mathcal{S}}^{v}_{d4}$.
For $t<t_{0}$ and for each domain
${\mathcal{S}}_{D}$, which is different from ${\mathcal{S}}^{v}_{d2}$ and
${\mathcal{S}}^{v}_{d4}$,
there is a nonempty two-dimensional open domain
${\mathcal{S}}_{D}^{\infty}$ that is contained in the
domain ${\mathcal{S}}_{D}$ and such that in
${\mathcal{S}}_{D}^{\infty}$ the set of ground-state
configurations coincides with ${\mathcal{S}}_{D}$.
Moreover, in comparison with the critical values of the anisotropy
parameter $\beta_2$, determined according to the fourth-order
effective Hamiltonians, the corresponding critical values of the
quantum systems described by Hamiltonians $H_{0}$ differ by $O(t^2)$,
i.e. for a quantum system $\gamma = 1 - \beta_2 t^2 + O(t^4)$.

Remarkably, in the fourth-order the hole-particle symmetric phase
diagrams of fermions and of hardcore bosons are geometrically
similar. That is, a phase diagram of hard core bosons, with any
$\beta_2 \geq 0$, can be obtained from a phase diagram of
fermions, with $\beta_2 \geq 5$, by the translation whose vector
reads: $\omega =-7$, $\varepsilon =-3$, and $\beta_{2} =-5$. The
existence of this translation vector is related to the fact that
for both kinds of systems there is one critical value of the
anisotropy parameter. Additionally, for fermions with $\beta_2 <
7$, it is necessary to replace the phase ${\mathcal{S}}_{dd}$ in
the central domain by ${\mathcal{S}}_{pcb}$. Off the hole-particle
symmetry, the relation between the bosonic and fermionic phase
diagrams is not that close. For fermions, there are three critical
values of $\beta_2$, while for hardcore bosons there is no
critical values. Thus, the system with hopping hardcore bosons is
less sensitive to the anisotropy of hopping, than the system with
hopping fermions. Nevertheless, if $\varepsilon =0$, then the
phase diagrams of both kinds of systems are topologically similar,
except that in the bosonic phase diagrams the phases
${\mathcal{S}}_{d3}^{v}$ and ${\mathcal{S}}_{2}$ are missing.
However, we know from Ref.~\cite{DJ1} that this deficiency can be
removed by switching on the n.n.n. interactions with negative
$\varepsilon$.

In the fourth-order effective Hamiltonians (\ref{E4sd}),
the weakest anisotropy of n.n. hopping assumes the form of
a fourth-order attractive n.n. interaction in vertical direction
(i.e. the direction of a weaker hopping). This interaction
favors n.n pairs of occupied or empty sites that are oriented vertically.
As a result, the dimeric and axial-stripe phases oriented vertically
are stabilized for any value of the anisotropy parameter $\beta_2$, while
${\mathcal{S}}_{pcb}$ and ${\mathcal{S}}_{dd}$ are replaced by
${\mathcal{S}}_{v2}^{v}$ above a critical value of $\beta_2$.
Note however, that at any higher order $2k$, $k=3,4 \ldots$
the weakest anisotropy of n.n. hopping will cause the same effects,
in the effective Hamiltonians as well as in the corresponding phase diagrams.
This implies that in the quantum systems described by $H_0$,
any anisotropy of n.n. hopping orients the dimeric and axial-stripe phases
in the direction of a weaker hopping.

\section{Summary}

We have considered two systems of correlated quantum particles,
described by extended Falicov-Kimball Hamiltonians,
with the hopping particles being either spinless fermions or hardcore bosons.
The both system have been studied in the regime of strong coupling and half-filling,
where the stability of some charge-stripe phases can be proved \cite{DJ2}.
Two main conclusions have been drawn.
Firstly, any anisotropy of nearest-neighbor hopping
orients the dimeric and axial-stripe phases in the direction of a weaker hopping.
Secondly, even a weak anisotropy of hopping reveals a tendency of
fermionic phase diagrams to become similar to the bosonic ones.

\section*{Acknowledgments}

V.D. is grateful to the University of Wroc{\l}aw for Scientific
Research Grant 2479/W/IFT, and to the Institute of Theoretical
Physics for financial support. V.D. gratefully acknowledges
the Max Born Scholarship (Wroc{\l}aw, Poland).

\section*{Appendix A}
Here we present some details pertaining to the $m$-potential method,
which is used in this paper to construct ground-state phase diagrams.

\begin{figure}[ph]
\centering \includegraphics[width=1\textwidth]{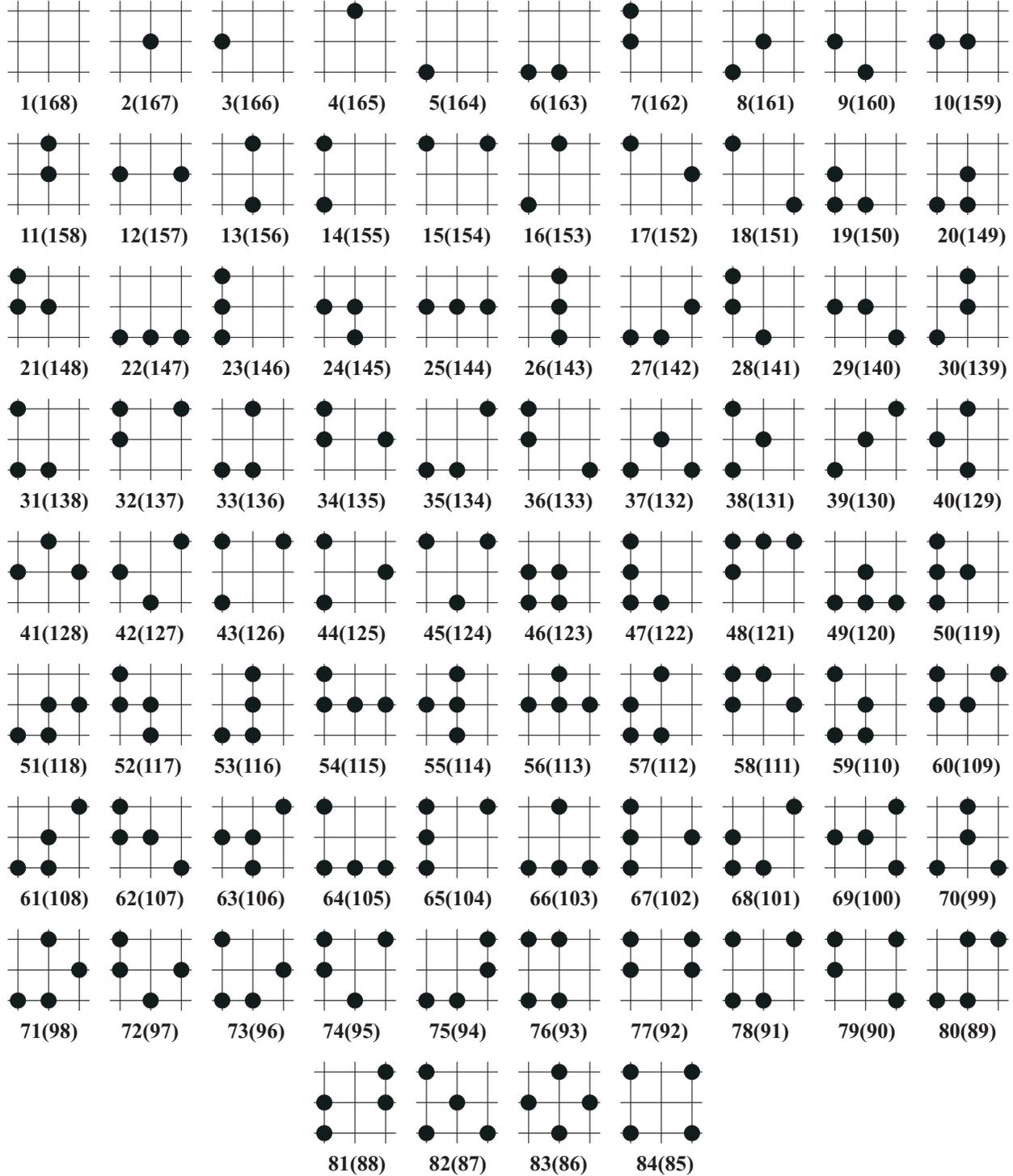}
\caption{\scriptsize{All the $T$-plaquette configurations in the
case of anisotropic interactions, up to rotations by $\pi$ and reflections
in lattice lines parallel to the axes.
The configurations that can be obtained from the displayed ones
by the hole-particle transformation are not shown,
only the numbers assigned to them are given in the brackets.}}
\label{bc168}
\end{figure}

In comparison to the isotropic case, the presence of the hopping anisotropy
lowers the symmetry of the Hamiltonians. Therefore, the zero-potentials
$K_{{\rm{T}}}^{(4)}$ and the set of $T$-plaquette configurations have to be
modified suitably. The zero-potentials, satisfying
\begin{eqnarray}
\label{zeropot} \sum\limits_{T} K_{{\rm{T}}}^{(4)} =0 ,
\end{eqnarray}
can be chosen in the form:
\begin{eqnarray}
K_{{\rm{T}}}^{(4)} = \sum\limits^{9}_{i=1} \alpha_{i}
k^{(i)}_{{\rm{T}}} ,
\end{eqnarray}
where the coefficients $\alpha_{i}$ have to be determined in the process
of constructing a phase diagram, and the potentials $k^{(i)}_{{\rm{T}}}$,
invariant with respect to the spatial symmetries of $H_0$ and fulfilling
condition (\ref{zeropot}), read:
\begin{align*}
k_{{\rm{T}}}^{(1)} & = s_{1} +s_{3} +s_{7} +s_{9} -4s_{5}, \displaybreak[0] \\
k_{{\rm{T}}}^{(2)} & = s_{2} +s_{8} -2s_{5}, \displaybreak[0] \\
k_{{\rm{T}}}^{(3)} & = s_{4} +s_{6} -2s_{5}, \displaybreak[0] \\
k_{{\rm{T}}}^{(4)} & = s_{1}s_{2} +s_{2}s_{3} +s_{7}s_{8}
+s_{8}s_{9} -2s_{4}s_{5}-2s_{5}s_{6},\displaybreak[0] \\
k_{{\rm{T}}}^{(5)} & = s_{1}s_{4} +s_{3}s_{6} +s_{4}s_{7}
+s_{6}s_{9} -2s_{2}s_{5}-2s_{5}s_{8},\displaybreak[0] \\
k_{{\rm{T}}}^{(6)} & = s_{1}s_{5} +s_{3}s_{5} +s_{5}s_{9}
+s_{5}s_{7} -s_{2}s_{4} -s_{4}s_{8} -s_{8}s_{6} -s_{2}s_{6},\displaybreak[0] \\
k_{{\rm{T}}}^{(7)} & = s_{1}s_{3} +s_{7}s_{9}-2s_{4}s_{6},\displaybreak[0] \\
k_{{\rm{T}}}^{(8)} & = s_{1}s_{7} +s_{3}s_{9}-2s_{2}s_{8},\displaybreak[0] \\
k_{{\rm{T}}}^{(9)} & = s_{1}s_{2}s_{4} +s_{6}s_{8}s_{9}
+s_{2}s_{3}s_{6} +s_{4}s_{7}s_{8} - s_{2}s_{4}s_{5} -s_{5}s_{6}s_{8}
- s_{2}s_{5}s_{6} -s_{4}s_{5}s_{8}.
\end{align*}
In the above expressions the sites of a $T$-plaquette have been
labeled $1,\ldots,9$, from left to right, starting at the bottom
left corner and ending in the upper right one. In comparison with the
isotropic case, the set of $T$-plaquette configurations is
considerably larger and consists of 168 configurations (see
Fig.~\ref{bc168}).

It turns up, that the coefficients $\alpha_{i}$ can be chosen as
affine functions of the parameters that enter linearly into the
truncated effective Hamiltonian, i.e. energy parameters and the
anisotropy parameter. Then, the energies of $T$-plaquette
configurations become affine functions of these parameters.
Consequently, the phase domains, being the solutions of a finite
number of weak inequalities between the $T$-plaquette energies,
are polyhedral convex sets in the space of energy and anisotropy
parameters. Any point of a polyhedron is a convex combination of a
finite number of points and directions (generating points and
directions) \cite{rockafellar}. The coefficients $\alpha_{i}$ determined
at the generating points and half-lines being boundaries of unbounded
domains, enable one to find their values at any other point by taking
suitable convex combinations.
This is the content of the tables presented in Appendix C.

\section*{Appendix B}

Here we provide some details referring to the presented phase diagrams:
the sets of admissible $T$-plaquette configurations,
used in constructing phase diagrams, equations of line boundaries between
the phase domains, and coordinates of the crossing points of the line boundaries.
The symbols like  ${\mathcal{S}}_{1} | {\mathcal{S}}_{2}$ stand for the line boundary
between the phases ${\mathcal{S}}_{1}$ and ${\mathcal{S}}_{2}$, etc.

\clearpage

\clearpage

\renewcommand\baselinestretch{1,5}\small\normalsize
\begin{table}[p]
\begin{center}
\small \caption{Domain boundaries of the phase diagrams for fermions,
shown in Fig.~\ref{sppd}.}
\label{tb1}
\begin{tabular}{|l|c|c|}
\hline Lines & $0<\beta_{2}\leqslant7$ & $7\leqslant \beta_{2}\leqslant10$
\\
\hline ${\mathcal{S}}_{seg} | {\mathcal{S}}_{cb}$ &
\multicolumn{2}{|c|}{$\omega=\beta_{2}+\frac{9}{2}$}
\\
\hline ${\mathcal{S}}_{seg} | {\mathcal{S}}_{dd}$ &
$\varepsilon=2\omega-2\beta_{2}+3$ & ---
\\
\hline ${\mathcal{S}}_{dd} | {\mathcal{S}}_{cb}$ &
$\varepsilon=-2\omega+2\beta_{2}+21$ & ---
\\
\hline ${\mathcal{S}}_{seg} | {\mathcal{S}}^{v}_{v1}$ &
\multicolumn{2}{|c|}{$\varepsilon=\omega+\frac{1}{2}$}
\\
\hline ${\mathcal{S}}^{v}_{d1} | {\mathcal{S}}_{cb}$ &
\multicolumn{2}{|c|}{$\varepsilon=-\omega+2\beta_{2}+\frac{19}{2}$}
\\
\hline ${\mathcal{S}}^{v}_{v1} | {\mathcal{S}}_{dd}$ &
$\omega=2\beta_{2}-\frac{5}{2}$ & ---
\\
\hline ${\mathcal{S}}_{dd} | {\mathcal{S}}^{v}_{d1}$ &
$\omega=\frac{23}{2}$ & ---
\\
\hline ${\mathcal{S}}^{v}_{v2} | {\mathcal{S}}_{dd}$ &
$\varepsilon=2\beta_{2}-6$ & ---
\\
\hline ${\mathcal{S}}^{v}_{v1} | {\mathcal{S}}^{v}_{v2}$ &
\multicolumn{2}{|c|}{$\varepsilon=\omega-\frac{7}{2}$}
\\
\hline ${\mathcal{S}}^{v}_{v2} | {\mathcal{S}}^{v}_{d1}$ &
\multicolumn{2}{|c|}{$\varepsilon=-\omega+2\beta_{2}+\frac{11}{2}$}
\\
\hline ${\mathcal{S}}^{v}_{v1} | {\mathcal{S}}^{v}_{d1}$ & --- &
$\omega=\beta_{2}+\frac{9}{2}$
\\
\hline
\end{tabular}
\hspace{10mm}
\begin{tabular}{|l|c|c|}
\hline Points & $0<\beta_{2}\leqslant7$ & $7 \leqslant \beta_{2}\leqslant10$
\\
\hline ${\bf a}$ & $\left( \beta_{2}+\frac{9}{2},12 \right)$ &
$\left( \beta_{2}+\frac{9}{2},\beta_{2}+5 \right)$
\\
\hline ${\bf b}$ & $\left( 2\beta_{2}-\frac{5}{2},2\beta_{2}-2
\right)$ & $\left( \beta_{2}+\frac{9}{2},\beta_{2}+1 \right)$
\\
\hline ${\bf c}$ & $\left( \frac{23}{2},2\beta_{2}-2 \right)$ & ---
\\
\hline ${\bf d}$ & $\left( 2\beta_{2}-\frac{5}{2},2\beta_{2}-6
\right)$ & ---
\\
\hline ${\bf e}$ & $\left( \frac{23}{2},2\beta_{2}-6 \right)$ & ---
\\
\hline
\end{tabular}
\end{center}
\end{table}

\begin{table}[ph]
\begin{center}
\small \caption{Domain boundaries of the phase diagrams for
hardcore bosons, shown in  Fig.~\ref{sppd}.}
\label{tb2}
\begin{tabular}{|l|c|c|}
\hline Lines & $0<\beta_{2}\leqslant2$ & $2<\beta_{2}\leqslant5$
\\
\hline ${\mathcal{S}}_{seg} | {\mathcal{S}}_{cb}$ &
\multicolumn{2}{|c|}{$\omega=\beta_{2}+\frac{5}{2}$}
\\
\hline ${\mathcal{S}}_{seg} | {\mathcal{S}}_{pcb}$ &
$\varepsilon=2\omega-2\beta_{2}+4$ & ---
\\
\hline ${\mathcal{S}}_{pcb} | {\mathcal{S}}_{cb}$ &
$\varepsilon=-2\omega+2\beta_{2}+14$ & ---
\\
\hline ${\mathcal{S}}_{seg} | {\mathcal{S}}^{v}_{v1}$ &
\multicolumn{2}{|c|}{$\varepsilon=\omega+\frac{9}{2}$}
\\
\hline ${\mathcal{S}}^{v}_{d1} | {\mathcal{S}}_{cb}$ &
\multicolumn{2}{|c|}{$\varepsilon=-\omega+2\beta_{2}+\frac{19}{2}$}
\\
\hline ${\mathcal{S}}^{v}_{v1} | {\mathcal{S}}_{pcb}$ &
$\omega=2\beta_{2}+\frac{1}{2}$ & ---
\\
\hline ${\mathcal{S}}_{pcb} | {\mathcal{S}}^{v}_{d1}$ &
$\omega=\frac{9}{2}$ & ---
\\
\hline ${\mathcal{S}}^{v}_{v2} | {\mathcal{S}}_{pcb}$ &
$\varepsilon=2\beta_{2}+1$ & ---
\\
\hline ${\mathcal{S}}^{v}_{v1} | {\mathcal{S}}^{v}_{v2}$ &
\multicolumn{2}{|c|}{$\varepsilon=\omega+\frac{1}{2}$}
\\
\hline ${\mathcal{S}}^{v}_{v2} | {\mathcal{S}}^{v}_{d1}$ &
\multicolumn{2}{|c|}{$\varepsilon=-\omega+2\beta_{2}+\frac{11}{2}$}
\\
\hline ${\mathcal{S}}^{v}_{v1} | {\mathcal{S}}^{v}_{d1}$ & --- &
$\omega=\beta_{2}+\frac{5}{2}$
\\
\hline
\end{tabular}
\hspace{10mm}
\begin{tabular}{|l|c|c|}
\hline Points & $0<\beta_{2}\leqslant2$ & $2<\beta_{2}\leqslant5$
\\
\hline ${\bf a}$ & $\left( \beta_{2}+\frac{5}{2},9 \right)$ &
$\left( \beta_{2}+\frac{5}{2},\beta_{2}+7 \right)$
\\
\hline ${\bf b}$ & $\left( 2\beta_{2}+\frac{1}{2},2\beta_{2}+5
\right)$ & $\left( \beta_{2}+\frac{5}{2},\beta_{2}+3 \right)$
\\
\hline ${\bf c}$ & $\left( \frac{9}{2},2\beta_{2}+5 \right)$ & ---
\\
\hline ${\bf d}$ & $\left( 2\beta_{2}+\frac{1}{2},2\beta_{2}+1
\right)$ & ---
\\
\hline ${\bf e}$ & $\left( \frac{9}{2},2\beta_{2}+1 \right)$ & ---
\\
\hline
\end{tabular}
\end{center}
\end{table}

\begin{table}[h]
\begin{center}
\scriptsize \caption{Domain boundaries of the phase diagrams
shown in  Fig.~\ref{epspdf}.}
\label{tb3}
\begin{tabular}{|l|c|c|c|c|}
\hline Lines & $0<\beta_{2}\leqslant1$ & $1<\beta_{2}\leqslant2$ &
$2<\beta_{2}\leqslant3$ & $3<\beta_{2}\leqslant6$
\\
\hline ${\mathcal{S}}_{+} | {\mathcal{S}}_{-}$ &
\multicolumn{4}{|c|}{$\delta=0$}
\\
\hline ${\mathcal{S}}_{dd} | {\mathcal{S}}_{-}$ &
\multicolumn{2}{|c|}{$\delta=-\frac{\omega}{2}+\frac{\beta_{2}}{2}-\frac{3}{4}$}
& \multicolumn{2}{|c|}{---}
\\
\hline ${\mathcal{S}}_{dd} | {\mathcal{S}}_{1}$ &
\multicolumn{2}{|c|}{
$\delta=-\frac{\omega}{6}+\frac{\beta_{2}}{6}-\frac{11}{12}$} &
\multicolumn{2}{|c|}{---}
\\
\hline ${\mathcal{S}}_{dd} | {\mathcal{S}}_{2}$ &
\multicolumn{2}{|c|}{$\delta=-\frac{3}{2}$} &
\multicolumn{2}{|c|}{---}
\\
\hline ${\mathcal{S}}_{dd} | {\mathcal{S}}_{3}$ &
\multicolumn{2}{|c|}{$\delta=\frac{\omega}{2}-\frac{\beta_{2}}{2}-\frac{21}{4}$}
& \multicolumn{2}{|c|}{---}
\\
\hline ${\mathcal{S}}_{-} | {\mathcal{S}}_{1}$ &
\multicolumn{4}{|c|}{$\delta=-\omega+\beta_{2}-\frac{1}{2}$}
\\
\hline ${\mathcal{S}}_{1} | {\mathcal{S}}_{2}$ &
\multicolumn{4}{|c|}{$\delta=-\omega+\beta_{2}+2$}
\\
\hline
\end{tabular}
\end{center}
\end{table}

\begin{table}[h]
\begin{center}
\begin{tabular}{|l|c|c|c|c|}
\hline Points & $0<\beta_{2}\leqslant1$ & $1<\beta_{2}\leqslant2$ &
$2<\beta_{2}\leqslant3$ & $3<\beta_{2}\leqslant6$
\\
\hline {\bf a} & $\left( \beta_{2}-\frac{3}{2},0 \right)$ &
\multicolumn{3}{|c|}{$\left( -\frac{1}{2},0 \right)$}
\\
\hline {\bf b} & $\left( \beta_{2}+\frac{1}{2},-1 \right)$ & $\left(
2\beta_{2}-\frac{5}{2},-\frac{\beta_{2}}{2}+\frac{1}{2} \right)$ &
$\left( 2\beta_{2}-\frac{5}{2},\frac{\beta_{2}}{2}-\frac{3}{2}
\right)$ & $\left( \frac{7}{2},0 \right)$
\\
\hline {\bf c} & $\left( \beta_{2}+\frac{7}{2},-\frac{3}{2} \right)$
& $\left( \frac{23}{2},-\frac{\beta_{2}}{2}+\frac{1}{2} \right)$ &
$\left( \frac{23}{2},\frac{\beta_{2}}{2}-\frac{3}{2} \right)$ &
$\left( \frac{3}{2},-\frac{1}{2} \right)$
\\
\hline {\bf d} & $\left( \beta_{2}+\frac{15}{2},-\frac{3}{2}
\right)$ & $\left( 2\beta_{2}+\frac{19}{2},-\beta_{2}+1 \right)$ &
$\left( \frac{3}{2},-\frac{1}{2} \right)$ & $\left(
2\beta_{2}+\frac{11}{2},-\beta_{2}+3 \right)$
\\
\hline {\bf e} & $\left( \beta_{2}+\frac{21}{2},0 \right)$ & $\left(
\frac{3}{2},\frac{\beta_{2}}{2}-\frac{3}{2} \right)$ & $\left(
4\beta_{2}-\frac{13}{2},-\frac{\beta_{2}}{2}+\frac{1}{2} \right)$ &
$\left( 2\beta_{2}-\frac{1}{2},-\beta_{2}+2 \right)$
\\
\hline {\bf f} & --- & $\left(
4\beta_{2}-\frac{5}{2},-\frac{\beta_{2}}{2}-\frac{1}{2} \right)$ &
$\left( 2\beta_{2}-\frac{5}{2},-\beta_{2}+\frac{3}{2} \right)$ &
$\left( 2\beta_{2}+\frac{7}{2},-\beta_{2}+2 \right)$
\\
\hline {\bf g} & --- & $\left( 2\beta_{2}-\frac{1}{2},-\beta_{2}
\right)$ & $\left( 2\beta_{2}+\frac{15}{2},-\beta_{2}+\frac{3}{2}
\right)$ & $\left( 2\beta_{2}-\frac{5}{2},-\beta_{2}+\frac{3}{2}
\right)$
\\
\hline
\end{tabular}
\end{center}
\end{table}
\clearpage

\begin{table}[tph]
\begin{center}
\small \caption{Domain boundaries of the phase diagrams
shown in  Fig.~\ref{epspdb}.}
\label{tb4}
\begin{tabular}{|l|c|}
\hline Lines & $0<\beta_{2}\leqslant3$
\\
\hline ${\mathcal{S}}_{-} | {\mathcal{S}}_{+}$ & $\delta=0$
\\
\hline ${\mathcal{S}}_{-} | {\mathcal{S}}^{v}_{v1}$ &
$\delta=-\frac{\omega}{4}-\frac{9}{8}$
\\
\hline ${\mathcal{S}}^{v}_{v1}|{\mathcal{S}}^{v}_{v3}$ &
$\delta=\frac{\omega}{4}+\frac{1}{8}$
\\
\hline ${\mathcal{S}}^{v}_{v3}|{\mathcal{S}}^{v}_{v2}$ &
$\delta=-\frac{\omega}{2}-\frac{1}{4}$
\\
\hline ${\mathcal{S}}_{-}|{\mathcal{S}}^{v}_{v3}$ &
$\delta=-\frac{\omega}{2}-\frac{7}{4}$
\\
\hline ${\mathcal{S}}^{v}_{v3}|{\mathcal{S}}^{v}_{d2}$ &
$\delta=\frac{\omega}{4}-\frac{3\beta_{2}}{2}-\frac{23}{8}$
\\
\hline ${\mathcal{S}}^{v}_{d2}|{\mathcal{S}}^{v}_{d3}$ &
$\delta=-\frac{3\omega}{4}+\frac{\beta_{2}}{2}+\frac{5}{8}$
\\
\hline ${\mathcal{S}}_{-}|{\mathcal{S}}^{v}_{d2}$ &
$\delta=-\frac{3\omega}{4}+\frac{\beta_{2}}{2}-\frac{11}{8}$
\\
\hline ${\mathcal{S}}^{v}_{d2}|{\mathcal{S}}_{1}$ &
$\delta=\frac{\omega}{4}-\frac{3\beta_{2}}{2}-\frac{39}{8}$
\\
\hline ${\mathcal{S}}^{v}_{v2}|{\mathcal{S}}^{v}_{d3}$ &
$\delta=\frac{\omega}{2}-2\beta_{2}-\frac{15}{4}$
\\
\hline ${\mathcal{S}}^{v}_{d3}|{\mathcal{S}}_{3}$ &
$\delta=\frac{\omega}{2}-2\beta_{2}-\frac{25}{4}$
\\
\hline ${\mathcal{S}}^{v}_{d3}|{\mathcal{S}}^{v}_{d1}$ &
$\delta=-\frac{3\omega}{4}+\frac{\beta_{2}}{2}+\frac{25}{8}$
\\
\hline ${\mathcal{S}}^{v}_{d1}|{\mathcal{S}}_{3}$ &
$\delta=-\frac{\omega}{4}-\frac{\beta_{2}}{2}-\frac{5}{8}$
\\
\hline ${\mathcal{S}}^{v}_{v2}|{\mathcal{S}}^{v}_{d1}$ &
$\omega=2\beta_{2}+\frac{11}{2}$
\\
\hline ${\mathcal{S}}^{v}_{d1}|{\mathcal{S}}_{cb}$ &
$\omega=2\beta_{2}+\frac{19}{2}$
\\
\hline ${\mathcal{S}}_{-}|{\mathcal{S}}_{1}$ &
$\delta=-\omega+\beta_{2}-\frac{1}{2}$
\\
\hline ${\mathcal{S}}_{1}|{\mathcal{S}}_{3}$ &
$\delta=-\omega+\beta_{2}+2$
\\
\hline ${\mathcal{S}}_{3}|{\mathcal{S}}_{cb}$ &
$\delta=-\omega+\beta_{2}+\frac{13}{2}$
\\
\hline
\end{tabular}
\hspace{10mm}
\begin{tabular}{|l|c|}
\hline Points & $0<\beta_{2}\leqslant3$
\\
\hline ${\bf a}$ & $\left( -\frac{9}{2},0 \right)$
\\
\hline ${\bf b}$ & $\left( -\frac{1}{2},0 \right)$
\\
\hline ${\bf c}$ & $\left( -\frac{5}{2}, -\frac{1}{2} \right)$
\\
\hline ${\bf d}$ & $\left( 2\beta_{2}+\frac{11}{2}, -\beta_{2}-1 \right)$
\\
\hline ${\bf e}$ & $\left( 2\beta_{2}+\frac{7}{2}, -\beta_{2}-2 \right)$
\\
\hline ${\bf f}$ & $\left( 2\beta_{2}+\frac{3}{2}, -\beta_{2}-\frac{5}{2}
\right)$
\\
\hline ${\bf g}$ & $\left( 2\beta_{2}+\frac{15}{2}, -\beta_{2}-\frac{5}{2}
\right)$
\\
\hline ${\bf h}$ & $\left( 2\beta_{2}+\frac{19}{2}, -\beta_{2}-3 \right)$
\\
\hline ${\bf i}$ & $\left( 2\beta_{2}+\frac{11}{2}, -\beta_{2}-7/2 \right)$
\\
\hline ${\bf j}$ & $\left( 2\beta_{2}+\frac{7}{2}, -\beta_{2}-4 \right)$
\\
\hline
\end{tabular}
\end{center}
\end{table}

\begin{table}[ph]
\begin{center}
\small \caption{Domain boundaries of the phase diagrams shown in
Fig.~\ref{p1fmnd0} and Fig.~\ref{p1hcbd0}.}
\label{tb5}
\begin{tabular}{|l|c|c|}
\hline Lines & fermions & bosons
\\
\hline ${\mathcal{S}}_{seg} | {\mathcal{S}}_{cb}$ &
$\omega=\frac{9}{2}$ & $\omega=\frac{5}{2}$
\\
\hline ${\mathcal{S}}^{v}_{v1} | {\mathcal{S}}^{v}_{d1}$ &
$\omega=\frac{9}{2}$ & $\omega=\frac{5}{2}$
\\
\hline ${\mathcal{S}}^{v}_{v1}|{\mathcal{S}}_{seg}$ &
$\varepsilon=\omega+\frac{1}{2}$ & $\varepsilon=\omega+\frac{9}{2}$
\\
\hline ${\mathcal{S}}^{v}_{d1}|{\mathcal{S}}_{cb}$ &
\multicolumn{2}{|c|}{$\varepsilon=-\omega+\frac{19}{2}$}
\\
\hline ${\mathcal{S}}^{v}_{v2}|{\mathcal{S}}^{v}_{v1}$ &
$\varepsilon=\omega-\frac{7}{2}$ & $\varepsilon=\omega+\frac{1}{2}$
\\
\hline ${\mathcal{S}}^{v}_{v2}|{\mathcal{S}}^{v}_{d1}$ &
\multicolumn{2}{|c|}{$\varepsilon=-\omega+\frac{11}{2}$}
\\
\hline
\end{tabular}
\hspace{10mm}
\begin{tabular}{|l|c|c|}
\hline Points & fermions & bosons
\\
\hline ${\bf a}$ & $\left( \frac{9}{2},5 \right)$ & $\left(
\frac{5}{2},7 \right)$
\\
\hline ${\bf b}$ & $\left( \frac{9}{2},1 \right)$ & $\left(
\frac{5}{2},3 \right)$
\\
\hline
\end{tabular}
\end{center}
\end{table}

\begin{table}[pb]
\begin{center}
\small \caption{Domain boundaries of the phase diagrams shown in
Fig.~\ref{p1fmne0} and Fig.~\ref{p1hcbe0}.}
\label{tb6}
\begin{tabular}{|l|c|c|}
\hline Lines & fermions & bosons
\\
\hline ${\mathcal{S}}_{-} | {\mathcal{S}}_{+}$ &
\multicolumn{2}{|c|}{$\delta=0$}
\\
\hline ${\mathcal{S}}^{v}_{v1} | {\mathcal{S}}_{-}$ &
$\delta=-\frac{\omega}{4}-\frac{1}{8}$ &
$\delta=-\frac{\omega}{4}-\frac{9}{8}$
\\
\hline ${\mathcal{S}}^{v}_{v1}|{\mathcal{S}}^{v}_{v3}$ &
$\delta=\frac{\omega}{4}-\frac{7}{8}$ &
$\delta=\frac{\omega}{4}+\frac{1}{8}$
\\
\hline ${\mathcal{S}}^{v}_{v3}|{\mathcal{S}}_{-}$ &
$\delta=-\frac{\omega}{2}+\frac{1}{4}$ &
$\delta=-\frac{\omega}{2}-\frac{7}{4}$
\\
\hline ${\mathcal{S}}^{v}_{v3}|{\mathcal{S}}^{v}_{v2}$ &
$\delta=-\frac{\omega}{2}+\frac{7}{4}$ &
$\delta=-\frac{\omega}{2}-\frac{1}{4}$
\\
\hline
\end{tabular}
\hspace{10mm}
\begin{tabular}{|l|c|c|}
\hline Points & fermions & bosons
\\
\hline ${\bf a}$ & $\left( -\frac{1}{2},0 \right)$ & $\left(
-\frac{9}{2},0 \right)$
\\
\hline ${\bf b}$ & $\left( \frac{7}{2},0 \right)$ & $\left(
-\frac{1}{2},0 \right)$
\\
\hline ${\bf c}$ & $\left( \frac{3}{2},-\frac{1}{2} \right)$ &
$\left( -\frac{5}{2},-\frac{1}{2} \right)$
\\
\hline
\end{tabular}
\end{center}
\end{table}

\begin{table}[ph]
\begin{center}
\small \caption{Domain boundaries of the phase diagrams shown in
Fig.~\ref{p2fmne0} and Fig.~\ref{p2hcbe0}.}
\label{tb7}
\begin{tabular}{|l|c|c|}
\hline Lines & fermions & bosons
\\
\hline ${\mathcal{S}}^{v}_{v3} | {\mathcal{S}}_{-}$ &
$\delta=-\frac{\omega}{2}+\frac{1}{4}$ &
$\delta=-\frac{\omega}{2}-\frac{7}{4}$
\\
\hline ${\mathcal{S}}^{v}_{v3} | {\mathcal{S}}^{v}_{v2}$ &
$\delta=-\frac{\omega}{2}+\frac{7}{4}$ &
$\delta=-\frac{\omega}{2}-\frac{1}{4}$
\\
\hline ${\mathcal{S}}^{v}_{v3}|{\mathcal{S}}^{v}_{d2}$ &
$\delta=\frac{\omega}{4}+\frac{17}{8}$ &
$\delta=\frac{\omega}{4}-\frac{23}{8}$
\\
\hline ${\mathcal{S}}^{v}_{d2}|{\mathcal{S}}_{-}$ &
$\delta=-\frac{3\omega}{4}-\frac{3}{8}$ &
$\delta=-\frac{3\omega}{4}-\frac{11}{8}$
\\
\hline ${\mathcal{S}}^{v}_{d2}|{\mathcal{S}}^{v}_{d4}$ &
$\delta=-\frac{3\omega}{4}+\frac{13}{8}$ & ---
\\
\hline ${\mathcal{S}}^{v}_{d2}|{\mathcal{S}}^{v}_{d3}$ & --- &
$\delta=-\frac{3\omega}{4}+\frac{5}{8}$
\\
\hline ${\mathcal{S}}^{v}_{d2}|{\mathcal{S}}_{1}$ &
$\delta=\frac{\omega}{4}+\frac{1}{8}$ &
$\delta=\frac{\omega}{4}-\frac{39}{8}$
\\
\hline
\end{tabular}
\hspace{10mm}
\begin{tabular}{|l|c|c|}
\hline Points & fermions & bosons
\\
\hline ${\bf a}$ & $\left( \frac{11}{2}, 3 \right)$ & $\left(
\frac{11}{2}, -1 \right)$
\\
\hline ${\bf b}$ & $\left( -\frac{1}{2}, 2 \right)$ & $\left(
\frac{7}{2}, -2 \right)$
\\
\hline ${\bf c}$ & $\left( \frac{7}{2}, 2 \right)$ & $\left(
\frac{3}{2}, -\frac{5}{2} \right)$
\\
\hline ${\bf d}$ & $\left( -\frac{5}{2}, \frac{3}{2} \right)$ &
$\left( \frac{15}{2}, -\frac{5}{2} \right)$
\\
\hline ${\bf e}$ & $\left( \frac{15}{2}, \frac{3}{2} \right)$ &
$\left( \frac{19}{2}, -3 \right)$
\\
\hline ${\bf f}$ & $\left( \frac{19}{2}, 1 \right)$ & $\left(
\frac{11}{2}, -\frac{7}{2} \right)$
\\
\hline ${\bf g}$ & $\left( \frac{3}{2}, \frac{1}{2} \right)$ &
$\left( \frac{7}{2}, -4 \right)$
\\
\hline ${\bf h}$ & $\left( \frac{11}{2}, \frac{1}{2} \right)$ & ---
\\
\hline ${\bf i}$ & $\left( -\frac{1}{2}, 0 \right)$ & ---
\\
\hline
\end{tabular}
\end{center}
\end{table}
\renewcommand\baselinestretch{1}\small\normalsize
\clearpage

\section*{Appendix C}

Below, in a series of tables, we provide the coefficients $\{\alpha_i \}$, $i=1,
\ldots, 9$, of the zero-potentials for the phase diagrams presented in this paper.
The coefficients that are missing in a table are equal to zero. As in Appendix B,
the symbol ${\mathcal{S}}_{1} | {\mathcal{S}}_{2}$ denotes the boundary between
the phases ${\mathcal{S}}_{1}$ and ${\mathcal{S}}_{2}$, etc.
\renewcommand\baselinestretch{1,5}\small\normalsize
\begin{table}[h]
\begin{center}
\small \caption{Zero-potentials coefficients for the phase diagram of fermions,
shown in Fig.~\ref{sppd}, for $\beta_{2}=0$.}
\label{tb8}
\begin{tabular}{|l|c|c|c|c|c|}
\hline $\{ \alpha_{i} \}$ & ${\mathcal{S}}_{seg} |
{\mathcal{S}}_{cb}$ & ${\mathcal{S}}_{seg} | {\mathcal{S}}_{v1}$ &
${\mathcal{S}}_{cb} | {\mathcal{S}}_{d1}$ & ${\mathcal{S}}_{v2} |
{\mathcal{S}}_{v1}$ & ${\mathcal{S}}_{v2} | {\mathcal{S}}_{d1}$
\\
\hline $\alpha_{1}$ & \multicolumn{5}{|c|}{$0$}
\\
\hline $\alpha_{2}$ & \multicolumn{5}{|c|}{$0$}
\\
\hline $\alpha_{3}$ & \multicolumn{5}{|c|}{$0$}
\\
\hline $\alpha_{4}$ &
\multicolumn{3}{|c|}{$-\frac{\omega}{96}+\frac{3}{64}$} &
$-\frac{\omega}{96}+\frac{7}{64}$ &
$-\frac{\omega}{96}-\frac{1}{64}$
\\
\hline $\alpha_{5}$ &
\multicolumn{3}{|c|}{$-\frac{\omega}{96}+\frac{3}{64}$} &
$-\frac{\omega}{96}+\frac{7}{64}$ &
$-\frac{\omega}{96}-\frac{1}{64}$
\\
\hline $\alpha_{6}$ & \multicolumn{3}{|c|}{$0$} &
\multicolumn{2}{|c|}{$-\frac{1}{8}$}
\\
\hline $\alpha_{7}$ & \multicolumn{3}{|c|}{$-\frac{1}{48}$} &
\multicolumn{2}{|c|}{$-\frac{1}{12}$}
\\
\hline $\alpha_{8}$ & \multicolumn{3}{|c|}{$-\frac{1}{48}$} &
\multicolumn{2}{|c|}{$-\frac{1}{12}$}
\\
\hline
\end{tabular}
\end{center}
\end{table}
\begin{table}[h]
\begin{center}
\small \caption{Zero-potentials coefficients for the phase diagram of fermions,
shown in Fig.~\ref{sppd}, for $\beta_{2}=7/2$.}
\label{tb9}
\begin{tabular}{|l|c|c|c|c|c|}
\hline $\{ \alpha_{i} \}$ & ${\mathcal{S}}_{seg} |
{\mathcal{S}}_{cb}$ & ${\mathcal{S}}_{seg} | {\mathcal{S}}^{v}_{v1}$
& ${\mathcal{S}}_{cb} | {\mathcal{S}}^{v}_{d1}$ &
${\mathcal{S}}^{v}_{v2} | {\mathcal{S}}^{v}_{v1}$ &
${\mathcal{S}}^{v}_{v2} | {\mathcal{S}}^{v}_{d1}$
\\
\hline $\alpha_{1}$ & \multicolumn{5}{|c|}{$0$}
\\
\hline $\alpha_{2}$ & \multicolumn{5}{|c|}{$0$}
\\
\hline $\alpha_{3}$ & \multicolumn{5}{|c|}{$0$}
\\
\hline $\alpha_{4}$ &
\multicolumn{5}{|c|}{$-\frac{\omega}{96}+\frac{3}{64}$}
\\
\hline $\alpha_{5}$ &
\multicolumn{5}{|c|}{$-\frac{\omega}{96}+\frac{23}{192}$}
\\
\hline $\alpha_{6}$ & \multicolumn{5}{|c|}{$0$}
\\
\hline $\alpha_{7}$ & \multicolumn{5}{|c|}{$-\frac{1}{48}$}
\\
\hline $\alpha_{8}$ & \multicolumn{5}{|c|}{$-\frac{1}{48}$}
\\
\hline
\end{tabular}
\end{center}
\end{table}

\begin{table}[ph]
\begin{center}
\small \caption{Zero-potentials coefficients for the phase diagram of fermions,
shown in Fig.~\ref{sppd}, for $\beta_{2}=7$.}
\label{tb10}
\begin{tabular}{|l|c|c|c|c|c|}
\hline $\{ \alpha_{i} \}$ & ${\mathcal{S}}_{seg} |
{\mathcal{S}}_{cb}$ & ${\mathcal{S}}_{seg} | {\mathcal{S}}^{v}_{v1}$
& ${\mathcal{S}}_{cb} | {\mathcal{S}}^{v}_{d1}$ &
${\mathcal{S}}^{v}_{v2} | {\mathcal{S}}^{v}_{v1}$ &
${\mathcal{S}}^{v}_{v2} | {\mathcal{S}}^{v}_{d1}$
\\
\hline $\alpha_{1}$ & \multicolumn{5}{|c|}{$0$}
\\
\hline $\alpha_{2}$ & \multicolumn{5}{|c|}{$0$}
\\
\hline $\alpha_{3}$ & \multicolumn{5}{|c|}{$0$}
\\
\hline $\alpha_{4}$ &
\multicolumn{3}{|c|}{$-\frac{\omega}{96}-\frac{1}{64}$} &
\multicolumn{2}{|c|}{$-\frac{\omega}{96}+\frac{3}{64}$}
\\
\hline $\alpha_{5}$ &
\multicolumn{3}{|c|}{$-\frac{\omega}{96}+\frac{49}{192}$} &
\multicolumn{2}{|c|}{$-\frac{\omega}{96}+\frac{37}{192}$}
\\
\hline $\alpha_{6}$ & \multicolumn{3}{|c|}{$\frac{1}{8}$} &
\multicolumn{2}{|c|}{$0$}
\\
\hline $\alpha_{7}$ & \multicolumn{3}{|c|}{$-\frac{1}{12}$} &
\multicolumn{2}{|c|}{$-\frac{1}{48}$}
\\
\hline $\alpha_{8}$ & \multicolumn{3}{|c|}{$-\frac{1}{12}$} &
\multicolumn{2}{|c|}{$-\frac{1}{48}$}
\\
\hline
\end{tabular}
\end{center}
\end{table}

\begin{table}[ph]
\begin{center}
\small \caption{Zero-potentials coefficients for the phase diagram of fermions,
shown in Fig.~\ref{sppd}, for $\beta_{2}=10$.}
\label{tb11}
\begin{tabular}{|l|c|c|c|c|c|}
\hline $\{ \alpha_{i} \}$ & ${\mathcal{S}}_{seg} |
{\mathcal{S}}_{cb}$ & ${\mathcal{S}}_{seg} | {\mathcal{S}}^{v}_{v1}$
& ${\mathcal{S}}_{cb} | {\mathcal{S}}^{v}_{d1}$ &
${\mathcal{S}}^{v}_{v2} | {\mathcal{S}}^{v}_{v1}$ &
${\mathcal{S}}^{v}_{v2} | {\mathcal{S}}^{v}_{d1}$
\\
\hline $\alpha_{1}$ & \multicolumn{5}{|c|}{$0$}
\\
\hline $\alpha_{2}$ & \multicolumn{5}{|c|}{$0$}
\\
\hline $\alpha_{3}$ & \multicolumn{5}{|c|}{$0$}
\\
\hline $\alpha_{4}$ & \multicolumn{3}{|c|}{$-\frac{1}{6}$} &
\multicolumn{2}{|c|}{$-\frac{\omega}{96}+\frac{3}{64}$}
\\
\hline $\alpha_{5}$ & \multicolumn{3}{|c|}{$\frac{1}{6}$} &
\multicolumn{2}{|c|}{$-\frac{\omega}{96}+\frac{49}{192}$}
\\
\hline $\alpha_{6}$ & \multicolumn{3}{|c|}{$\frac{1}{8}$} &
\multicolumn{2}{|c|}{$0$}
\\
\hline $\alpha_{7}$ & \multicolumn{3}{|c|}{$-\frac{1}{12}$} &
\multicolumn{2}{|c|}{$-\frac{1}{48}$}
\\
\hline $\alpha_{8}$ & \multicolumn{3}{|c|}{$-\frac{1}{12}$} &
\multicolumn{2}{|c|}{$-\frac{1}{48}$}
\\
\hline
\end{tabular}
\end{center}
\end{table}

\begin{table}[ph]
\begin{center}
\small \caption{Zero-potentials coefficients for the phase diagram of hardcore
bosons, shown in Fig.~\ref{sppd}, for $\beta_{2}=0$.}
\label{tb12}
\begin{tabular}{|l|c|c|c|c|c|}
\hline $\{ \alpha_{i} \}$ & ${\mathcal{S}}_{seg} |
{\mathcal{S}}_{cb}$ & ${\mathcal{S}}_{seg} | {\mathcal{S}}_{v1}$ &
${\mathcal{S}}_{cb} | {\mathcal{S}}_{d1}$ & ${\mathcal{S}}_{v2} |
{\mathcal{S}}_{v1}$ & ${\mathcal{S}}_{v2} | {\mathcal{S}}_{d1}$
\\
\hline $\alpha_{1}$ & \multicolumn{5}{|c|}{$0$}
\\
\hline $\alpha_{2}$ & \multicolumn{5}{|c|}{$0$}
\\
\hline $\alpha_{3}$ & \multicolumn{5}{|c|}{$0$}
\\
\hline $\alpha_{4}$ &
\multicolumn{3}{|c|}{$-\frac{\omega}{96}+\frac{5}{192}$} &
$-\frac{\omega}{96}+\frac{17}{192}$ &
$-\frac{\omega}{96}-\frac{7}{192}$
\\
\hline $\alpha_{5}$ &
\multicolumn{3}{|c|}{$-\frac{\omega}{96}+\frac{5}{192}$} &
$-\frac{\omega}{96}+\frac{17}{192}$ &
$-\frac{\omega}{96}-\frac{7}{192}$
\\
\hline $\alpha_{6}$ & \multicolumn{3}{|c|}{$0$} &
\multicolumn{2}{|c|}{$-\frac{1}{8}$}
\\
\hline $\alpha_{7}$ & \multicolumn{3}{|c|}{$-\frac{1}{48}$} &
\multicolumn{2}{|c|}{$-\frac{1}{12}$}
\\
\hline $\alpha_{8}$ & \multicolumn{3}{|c|}{$-\frac{1}{48}$} &
\multicolumn{2}{|c|}{$-\frac{1}{12}$}
\\
\hline
\end{tabular}
\end{center}
\end{table}

\begin{table}[ph]
\begin{center}
\small \caption{Zero-potentials coefficients for the phase diagram of hardcore
bosons, shown in Fig.~\ref{sppd}, for $\beta_{2}=1$.}
\label{tb13}
\begin{tabular}{|l|c|c|c|c|c|}
\hline $\{ \alpha_{i} \}$ & ${\mathcal{S}}_{seg} |
{\mathcal{S}}_{cb}$ & ${\mathcal{S}}_{seg} | {\mathcal{S}}^{v}_{v1}$
& ${\mathcal{S}}_{cb} | {\mathcal{S}}^{v}_{d1}$ &
${\mathcal{S}}^{v}_{v2} | {\mathcal{S}}^{v}_{v1}$ &
${\mathcal{S}}^{v}_{v2} | {\mathcal{S}}^{v}_{d1}$
\\
\hline $\alpha_{1}$ & \multicolumn{5}{|c|}{$0$}
\\
\hline $\alpha_{2}$ & \multicolumn{5}{|c|}{$0$}
\\
\hline $\alpha_{3}$ & \multicolumn{5}{|c|}{$0$}
\\
\hline $\alpha_{4}$ & $-\frac{1}{24}$ &
\multicolumn{4}{|c|}{$-\frac{\omega}{96}+\frac{5}{192}$}
\\
\hline $\alpha_{5}$ & $\frac{1}{24}$ &
\multicolumn{4}{|c|}{$-\frac{\omega}{96}+\frac{3}{64}$}
\\
\hline $\alpha_{6}$ & \multicolumn{5}{|c|}{$0$}
\\
\hline $\alpha_{7}$ & \multicolumn{5}{|c|}{$-\frac{1}{48}$}
\\
\hline $\alpha_{8}$ & \multicolumn{5}{|c|}{$-\frac{1}{48}$}
\\
\hline
\end{tabular}
\end{center}
\end{table}
\clearpage

\begin{table}[ph]
\begin{center}
\small \caption{Zero-potentials coefficients for the phase diagram of hardcore
bosons, shown in Fig.~\ref{sppd}, for $\beta_{2}=2$.}
\label{tb14}
\begin{tabular}{|l|c|c|c|c|c|}
\hline $\{ \alpha_{i} \}$ & ${\mathcal{S}}_{seg} |
{\mathcal{S}}_{cb}$ & ${\mathcal{S}}_{seg} | {\mathcal{S}}^{v}_{v1}$
& ${\mathcal{S}}_{cb} | {\mathcal{S}}^{v}_{d1}$ &
${\mathcal{S}}^{v}_{v2} | {\mathcal{S}}^{v}_{v1}$ &
${\mathcal{S}}^{v}_{v2} | {\mathcal{S}}^{v}_{d1}$
\\
\hline $\alpha_{1}$ & \multicolumn{5}{|c|}{$0$}
\\
\hline $\alpha_{2}$ & \multicolumn{5}{|c|}{$0$}
\\
\hline $\alpha_{3}$ & \multicolumn{5}{|c|}{$0$}
\\
\hline $\alpha_{4}$ & $-\frac{1}{12}$ &
\multicolumn{2}{|c|}{$-\frac{\omega}{96}-\frac{1}{192}$} &
\multicolumn{2}{|c|}{$-\frac{\omega}{96}+\frac{1}{24}$}
\\
\hline $\alpha_{5}$ & $\frac{1}{12}$ &
\multicolumn{4}{|c|}{$-\frac{\omega}{96}+\frac{19}{192}$}
\\
\hline $\alpha_{6}$ & $\frac{1}{8}$ &
\multicolumn{2}{|c|}{$\frac{1}{32}$} &
\multicolumn{2}{|c|}{$-\frac{1}{32}$}
\\
\hline $\alpha_{7}$ & $-\frac{1}{12}$ &
\multicolumn{2}{|c|}{$-\frac{1}{48}$} &
\multicolumn{2}{|c|}{$-\frac{1}{192}$}
\\
\hline $\alpha_{8}$ & $-\frac{1}{12}$ &
\multicolumn{2}{|c|}{$-\frac{1}{48}$} &
\multicolumn{2}{|c|}{$-\frac{5}{96}$}
\\
\hline
\end{tabular}
\end{center}
\end{table}

\begin{table}[ph]
\begin{center}
\small \caption{Zero-potentials coefficients for the phase diagram of hardcore
bosons, shown in Fig.~\ref{sppd}, for $\beta_{2}=5$.}
\label{tb15}
\begin{tabular}{|l|c|c|c|c|c|}
\hline $\{ \alpha_{i} \}$ & ${\mathcal{S}}_{seg} |
{\mathcal{S}}_{cb}$ & ${\mathcal{S}}_{seg} | {\mathcal{S}}^{v}_{v1}$
& ${\mathcal{S}}_{cb} | {\mathcal{S}}^{v}_{d1}$ &
${\mathcal{S}}^{v}_{v2} | {\mathcal{S}}^{v}_{v1}$ &
${\mathcal{S}}^{v}_{v2} | {\mathcal{S}}^{v}_{d1}$
\\
\hline $\alpha_{1}$ & \multicolumn{5}{|c|}{$0$}
\\
\hline $\alpha_{2}$ & \multicolumn{5}{|c|}{$0$}
\\
\hline $\alpha_{3}$ & \multicolumn{5}{|c|}{$0$}
\\
\hline $\alpha_{4}$ & \multicolumn{3}{|c|}{$-\frac{11}{96}$} &
\multicolumn{2}{|c|}{$-\frac{\omega}{96}+\frac{1}{24}$}
\\
\hline $\alpha_{5}$ & \multicolumn{3}{|c|}{$\frac{11}{96}$} &
\multicolumn{2}{|c|}{$-\frac{\omega}{96}+\frac{19}{192}$}
\\
\hline $\alpha_{6}$ & \multicolumn{3}{|c|}{$\frac{1}{8}$} &
\multicolumn{2}{|c|}{$-\frac{1}{32}$}
\\
\hline $\alpha_{7}$ & \multicolumn{3}{|c|}{$-\frac{1}{12}$} &
\multicolumn{2}{|c|}{$-\frac{1}{192}$}
\\
\hline $\alpha_{8}$ & \multicolumn{3}{|c|}{$-\frac{1}{12}$} &
\multicolumn{2}{|c|}{$-\frac{5}{96}$}
\\
\hline
\end{tabular}
\end{center}
\end{table}

\begin{table}[ph]
\begin{center}
\small \caption{Zero-potentials coefficients for the phase diagram
shown in Fig.~\ref{epspdf}, for $\beta_{2}=0$.}
\label{tb16}
\begin{tabular}{|l|c|c|c|c|c|c|}
\hline $\{ \alpha_{i} \}$ & ${\mathcal{S}}_{+} | {\mathcal{S}}_{-}$
& ${\mathcal{S}}_{-} | {\mathcal{S}}_{1}$ & ${\mathcal{S}}_{1} |
{\mathcal{S}}_{2}$ & ${\mathcal{S}}_{2} | {\mathcal{S}}_{3}$ &
${\mathcal{S}}_{3} | {\mathcal{S}}_{cb}$ & $\delta=0$,
$\omega\geqslant\frac{21}{2}$
\\
\hline $\alpha_{1}$ & $0$ & $\frac{\omega}{16}+\frac{3}{32}$ &
$\frac{\omega}{16}-\frac{3}{32}$ & $\frac{\omega}{16}-\frac{9}{32}$
& $\frac{\omega}{16}-\frac{21}{32}$ & $0$
\\
\hline $\alpha_{2}$ & $0$ & $\frac{\omega}{8}+\frac{1}{16}$ &
$\frac{\omega}{8}-\frac{3}{16}$ & $\frac{\omega}{8}-\frac{3}{4}$ &
$\frac{\omega}{8}-\frac{21}{16}$ & $0$
\\
\hline $\alpha_{3}$ & $0$ & $\frac{\omega}{8}+\frac{1}{16}$ &
$\frac{\omega}{8}-\frac{3}{16}$ & $\frac{\omega}{8}-\frac{3}{4}$ &
$\frac{\omega}{8}-\frac{21}{16}$ & $0$
\\
\hline $\alpha_{4}$ &
\multicolumn{3}{|c|}{$-\frac{\omega}{96}+\frac{3}{64}$} &
$-\frac{\omega}{96}+\frac{1}{16}$ &
\multicolumn{2}{|c|}{$-\frac{\omega}{96}+\frac{3}{64}$}
\\
\hline $\alpha_{5}$ &
\multicolumn{3}{|c|}{$-\frac{\omega}{96}+\frac{3}{64}$} &
$-\frac{\omega}{96}+\frac{1}{16}$ &
\multicolumn{2}{|c|}{$-\frac{\omega}{96}+\frac{3}{64}$}
\\
\hline $\alpha_{6}$ & \multicolumn{2}{|c|}{$0$} & $-\frac{1}{16}$ &
$-\frac{1}{32}$ & \multicolumn{2}{|c|}{$0$}
\\
\hline $\alpha_{7}$ & \multicolumn{3}{|c|}{$-\frac{1}{48}$} &
$\frac{1}{96}$ & \multicolumn{2}{|c|}{$-\frac{1}{48}$}
\\
\hline $\alpha_{8}$ & \multicolumn{3}{|c|}{$-\frac{1}{48}$} &
$\frac{1}{96}$ & \multicolumn{2}{|c|}{$-\frac{1}{48}$}
\\
\hline
\end{tabular}
\end{center}
\end{table}

\begin{table}[h]
\begin{center}
\small \caption{Zero-potentials coefficients for the phase diagram
shown in Fig.~\ref{epspdf}, for $\beta_{2}=1$.}
\label{tb17}
\begin{tabular}{|l|c|c|c|c|c|c|}
\hline $\{ \alpha_{i} \}$ & ${\mathcal{S}}_{+} | {\mathcal{S}}_{-}$
& ${\mathcal{S}}_{-} | {\mathcal{S}}_{1}$ & ${\mathcal{S}}_{1} |
{\mathcal{S}}_{2}$ & ${\mathcal{S}}_{2} | {\mathcal{S}}_{3}$ &
${\mathcal{S}}_{3} | {\mathcal{S}}_{cb}$ & $\delta=0$,
$\omega\geqslant\frac{23}{2}$
\\
\hline $\alpha_{1}$ & $0$ & $\frac{\omega}{16}+\frac{1}{32}$ &
$\frac{\omega}{16}-\frac{3}{32}$ & $\frac{\omega}{16}-\frac{11}{32}$
& $\frac{\omega}{16}-\frac{23}{32}$ & $0$
\\
\hline $\alpha_{2}$ & $0$ & $\frac{\omega}{8}-\frac{1}{16}$ &
$\frac{\omega}{8}-\frac{11}{32}$ & $\frac{\omega}{8}-\frac{41}{48}$
& $\frac{\omega}{8}-\frac{23}{16}$ & $0$
\\
\hline $\alpha_{3}$ & $0$ & $\frac{\omega}{8}-\frac{1}{16}$ &
$\frac{\omega}{8}-\frac{13}{32}$ & $\frac{\omega}{8}-\frac{43}{48}$
& $\frac{\omega}{8}-\frac{23}{16}$ & $0$
\\
\hline $\alpha_{4}$ &
\multicolumn{2}{|c|}{$-\frac{\omega}{96}+\frac{3}{64}$} &
$-\frac{\omega}{96}+\frac{9}{128}$ &
$-\frac{\omega}{96}+\frac{13}{192}$ &
\multicolumn{2}{|c|}{$-\frac{\omega}{96}+\frac{3}{64}$}
\\
\hline $\alpha_{5}$ &
\multicolumn{2}{|c|}{$-\frac{\omega}{96}+\frac{13}{192}$} &
$-\frac{\omega}{96}+\frac{29}{384}$ &
$-\frac{\omega}{96}+\frac{5}{64}$ &
\multicolumn{2}{|c|}{$-\frac{\omega}{96}+\frac{13}{192}$}
\\
\hline $\alpha_{6}$ & \multicolumn{2}{|c|}{$0$} &
\multicolumn{2}{|c|}{$-\frac{1}{32}$} & \multicolumn{2}{|c|}{$0$}
\\
\hline $\alpha_{7}$ & \multicolumn{2}{|c|}{$-\frac{1}{48}$} &
$\frac{1}{96}$ & $-\frac{1}{192}$ &
\multicolumn{2}{|c|}{$-\frac{1}{48}$}
\\
\hline $\alpha_{8}$ & \multicolumn{2}{|c|}{$-\frac{1}{48}$} &
$\frac{1}{96}$ & $\frac{5}{192}$ &
\multicolumn{2}{|c|}{$-\frac{1}{48}$}
\\
\hline
\end{tabular}
\end{center}
\end{table}

\begin{table}[h]
\begin{center}
\small \caption{Zero-potentials coefficients for the phase diagram
shown in Fig.~\ref{epspdf}, for $\beta_{2}=3/2$.}
\label{tb18}
\begin{tabular}{|l|c|c|c|c|c|c|c|c|c|}
\hline lines \& points & $\alpha_{1}$ & $\alpha_{2}$ & $\alpha_{3}$
& $\alpha_{4}$ & $\alpha_{5}$ & $\alpha_{6}$ & $\alpha_{7}$ &
$\alpha_{8}$ & $\alpha_{9}$
\\
\hline ${\mathcal{S}}_{+} | {\mathcal{S}}_{-}$ & $0$ & $0$ & $0$ &
$-\frac{\omega}{96}+\frac{3}{64}$ &
$-\frac{\omega}{96}+\frac{5}{64}$ & $\frac{1}{64}$ &
$-\frac{7}{192}$ & $-\frac{1}{48}$ & 0
\\
\hline ${\mathcal{S}}_{-} | {\mathcal{S}}_{1}$ & $\frac{\omega}{16}$
& $\frac{\omega}{8}-\frac{1}{8}$ & $\frac{\omega}{8}-\frac{3}{32}$ &
$-\frac{\omega}{96}+\frac{15}{256}$ &
$-\frac{\omega}{96}+\frac{5}{64}$ & $-\frac{1}{256}$ &
$-\frac{17}{384}$ & $-\frac{1}{192}$ & 0
\\
\hline ${\mathcal{S}}_{1} | {\mathcal{S}}_{2}$ &
$\frac{\omega}{16}-\frac{49}{256}$ &
$\frac{\omega}{8}-\frac{31}{64}$ & $\frac{\omega}{8}-\frac{13}{32}$
& $-\frac{\omega}{96}+\frac{7}{128}$ &
$-\frac{\omega}{96}+\frac{27}{256}$ & $-\frac{7}{256}$ &
$-\frac{23}{384}$ & $-\frac{23}{384}$ & $\frac{19}{256}$
\\
\hline ${\mathcal{S}}_{2} | {\mathcal{S}}_{3}$ &
$\frac{\omega}{16}-\frac{105}{256}$ &
$\frac{\omega}{8}-\frac{115}{128}$ &
$\frac{\omega}{8}-\frac{29}{32}$ &
$-\frac{\omega}{96}+\frac{7}{128}$ &
$-\frac{\omega}{96}+\frac{43}{512}$ & $-\frac{25}{512}$ &
$-\frac{157}{3072}$ &  $\frac{113}{3072}$ & $0$
\\
\hline ${\mathcal{S}}_{3} | {\mathcal{S}}_{cb}$ &
$\frac{\omega}{16}-\frac{3}{4}$ & $\frac{\omega}{8}-\frac{3}{2}$ &
$\frac{\omega}{8}-\frac{3}{2}$ &  $-\frac{\omega}{96}+\frac{3}{64}$
& $-\frac{\omega}{96}+\frac{5}{64}$ & $0$ & $-\frac{1}{48}$ &
$-\frac{1}{48}$ & $0$
\\
\hline $\delta=0$, $\omega\geqslant\frac{25}{2}$ & $0$ & $0$ & $0$ &
$-\frac{\omega}{96}+\frac{3}{64}$ &
$-\frac{\omega}{96}+\frac{5}{64}$ & $\frac{1}{64}$ & $-\frac{1}{48}$
& $-\frac{7}{192}$ & $0$
\\
\hline $\left( \frac{1}{2},0 \right)$ & $0$ & $0$ & $0$ &
$\frac{1}{24}$ & $\frac{7}{96}$ & $0$ & $-\frac{1}{48}$ &
$-\frac{1}{48}$ & $0$
\\
\hline
\end{tabular}
\end{center}
\end{table}

\begin{table}[h]
\begin{center}
\small \caption{Zero-potentials coefficients for the phase diagram
shown in Fig.~\ref{epspdf}, for $\beta_{2}=2$.}
\label{tb19}
\begin{tabular}{|l|c|c|c|c|c|c|c|c|c|}
\hline lines \& points & $\alpha_{1}$ & $\alpha_{2}$ & $\alpha_{3}$
& $\alpha_{4}$ & $\alpha_{5}$ & $\alpha_{6}$ & $\alpha_{7}$ &
$\alpha_{8}$ & $\alpha_{9}$
\\
\hline ${\mathcal{S}}_{+} | {\mathcal{S}}_{-}$ & $0$ & $0$ & $0$ &
$-\frac{\omega}{96}+\frac{3}{64}$ &
$-\frac{\omega}{96}+\frac{17}{192}$ & $\frac{1}{32}$ &
$-\frac{5}{96}$ & $-\frac{1}{48}$ & 0
\\
\hline ${\mathcal{S}}_{-} | {\mathcal{S}}_{1}$ &
$\frac{\omega}{16}-\frac{1}{32}$ & $\frac{\omega}{8}-\frac{3}{16}$ &
$\frac{\omega}{8}-\frac{3}{16}$ & $-\frac{\omega}{96}+\frac{3}{64}$
& $-\frac{\omega}{96}+\frac{17}{192}$ & $0$ & $-\frac{1}{48}$ &
$-\frac{1}{48}$ & 0
\\
\hline ${\mathcal{S}}_{1} | {\mathcal{S}}_{2}$ &
$\frac{\omega}{16}-\frac{1}{4}$ & $\frac{\omega}{8}-\frac{9}{16}$ &
$\frac{\omega}{8}-\frac{7}{16}$ & $-\frac{\omega}{96}+\frac{3}{64}$
& $-\frac{\omega}{96}+\frac{23}{192}$ & $-\frac{1}{32}$ &
$-\frac{1}{12}$ & $-\frac{1}{12}$ & $\frac{3}{32}$
\\
\hline ${\mathcal{S}}_{2} | {\mathcal{S}}_{3}$ &
$\frac{\omega}{16}-\frac{15}{32}$ & $\frac{\omega}{8}-\frac{15}{16}$
& $\frac{\omega}{8}-\frac{15}{16}$ &
$-\frac{\omega}{96}+\frac{3}{64}$ &
$-\frac{\omega}{96}+\frac{17}{192}$ & $-\frac{1}{16}$ &
$-\frac{1}{12}$ &  $\frac{1}{24}$ & $0$
\\
\hline ${\mathcal{S}}_{3} | {\mathcal{S}}_{cb}$ &
$\frac{\omega}{16}-\frac{25}{32}$ & $\frac{\omega}{8}-\frac{25}{16}$
& $\frac{\omega}{8}-\frac{25}{16}$ &
$-\frac{\omega}{96}+\frac{3}{64}$ &
$-\frac{\omega}{96}+\frac{17}{192}$ & $0$ & $-\frac{1}{48}$ &
$-\frac{1}{48}$ & $0$
\\
\hline $\delta=0$, $\omega\geqslant\frac{27}{2}$ & $0$ & $0$ & $0$ &
$-\frac{\omega}{96}+\frac{3}{64}$ &
$-\frac{\omega}{96}+\frac{17}{192}$ & $\frac{1}{32}$ &
$-\frac{1}{48}$ & $-\frac{5}{96}$ & $0$
\\
\hline $\left( \frac{3}{2},0 \right)$ & $0$ & $0$ & $0$ &
$\frac{1}{32}$ & $\frac{7}{96}$ & $0$ & $-\frac{1}{48}$ &
$-\frac{1}{48}$ & $0$
\\
\hline
\end{tabular}
\end{center}
\end{table}

\begin{table}[h]
\begin{center}
\small \caption{Zero-potentials coefficients for the phase diagram
shown in Fig.~\ref{epspdf}, for $\beta_{2}=5/2$.}
\label{tb20}
\begin{tabular}{|l|c|c|c|c|c|c|c|c|c|}
\hline lines \& points & $\alpha_{1}$ & $\alpha_{2}$ & $\alpha_{3}$
& $\alpha_{4}$ & $\alpha_{5}$ & $\alpha_{6}$ & $\alpha_{7}$ &
$\alpha_{8}$ & $\alpha_{9}$
\\
\hline ${\mathcal{S}}_{+} | {\mathcal{S}}_{-}$ & $0$ & $0$ & $0$ &
$-\frac{\omega}{96}+\frac{3}{64}$ &
$-\frac{\omega}{96}+\frac{19}{192}$ & $\frac{1}{32}$ &
$-\frac{5}{96}$ & $-\frac{1}{48}$ & 0
\\
\hline ${\mathcal{S}}_{-} | {\mathcal{S}}_{1}$ &
$\frac{\omega}{16}-\frac{1}{16}$ & $\frac{\omega}{8}-\frac{1}{4}$ &
$\frac{\omega}{8}-\frac{1}{4}$ & $-\frac{\omega}{96}+\frac{3}{64}$ &
$-\frac{\omega}{96}+\frac{19}{192}$ & $0$ & $-\frac{1}{48}$ &
$-\frac{1}{48}$ & 0
\\
\hline ${\mathcal{S}}_{1} | {\mathcal{S}}_{2}$ &
$\frac{\omega}{16}-\frac{9}{32}$ & $\frac{\omega}{8}-\frac{5}{8}$ &
$\frac{\omega}{8}-\frac{1}{2}$ & $-\frac{\omega}{96}+\frac{3}{64}$ &
$-\frac{\omega}{96}+\frac{25}{192}$ & $-\frac{1}{32}$ &
$-\frac{1}{12}$ & $-\frac{1}{12}$ & $\frac{3}{32}$
\\
\hline ${\mathcal{S}}_{2} | {\mathcal{S}}_{3}$ &
$\frac{\omega}{16}-\frac{33}{64}$ & $\frac{\omega}{8}-\frac{31}{32}$
& $\frac{\omega}{8}-1$ & $-\frac{\omega}{96}+\frac{3}{64}$ &
$-\frac{\omega}{96}+\frac{35}{384}$ & $-\frac{9}{128}$ &
$-\frac{1}{12}$ &  $\frac{5}{192}$ & $0$
\\
\hline ${\mathcal{S}}_{3} | {\mathcal{S}}_{cb}$ &
$\frac{\omega}{16}-\frac{13}{16}$ & $\frac{\omega}{8}-\frac{13}{8}$
& $\frac{\omega}{8}-\frac{13}{8}$ &
$-\frac{\omega}{96}+\frac{3}{64}$ &
$-\frac{\omega}{96}+\frac{19}{192}$ & $0$ & $-\frac{1}{48}$ &
$-\frac{1}{48}$ & $0$
\\
\hline $\delta=0$, $\omega\geqslant\frac{29}{2}$ & $0$ & $0$ & $0$ &
$-\frac{\omega}{96}+\frac{3}{64}$ &
$-\frac{\omega}{96}+\frac{19}{192}$ & $\frac{1}{32}$ &
$-\frac{1}{48}$ & $-\frac{5}{96}$ & $0$
\\
\hline $\left( \frac{5}{2},0 \right)$ & $0$ & $0$ & $0$ &
$\frac{1}{48}$ & $\frac{7}{96}$ & $0$ & $-\frac{1}{48}$ &
$-\frac{1}{48}$ & $0$
\\
\hline
\end{tabular}
\end{center}
\end{table}

\begin{table}[h]
\begin{center}
\small \caption{Zero-potentials coefficients for the phase diagram
shown in Fig.~\ref{epspdf}, for $\beta_{2}=3$.}
\label{tb21}
\begin{tabular}{|l|c|c|c|c|c|c|c|c|}
\hline lines \& points & $\alpha_{1}$ & $\alpha_{2}$ & $\alpha_{3}$
& $\alpha_{4}$ & $\alpha_{5}$ & $\alpha_{6}$ & $\alpha_{7}$ &
$\alpha_{8}$
\\
\hline ${\mathcal{S}}_{+} | {\mathcal{S}}_{-}$ & $0$ & $0$ & $0$ &
$-\frac{\omega}{96}+\frac{3}{64}$ &
$-\frac{\omega}{96}+\frac{7}{64}$ & $\frac{1}{32}$ & $-\frac{5}{96}$
& $-\frac{1}{48}$
\\
\hline ${\mathcal{S}}_{-} | {\mathcal{S}}_{1}$ &
$\frac{\omega}{16}-\frac{3}{32}$ & $\frac{\omega}{8}-\frac{5}{16}$ &
$\frac{\omega}{8}-\frac{5}{16}$ & $-\frac{\omega}{96}+\frac{3}{64}$
& $-\frac{\omega}{96}+\frac{7}{64}$ & $0$ & $-\frac{1}{48}$ &
$-\frac{1}{48}$
\\
\hline ${\mathcal{S}}_{1} | {\mathcal{S}}_{2}$ &
$\frac{\omega}{16}-\frac{11}{32}$ & $\frac{\omega}{8}-\frac{7}{16}$
& $\frac{\omega}{8}-\frac{11}{16}$ &
$-\frac{\omega}{96}+\frac{3}{64}$ &
$-\frac{\omega}{96}+\frac{3}{64}$ & $-\frac{1}{16}$ &
$-\frac{1}{12}$ & $-\frac{1}{12}$
\\
\hline ${\mathcal{S}}_{2} | {\mathcal{S}}_{3}$ &
$\frac{\omega}{16}-\frac{9}{16}$ & $\frac{\omega}{8}-1$ &
$\frac{\omega}{8}-\frac{17}{16}$ & $-\frac{\omega}{96}+\frac{3}{64}$
& $-\frac{\omega}{96}+\frac{3}{32}$ & $-\frac{5}{64}$ &
$-\frac{1}{12}$ &  $\frac{1}{96}$
\\
\hline ${\mathcal{S}}_{3} | {\mathcal{S}}_{cb}$ &
$\frac{\omega}{16}-\frac{27}{32}$ & $\frac{\omega}{8}-\frac{27}{16}$
& $\frac{\omega}{8}-\frac{27}{16}$ &
$-\frac{\omega}{96}+\frac{3}{64}$ &
$-\frac{\omega}{96}+\frac{7}{64}$ & $0$ & $-\frac{1}{48}$ &
$-\frac{1}{48}$
\\
\hline $\delta=0$, $\omega\geqslant\frac{31}{2}$ & $0$ & $0$ & $0$ &
$-\frac{\omega}{96}+\frac{3}{64}$ &
$-\frac{\omega}{96}+\frac{7}{64}$ & $\frac{1}{32}$ & $-\frac{1}{48}$
& $-\frac{5}{96}$
\\
\hline $\left( \frac{7}{2},0 \right)$ (${\bf{b}}$) & $0$ & $0$ & $0$
& $\frac{1}{96}$ & $\frac{7}{96}$ & $0$ & $-\frac{1}{48}$ &
$-\frac{1}{48}$
\\
\hline
\end{tabular}
\end{center}
\end{table}

\begin{table}[h]
\begin{center}
\small \caption{Zero-potentials coefficients for the phase diagram
shown in Fig.~\ref{epspdf}, for $\beta_{2}=6$.}
\label{tb22}
\begin{tabular}{|l|c|c|c|c|c|c|c|c|}
\hline lines \& points & $\alpha_{1}$ & $\alpha_{2}$ & $\alpha_{3}$
& $\alpha_{4}$ & $\alpha_{5}$ & $\alpha_{6}$ & $\alpha_{7}$ &
$\alpha_{8}$
\\
\hline ${\mathcal{S}}_{+} | {\mathcal{S}}_{-}$ & $0$ & $0$ & $0$ &
$-\frac{\omega}{96}+\frac{3}{64}$ &
$-\frac{\omega}{96}+\frac{11}{64}$ & $\frac{1}{32}$ &
$-\frac{5}{96}$ & $-\frac{1}{48}$
\\
\hline ${\mathcal{S}}_{-} | {\mathcal{S}}_{1}$ &
$\frac{\omega}{16}-\frac{9}{32}$ & $\frac{\omega}{8}-\frac{11}{16}$
& $\frac{\omega}{8}-\frac{11}{16}$ &
$-\frac{\omega}{96}+\frac{3}{64}$ &
$-\frac{\omega}{96}+\frac{11}{64}$ & $0$ & $-\frac{1}{48}$ &
$-\frac{1}{48}$
\\
\hline ${\mathcal{S}}_{1} | {\mathcal{S}}_{2}$ &
$\frac{\omega}{16}-\frac{19}{32}$ & $\frac{\omega}{8}-\frac{11}{16}$
& $\frac{\omega}{8}-\frac{15}{16}$ &
$-\frac{\omega}{96}+\frac{3}{64}$ &
$-\frac{\omega}{96}+\frac{7}{64}$ & $-\frac{1}{8}$ & $-\frac{1}{12}$
& $-\frac{1}{12}$
\\
\hline ${\mathcal{S}}_{2} | {\mathcal{S}}_{3}$ &
$\frac{\omega}{16}-\frac{25}{32}$ & $\frac{\omega}{8}-\frac{21}{16}$
& $\frac{\omega}{8}-\frac{23}{16}$ &
$-\frac{\omega}{96}+\frac{3}{64}$ &
$-\frac{\omega}{96}+\frac{9}{64}$ & $-\frac{3}{32}$ &
$-\frac{1}{12}$ &  $-\frac{1}{48}$
\\
\hline ${\mathcal{S}}_{3} | {\mathcal{S}}_{cb}$ &
$\frac{\omega}{16}-\frac{33}{32}$ & $\frac{\omega}{8}-\frac{33}{16}$
& $\frac{\omega}{8}-\frac{33}{16}$ &
$-\frac{\omega}{96}+\frac{3}{64}$ &
$-\frac{\omega}{96}+\frac{11}{64}$ & $0$ & $-\frac{1}{48}$ &
$-\frac{1}{48}$
\\
\hline $\delta=0$, $\omega\geqslant\frac{43}{2}$ & $0$ & $0$ & $0$ &
$-\frac{\omega}{96}+\frac{3}{64}$ &
$-\frac{\omega}{96}+\frac{11}{64}$ & $\frac{1}{32}$ &
$-\frac{1}{48}$ & $-\frac{5}{96}$
\\
\hline ${\bf{b}}$ & $0$ & $0$ & $0$ & $\frac{1}{96}$ &
$\frac{13}{96}$ & $-\frac{1}{32}$ & $-\frac{5}{96}$ &
$-\frac{1}{48}$
\\
\hline
\end{tabular}
\end{center}
\end{table}

\begin{table}[h]
\begin{center}
\small \caption{Zero-potentials coefficients for the phase diagram
shown in Fig.~\ref{epspdb}, for $\beta_{2}=0$.}
\label{tb23}
\begin{tabular}{|l|c|c|c|c|c|c|c|c|}
\hline lines \& points & $\alpha_{1}$ & $\alpha_{2}$ & $\alpha_{3}$
& $\alpha_{4}$ & $\alpha_{5}$ & $\alpha_{6}$ & $\alpha_{7}$ &
$\alpha_{8}$
\\
\hline ${\mathcal{S}}_{+} | {\mathcal{S}}_{-}$ & $0$ & $0$ & $0$ &
$-\frac{\omega}{96}+\frac{5}{192}$ &
$-\frac{\omega}{96}+\frac{5}{192}$ & $0$ & $-\frac{1}{48}$ &
$-\frac{1}{48}$
\\
\hline ${\mathcal{S}}_{-} | {\mathcal{S}}_{1}$ &
$\frac{\omega}{16}+\frac{3}{32}$ & $\frac{\omega}{8}+\frac{1}{16}$ &
$\frac{\omega}{8}+\frac{1}{16}$ & $-\frac{\omega}{96}+\frac{5}{192}$
& $-\frac{\omega}{96}+\frac{5}{192}$ & $0$ & $-\frac{1}{48}$ &
$-\frac{1}{48}$
\\
\hline ${\mathcal{S}}_{1} | {\mathcal{S}}_{3}$ &
$\frac{\omega}{16}-\frac{9}{32}$ & $\frac{\omega}{8}-\frac{1}{16}$ &
$\frac{\omega}{8}-\frac{1}{16}$ & $-\frac{\omega}{96}-\frac{7}{192}$
& $-\frac{\omega}{96}-\frac{7}{192}$ & $-\frac{1}{8}$ &
$-\frac{1}{12}$ &  $-\frac{1}{12}$
\\
\hline ${\mathcal{S}}_{3} | {\mathcal{S}}_{cb}$ &
$\frac{\omega}{16}-\frac{13}{32}$ & $\frac{\omega}{8}-\frac{13}{16}$
& $\frac{\omega}{8}-\frac{13}{16}$ &
$-\frac{\omega}{96}+\frac{5}{192}$ &
$-\frac{\omega}{96}+\frac{5}{192}$ & $0$ & $-\frac{1}{48}$ &
$-\frac{1}{48}$
\\
\hline $\delta=0$, $\omega\geqslant\frac{19}{2}$ & $0$ & $0$ & $0$ &
$-\frac{\omega}{96}+\frac{5}{192}$ &
$-\frac{\omega}{96}+\frac{5}{192}$ & $0$ & $-\frac{1}{48}$ &
$-\frac{1}{48}$
\\
\hline ${\bf{b}}$ & $0$ & $0$ & $0$ & $\frac{3}{32}$ &
$\frac{3}{32}$ & $-\frac{1}{8}$ & $-\frac{1}{12}$ & $-\frac{1}{12}$
\\
\hline $\left( \frac{11}{2},0 \right)$ & $0$ & $0$ & $0$ &
$-\frac{3}{32}$ & $-\frac{3}{32}$ & $-\frac{1}{8}$ & $-\frac{1}{12}$
& $-\frac{1}{12}$
\\
\hline
\end{tabular}
\end{center}
\end{table}

\begin{table}[h]
\begin{center}
\small \caption{Zero-potentials coefficients for the phase diagram
shown in Fig.~\ref{epspdb}, for $\beta_{2}=3$.}
\label{tb24}
\begin{tabular}{|l|c|c|c|c|c|c|c|c|}
\hline lines \& points & $\alpha_{1}$ & $\alpha_{2}$ & $\alpha_{3}$
& $\alpha_{4}$ & $\alpha_{5}$ & $\alpha_{6}$ & $\alpha_{7}$ &
$\alpha_{8}$
\\
\hline ${\mathcal{S}}_{+} | {\mathcal{S}}_{-}$ & $0$ & $0$ & $0$ &
$-\frac{\omega}{96}+\frac{5}{192}$ &
$-\frac{\omega}{96}+\frac{17}{192}$ & $\frac{1}{32}$ &
$-\frac{5}{96}$ & $-\frac{1}{48}$
\\
\hline ${\mathcal{S}}_{-} | {\mathcal{S}}_{1}$ &
$\frac{\omega}{16}-\frac{3}{32}$ & $\frac{\omega}{8}-\frac{5}{16}$ &
$\frac{\omega}{8}-\frac{5}{16}$ & $-\frac{\omega}{96}+\frac{5}{192}$
& $-\frac{\omega}{96}+\frac{17}{192}$ & $0$ & $-\frac{1}{48}$ &
$-\frac{1}{48}$
\\
\hline ${\mathcal{S}}_{1} | {\mathcal{S}}_{3}$ &
$\frac{\omega}{16}-\frac{15}{32}$ & $\frac{\omega}{8}-\frac{7}{16}$
& $\frac{\omega}{8}-\frac{7}{16}$ &
$-\frac{\omega}{96}-\frac{7}{192}$ &
$-\frac{\omega}{96}+\frac{5}{192}$ & $-\frac{1}{8}$ &
$-\frac{1}{12}$ &  $-\frac{1}{12}$
\\
\hline ${\mathcal{S}}_{3} | {\mathcal{S}}_{cb}$ &
$\frac{\omega}{16}-\frac{19}{32}$ & $\frac{\omega}{8}-\frac{19}{16}$
& $\frac{\omega}{8}-\frac{19}{16}$ &
$-\frac{\omega}{96}+\frac{5}{192}$ &
$-\frac{\omega}{96}+\frac{17}{192}$ & $0$ & $-\frac{1}{48}$ &
$-\frac{1}{48}$
\\
\hline $\delta=0$, $\omega\geqslant\frac{31}{2}$ & $0$ & $0$ & $0$ &
$-\frac{\omega}{96}+\frac{5}{192}$ &
$-\frac{\omega}{96}+\frac{17}{192}$ & $\frac{1}{32}$ &
$-\frac{1}{48}$ & $-\frac{5}{96}$
\\
\hline ${\bf{b}}$ & $0$ & $0$ & $0$ & $\frac{1}{32}$ &
$\frac{3}{32}$ & $-\frac{1}{32}$ & $-\frac{5}{96}$ & $-\frac{1}{48}$
\\
\hline $\left( \frac{23}{2},0 \right)$ & $0$ & $0$ & $0$ &
$-\frac{3}{32}$ & $-\frac{1}{32}$ & $-\frac{1}{32}$ &
$-\frac{1}{48}$ & $-\frac{5}{96}$
\\
\hline
\end{tabular}
\end{center}
\end{table}

\begin{table}[h]
\begin{center}
\small \caption{Zero-potentials coefficients for the phase diagram
shown in Fig.~\ref{p1fmne0}.}
\label{tb25}
\begin{tabular}{|l|c|}
\hline lines & $\alpha_{1}$
\\
\hline ${\mathcal{S}}_{+} | {\mathcal{S}}_{-}$ & $0$
\\
\hline ${\mathcal{S}}_{-} | {\mathcal{S}}_{v3}^{v}$ &
$\frac{\omega}{16}+\frac{1}{32}$
\\
\hline ${\mathcal{S}}_{v3}^{v} | {\mathcal{S}}_{v2}^{v}$ &
$\frac{\omega}{16}-\frac{7}{32}$
\\
\hline $\delta=0$, $\omega\geqslant\frac{7}{2}$ & $0$
\\
\hline
\end{tabular}
\end{center}
\end{table}

\begin{table}[h]
\begin{center}
\small \caption{Zero-potentials coefficients for the phase diagram
shown in  Fig.~\ref{p1hcbe0}.}
\label{tb26}
\begin{tabular}{|l|c|}
\hline lines & $\alpha_{1}$
\\
\hline ${\mathcal{S}}_{+} | {\mathcal{S}}_{-}$ & $0$
\\
\hline ${\mathcal{S}}_{-} | {\mathcal{S}}_{v3}^{v}$ &
$\frac{\omega}{16}+\frac{9}{32}$
\\
\hline ${\mathcal{S}}_{v3}^{v} | {\mathcal{S}}_{v2}^{v}$ &
$\frac{\omega}{16}+\frac{1}{32}$
\\
\hline $\delta=0$, $\omega\geqslant-\frac{1}{2}$ & $0$
\\
\hline
\end{tabular}
\end{center}
\end{table}

\begin{table}[h]
\begin{center}
\small \caption{Zero-potentials coefficients for the phase diagram
shown in Fig.~\ref{p2fmne0}.}
\label{tb27}
\begin{tabular}{|l|c|c|c|c|c|c|c|c|}
\hline lines \& points & $\alpha_{1}$ & $\alpha_{2}$ & $\alpha_{3}$
& $\alpha_{4}$ & $\alpha_{5}$ & $\alpha_{6}$ & $\alpha_{7}$ &
$\alpha_{8}$
\\
\hline ${\mathcal{S}}_{-} | {\mathcal{S}}_{v3}^{v}$ & $0$ & $0$ &
$\frac{\omega}{8}+\frac{1}{16}$ & $-\frac{\omega}{24}+\frac{1}{16}$
& $-\frac{\omega}{96}+\frac{3}{64}$ & $-\frac{1}{32}$ &
$-\frac{1}{48}$ & $-\frac{1}{48}$
\\
\hline ${\mathcal{S}}_{-} | {\mathcal{S}}_{1}$ & $0$ &
$-\frac{1}{8}$ & $\frac{\omega}{4}+\frac{1}{4}$ &
$-\frac{7\omega}{96}-\frac{3}{64}$ &
$-\frac{\omega}{96}+\frac{3}{64}$ & $0$ & $-\frac{1}{48}$ &
$-\frac{1}{48}$
\\
\hline ${\mathcal{S}}_{1} | {\mathcal{S}}_{2}$ &
$\frac{7\omega}{96}-\frac{17}{64}$ &
$\frac{7\omega}{48}-\frac{1}{32}$ &
$\frac{5\omega}{48}-\frac{3}{32}$ & $0$ &
$-\frac{\omega}{96}-\frac{1}{64}$ & $-\frac{1}{8}$ & $-\frac{1}{12}$
& $-\frac{1}{12}$
\\
\hline ${\mathcal{S}}_{2} | {\mathcal{S}}_{3}$ &
$\frac{7\omega}{96}-\frac{29}{64}$ &
$\frac{7\omega}{48}-\frac{21}{32}$ &
$\frac{5\omega}{48}-\frac{19}{32}$ & $0$ &
$-\frac{\omega}{96}+\frac{1}{64}$ & $-\frac{3}{32}$ &
$-\frac{1}{12}$ & $-\frac{1}{48}$
\\
\hline ${\mathcal{S}}_{3} | {\mathcal{S}}_{cb}$ & $0$ & $0$ &
$\frac{\omega}{4}-\frac{21}{8}$ &
$-\frac{7\omega}{96}+\frac{45}{64}$ &
$-\frac{\omega}{96}+\frac{3}{64}$ & $0$ & $-\frac{1}{48}$ &
$-\frac{1}{48}$
\\
\hline ${\mathcal{S}}_{d1}^{v} | {\mathcal{S}}_{cb}$ & $0$ & $0$ &
$-\frac{\delta}{4}$ & $\frac{\delta}{16}-\frac{5}{96}$ &
$-\frac{5}{96}$ & $0$ & $-\frac{1}{48}$ & $-\frac{1}{48}$
\\
\hline ${\mathcal{S}}_{v2}^{v} | {\mathcal{S}}_{d1}^{v}$ & $0$ & $0$
& $-\frac{\delta}{4}$ & $\frac{\delta}{16}-\frac{1}{96}$ &
$-\frac{5}{192}$ & $-\frac{1}{64}$ & $-\frac{1}{48}$ &
$-\frac{1}{48}$
\\
\hline
\end{tabular}
\end{center}
\end{table}
\begin{table}[h]
\begin{center}
\small \caption{Zero-potentials coefficients for the phase diagram
shown in Fig.~\ref{p2hcbe0}.}
\label{tb28}
\begin{tabular}{|l|c|c|c|c|c|c|c|c|}
\hline lines \& points & $\alpha_{1}$ & $\alpha_{2}$ & $\alpha_{3}$
& $\alpha_{4}$ & $\alpha_{5}$ & $\alpha_{6}$ & $\alpha_{7}$ &
$\alpha_{8}$
\\
\hline ${\mathcal{S}}_{-} | {\mathcal{S}}_{v3}^{v}$ & $0$ &
$-\frac{1}{4}$ & $\frac{\omega}{8}+\frac{9}{16}$ &
$-\frac{\omega}{24}-\frac{7}{48}$ &
$-\frac{\omega}{96}+\frac{5}{192}$ & $\frac{1}{32}$ &
$-\frac{1}{48}$ & $-\frac{1}{48}$
\\
\hline ${\mathcal{S}}_{-} | {\mathcal{S}}_{1}$ & $0$ &
$-\frac{1}{8}$ & $\frac{\omega}{4}+\frac{1}{4}$ &
$-\frac{7\omega}{96}-\frac{13}{192}$ &
$-\frac{\omega}{96}+\frac{5}{192}$ & $0$ & $-\frac{1}{48}$ &
$-\frac{1}{48}$
\\
\hline ${\mathcal{S}}_{1} | {\mathcal{S}}_{3}$ & $-\frac{1}{4}$ &
$0$ & $\frac{\omega}{4}-\frac{1}{8}$ &
$-\frac{7\omega}{96}-\frac{1}{192}$ &
$-\frac{\omega}{96}-\frac{7}{192}$ & $-\frac{1}{8}$ &
$-\frac{1}{12}$ & $-\frac{1}{12}$
\\
\hline ${\mathcal{S}}_{3} | {\mathcal{S}}_{cb}$ & $0$ & $0$ &
$\frac{\omega}{4}-\frac{13}{8}$ &
$-\frac{7\omega}{96}+\frac{83}{192}$ &
$-\frac{\omega}{96}+\frac{5}{192}$ & $0$ & $-\frac{1}{48}$ &
$-\frac{1}{48}$
\\
\hline ${\mathcal{S}}_{d1}^{v} | {\mathcal{S}}_{cb}$ & $0$ & $0$ &
$-\frac{\delta}{4}$ & $\frac{\delta}{16}-\frac{7}{96}$ &
$-\frac{7}{96}$ & $0$ & $-\frac{1}{48}$ & $-\frac{1}{48}$
\\
\hline ${\mathcal{S}}_{v2}^{v} | {\mathcal{S}}_{d1}^{v}$ & $0$ & $0$
& $-\frac{\delta}{4}$ & $\frac{\delta}{16}-\frac{1}{32}$ &
$-\frac{3}{64}$ & $-\frac{1}{64}$ & $-\frac{1}{48}$ &
$-\frac{1}{48}$
\\
\hline ${\mathcal{S}}_{v3}^{v} | {\mathcal{S}}_{v2}^{v}$ &
$-\frac{1}{8}$ & $0$ & $\frac{\omega}{8}+\frac{5}{16}$ &
$-\frac{\omega}{24}-\frac{5}{96}$ &
$-\frac{\omega}{96}+\frac{5}{192}$ & $-\frac{1}{8}$ &
$-\frac{1}{12}$ & $-\frac{1}{12}$
\\
\hline {\bf{d}} & $0$ & $0$ & $\frac{5}{8}$ & $-\frac{23}{96}$ &
$-\frac{13}{192}$ & $-\frac{1}{64}$ & $-\frac{1}{48}$ &
$-\frac{1}{48}$
\\
\hline
\end{tabular}
\end{center}
\end{table}
\renewcommand\baselinestretch{1}\small\normalsize

\end{document}